\documentclass[showpacs,aps,twocolumn]{revtex4}
\usepackage{graphicx}
\usepackage{amsmath,amssymb,amsfonts}
\usepackage{array}
\usepackage{url}
\usepackage{hyperref}
\usepackage{multirow}
\usepackage{float}
\usepackage{lineno}
\usepackage{xspace}
\usepackage{ulem}
\begin{document}

\title{X-ray emission spectrum for axion-photon conversion in magnetospheres of strongly magnetized neutron stars}
\author{Shubham Yadav}
\email{shubhamphy28@gmail.com}

\author{M.Mishra}
\email{madhukar@pilani.bits-pilani.ac.in}

\author{Tapomoy Guha Sarkar}
\email{tapomoy@pilani.bits-pilani.ac.in}

\affiliation{Department of Physics, Birla Institute of Technology and Science, Pilani, Pilani - 333031, Rajasthan, India}

\begin{abstract}
Detecting axionic dark matter (DM) could be possible in an X-ray spectrum from strongly magnetized neutron stars (NSs). We examine the possibility of axion-photon conversion in the magnetospheres of strongly magnetized NSs. In the current work, we investigate how the modified Tolman Oppenheimer Volkoff (TOV) system of equations (in the presence of a magnetic field) affects the energy spectrum of axions and axions-converted-photon flux. We have considered the distance-dependent magnetic field in the modified TOV system of equations. We employ three different equations of states (EoSs), namely APR, FPS, and SLY, to solve these equations. We obtain the axions emission rate by including the Cooper-pair-breaking formation process and Bremsstrahlung process in the core of NSs using the NSCool code. We primarily focus on Magnificient seven (M7) star RXJ 1856.5-3754. We further investigate the impact of the magnetic field on the actual observables, such as axion energy spectrum and axion-converted-photon flux at an axion mass in meV range by assuming mass $M_{NS} \sim 1.4M_{\odot}$.  We compare our calculated axion-converted-photon flux from all available archival data sets from PN+MOS+Chandra. We also study the variation of the energy spectrum at a fixed energy with varying central magnetic fields. Our predicted axion-converted-photon flux values as a function of axion energy closely follow the experimentally archival data, which allows us to put bounds on the axion mass for the three EoS.

\end{abstract}
\date{\today}
\maketitle

\section{Introduction}
\label{intro}
Compact objects such as NSs serve as an excellent laboratory for the search of light pseudoscalar particles like QCD axions and axion-like particles (ALPs)~\cite{potekhin2019thermal,potekhin2012thermonuclear,beznogov2016diffusive,potekhin2014atmospheres,1998nspt.conf..183P,Buschmann:2021juv,Buschmann:2019icd,chadha2022axion,DINE1983137,1968ApJ...153..807H,2021EPJC...81..698P,Psaltis_2014,Sedrakian:2006mq,gudmundsson1983structure,Zavlin_2007,BEZNOGOV20211}. At the end of a massive star's life, a core-collapse supernova explosion occurs. The remnant core collapses into tiny and superdense NSs. The cores of the NSs are relatively hotter, with temperatures between $1-10$ keV and surface temperature at $T\sim 0.1$ keV~\cite{Buschmann:2019pfp}. 
Recent searches~\cite{DEXHEIMER2017487,SINHA201343,Dexheimer:2017fhy,2013arXiv1307.7450G,10.1093/mnras/stu2706,Lattimer_2001,PhysRevD.32.3172,dexheimer2012hybrid,potekhin2007heat,potekhin2005magnetic} have reported new information about the potency of magnetic fields inside and on the surface of these compact astronomical objects.
The NSs that have an unusually high magnetic field are commonly called magnetars~\cite{potekhin2015neutron,Lopes_2015,2013arXiv1307.7450G,10.1093/mnras/stu2706,refId0,Kaminker_2006,Valyavin2014SuppressionOC,doi:10.1146/annurev-astro-081915-023329,Gomes_2017,refId0,braithwaite2004fossil}. Such strong magnetic fields can decay at relatively shorter times~\cite{bocquet1995rotating}. For such stars, the contributions from the toroidal fields can be much more potent than the poloidal fields~\cite{10.1111/j.1365-2966.2008.14034.x,braithwaite2006stable}.
NSs cool primarily through the neutrino cooling mechanism~\cite{Buschmann:2019pfp,Potekhin_2003,Kolomeitsev_2008,Leinson_2000,10.1093/mnras/sty776,PhysRevD.51.348}. There is continuous transport of heat from the interiors towards the surface of NSs.
Neutrino cooling predominates for an initial thousand years, and then eventually, photon emission surpasses neutrinos during the later ages of NSs. Along with neutrinos, axions may also be emitted from the cores of NSs~\cite{2001PhR...354....1Y,PhysRevD.34.843,nomoto1987cooling}.
Axions are hypothetical, pseudoscalar elementary particles proposed to solve specific strong CP problem~\cite{Peccei:1977hh,Peccei:1977ur,umeda1998axion,PhysRevLett.43.103,SHIFMAN1980493,Weinberg:1977ma,DINE1981199,PhysRevLett.40.279,PhysRevLett.123.021801,Dessert_2022} in particle physics and also considered as a compelling candidate for the search of the mysterious DM~\cite{Buschmann:2021juv,Duffy_2009,ABBOTT1983133,PRESKILL1983127,PhysRevLett.117.141801,PhysRevD.97.123006,Marsh_2017,PhysRevC.98.035802,beznogov2020thermal,yakovlev2003thermal,adams2022axion,backes2021quantum,chen2022phenomenology,arza2019photon,song2024polarization,tjemsland2024adiabatic}. These axions may be produced inside the cores of NSs thermally by Cooper pair breaking formation and N-N Bremsstrahlung process.
Furthermore, in the presence of strong magnetic fields surrounding the NSs, these produced axions (in the core) may be converted into X-rays due to their weak interaction with the matter. Such a conversion mechanism has been proposed as a possible explanation for hard X-ray excess from nearby NSs~\cite{Buschmann:2021juv}.
 
The QCD axion and axion-like particles (ALPs) are anticipated to pair derivatively to fermionic matter and electromagnetism, thus enabling their production inside the heated cores of NSs.
The effective Lagrangian of axions is given by: 
\begin{equation}
\mathcal L= -\frac{1}{2} \partial_{\mu}a\partial^{\mu}a+\mathcal L_{int}(\partial_{\mu} a,\psi_{f}).
\end{equation}
The second term in the expression denotes axion coupling to a fermionic field $\psi_{f}$ ~\cite{Buschmann:2019pfp}. According to the recent experimental results of the axionic DM experiment (ADMX)~\cite{PhysRevLett.120.151301,PhysRevLett.120.151301,PhysRevLett.124.101303,PhysRevLett.127.261803} and Large-scale microwave cavity search~\cite{PhysRevD.64.092003}, axions can solve the problem of mysterious DM.
Currently, there are two possible axion models, namely the Dean-Fischler-Srednitsky-Zhitnitsky - DFSZ model, which contains an additional coupling of axions with the charged leptons, and the Kim-Shifman-Weinstein-Zakharov (hadronic) - KSVZ model which is related to interaction of axions with photons and hadrons~\cite{Leinson_2019,Zhitnitsky:1980tq}. 
The production rate of axions depends on their emissivity, critical temperature profiles of the NSs, and proton and neutron Fermi momenta profiles.
%------------------------------------------------------------
There is an extensive literature on the axion mass bounds: plasma haloscopes and MADMAX ~\cite{brun2019new,PhysRevLett.123.141802} reveals an axion mass between $m_{a}\sim$ $40–400$ $\mu$eV, experimental data from HAYSTAC and ADMX reports an axion mass  $m_{a}\sim 1–100$ $\mu$eV, and CASPEr~\cite{PhysRevX.4.021030,PhysRevLett.126.141802,Garcon_2018}, DM-Radio~\cite{7750582} and ABRACADABRA~\cite{PhysRevLett.117.141801,PhysRevLett.122.121802,PhysRevLett.127.081801}, confirms that mass of axion cannot be in $\mu$eV range ($m_{a} \ll \mu$eV). 
The experimental data of ADMX (in context to Dean–Fischler–Srednitsky–
Zhitnitsky (DFSZ)) model presents an axion mass range $m_{a} \sim$ 2.7 $-$ 4.2 $\mu$eV~\cite{PhysRevLett.120.151301,PhysRevLett.124.101303,PhysRevLett.127.261803,DINE1981199,Zhitnitsky:1980tq}. Nevertheless, the black hole superradiance~\cite{PhysRevD.81.123530,PhysRevD.83.044026} reports mass of QCD axion as $m_{a}>2\times 10^{-11}$ eV.
Buschmann et al.~\cite{Buschmann:2021juv} (about Kim–Shifman–Weinstein–Zakharov (KSVZ) axion model), restricts the mass of axion at $m_{a}\leqslant 16$ meV. For the KSVZ model~\cite{PhysRevLett.43.103,SHIFMAN1980493} the experimental results from
the HAYSTAC~\cite{backes2021quantum} and ADMX~\cite{PhysRevD.64.092003} indicate that the range of axion mass is still unconfirmed within the parameter range of about ten powers of magnitude. Exploring the mass of axionic DM (that is, not in the meV range) in a laboratory is a bit challenging~\cite{Armengaud_2019,Liu_2022,Co_2020,10.21468/SciPostPhys.10.2.050,arvanitaki2014resonantly,carosi2020microwave}. Recent study~\cite{PhysRevD.107.103017,PhysRevD.109.023001,yadav2024thermal}, reports the axion mass bound as $m_{a}\leqslant$ $10$ meV.
%------------------------------------------------------------
 
Various studies describe the potential use of resonant conversion of axions, particularly in the  $\mu$eV–meV mass range, relevant to QCD (Quantum Chromodynamics) axionic DM. Resonant conversion refers to axion-photon conversion under plasma densities and intense magnetic fields in the magnetosphere of strongly magnetized NSs. The axion mass is crucial in determining the frequency of the associated photon signal. These features have drawn researchers' attention to conduct studies on axion-photons conversion in the NSs magnetosphere~\cite{Buschmann:2019pfp,hook2018radio,PhysRevD.37.1237}. They have been actively characterizing and searching for potential X-ray and radio signals. This area of research holds promise for advancing our understanding of DM and theories related to fundamental particle physics, provided that the predicted signals can be detected and verified through observational data. It has also been revealed in the literature~\cite{DEXHEIMER2017487,Dexheimer:2017fhy,Lopes_2015,2013arXiv1307.7450G,bocquet1995rotating,PhysRevLett.120.182701,Jonker_2007,Heinke_2009} that the distribution of a strong magnetic field affects both internal structure as well as the cooling rates of NSs. \\
%%%%%%%%%%%%%%%%%%%%%%%%%%%%%%%%%%%%%%%%%%%%%%%%%%%%%%%%%%%%%%%
Witte et al.~\cite{witte2021axion} performed a calculation to compute the radio signal from the NS magnetosphere by adopting a charge-symmetric Goldreich-Julian model. They accomplish this by employing a cutting-edge auto-differentiation ray tracing technique, which enables them to closely monitor photon absorption, energy exchanges with the surrounding plasma, axion-photon dephasing, and the properties of reflections and refraction. A. Hook et al.~\cite{hook2018radio} reported that axions can be detected through the narrow radio lines from axionic DM conversion in the NS magnetosphere. They solved the axion-photon mixing equations by including magnetized plasma. The conversion can be possible when plasma frequency matches the mass of axions. 
McDonald et al.~\cite{McDonald_2023} derived the essential transport equations in magnetized plasmas. Due to the resonant conversion of axions to photons, these equations describe the creation and propagation of photons in any 3-D medium. Millar et al.~\cite{millar2021axion} shows the axion-photon conversion in a highly magnetized anisotropic material in 3-D compared to the simplistic 1-D calculation mentioned in the literature. The results show orders of magnitude difference in conversion for the 3-D medium compared to 1-D. The findings will significantly impact the radio signal seen through a telescope.
%%%%%%%%%%%%%%%%%%%%%%%%%%%%%%%%%%%%%%%%%%%%%%%%%%%%%%%%%%%%%%%%%%%%%%%%%%%%%%%%%%%%%%%%%%
\noindent In the current work, we investigate how the modified Tolman Oppenheimer Volkoff (TOV) system of equations (in the presence of magnetic field)~\cite{PhysRev.55.364,PhysRev.55.374} affects the axions energy spectrum and axion-converted-photon flux at an axion mass in meV range by assuming mass $M_{NS} \sim 1.4M_{\odot}$ of the non-rotating NSs. Initially, we solved the modified TOV system of equations by employing three EoSs, namely APR, FPS, and SLY, and the distance-dependent magnetic field to build the profiles. We change the NSCool code by incorporating the N-N Bremsstrahlung process and the Cooper-pair-breaking formation process to determine the axion emission rate inside the core of NSs. In particular, our fiducial analysis includes Magnificient seven star, RXJ 1856.5-3754 (M7) with the characteristic age of $3.7\times 10^{6}$ yrs~\cite{Mignani_2013,10.1111/j.1365-2966.2011.19302.x}.   

Furthermore, we investigate the influence of the strong magnetic field on the fundamental observables, such as the axion-converted-photon flux and energy spectrum of axions. We also compare our computed axion-converted-photon flux with archival available data from PN+MOS+Chandra~\cite{Dessert_2020,Buschmann:2019pfp} by taking axion-photon coupling in the magnetosphere of NSs. We also obtained the axion mass bounds from the three mentioned EoSs and compared them with the existing bounds in the literature.
%%%%%%%%%%%%%%%%%%%%%%%%%%%%%%%%%%%%%%%%%%%%%%%%%%%%%%%%%%%%%
\noindent The paper is organized as follows. In Section~\ref{intro}, we begin with a brief Introduction. In Section~\ref{form}, we describe the modified TOV system of equations, Equation of states, Neutron star cooling, and the axion emission rates by Cooper-Breaking and formation and Bremsstrahlung process in the core of NSs, Energy Spectrum of Axions, Conversion Probability and Axion-converted-Photon Flux. In Section~\ref{result}, we explore the impact of magnetic fields on various observables, such as the energy spectrum of axions and
axion-converted-photon flux for different NSs at axion mass in the meV range for three different EoSs. Finally, Section~\ref{conc} summarizes and concludes the work.
 
\section{Formalism}
\label{form}
\subsection{Modified TOV equations for non-rotating neutron star}
The present work relies on the modified TOV system of equations (in the presence of a strong magnetic field)~\cite{chatterjee2019magnetic}.As discussed in~\cite{chatterjee2019magnetic,10.1093/mnras/stu2706,dexheimer2017self}, this would require us to solve the Einstein's field equations in the stellar interior including the full electromagnetic Energy-Momentum tensor~\cite{bonazzola1993axisymmetric}.
Perturbative solutions to study the neutron star's deformation developed in \cite{konno1999deformation} shall also be valid only for low magnetic fields. In this work, we adopt a more phenomenological approach to replicate the effects of the magnetic field (at least qualitatively). This is done by including a magnetic energy density to the usual fluid energy density in the TOV equations. Further, the conservation of the Energy Momentum tensor in the general relativistic framework~\cite{bonazzola1993axisymmetric} motivates us to add a Lorentz force term in the pressure gradient equation. This Lorentz force term, which imprints the effect of the magnetic field is also motivated by the anisotropic term in the pressure gradient equation studied in~\cite{bowers1974anisotropic}. To examine the effects of magnetic fields, we require the suitable EoS~\cite{10.1093/mnras/stu2706,PRAKASH19971,Lattimer_2001,beznogov2015statistical}, mass, pressure, and baryon density as a function of distance (profiles) and the magnetic field profile.   The TOV system of equations for mass, gravitational potential, and hydrostatic equilibrium, respectively, are given by: 
\begin{equation}
\frac{dm}{dr}=4\pi r^{2}\left (\epsilon +\frac{B^{2}}{2\mu _{0}c^2} \right )
\end{equation}
\begin{equation}
\frac{d\phi }{dr}=\frac{G\left ( m(r)+  4 \pi r ^3  P/c^{2} \right )}{r\left ( r\,c^2-2G\,m(r)\right )}
\end{equation}
\begin{equation}
\frac{dP}{dr}=-c^{2} \left ( \epsilon +\frac{B^2}{2\mu _{0}c^2}+\frac{P}{c^{2}} \right )\left ( \frac{d\phi }{dr}-{\Lambda}\left ( r \right ) \right ),
\end{equation}
where $\Lambda(r)$ corresponds to a Lorentz force term ~\cite{bonazzola1993axisymmetric}. 
Instead of solving the full Maxwell-Einstein system of equations, we have adopted the phenomenological fit function 
for $\Lambda(r)$ from \cite{chatterjee2019magnetic}. This parametrized fit function depends on  the star's mean radius and central magnetic field $B_{c}$, given as:
\begin{equation}
\frac{{\Lambda}\left ( r \right )}{ 10^{-41}} = B_{c}^2\left[-3.8 \left(\frac{r}{\bar r} \right ) + 8.1 \left(\frac{r}{\bar r} \right )^3  -1.6   \left(\frac{r}{\bar r} \right )^5 -2.3   \left(\frac{r}{\bar r} \right )^7 \right ]
\end{equation}
The detailed formalism is presented in our recent work~\cite{yadav2024thermal}.  
Various magnetic field profiles are available in the literature to describe the state of super-dense matter of NSs in strong magnetic fields.
Chatterjee et al.~\cite{chatterjee2019magnetic} explored the impact of the magnetic field using an axisymmetric numerical code by changing three parameters: mass, EoSs, and the strength of the magnetic field for non-rotating NSs. \\
%%%%%%%%%%%%%%%%%%%%%%%%%%%%%%%%
According to the azimuthal symmetry, the magnitude of the magnetic field $B = ({\vec B} . {\vec B})^{1/2}$ is given by the expression:
\begin{equation}
B(r, \theta) \approx \sum_{l=0}^{l_{max}} B_{l} (r)  Y_l^0 (\theta). 
\end{equation}

As per the numerical results, the magnetic monopole term dominates over all other terms. All the anisotropies (first approximation) are beyond our current work~\cite{yadav2024thermal}.
The simplified radial profile for the monopole term of $B$ is given as:
\begin{equation}
B_0(r) = B_c\left [ 1 - 1.6 \left(\frac{r}{\bar r} \right )^2  -  \left(\frac{r}{\bar r} \right )^4 + 4.2  \left(\frac{r}{\bar r} \right )^6 -2.4  \left(\frac{r}{\bar r} \right )^8 \right],
\end{equation}
here $\overline{r}$ is the mean radius, and $B_c$ corresponds to the value of a central magnetic field for the NSs. Finding a mean radius is difficult in the presence of a magnetic field. So, $\overline{r}$ value is taken more than the actual radius of the NSs. We have assumed the mean radius to be $\overline{r}= 13$ km and the central magnetic field  $B_c = 10^{18}$ Gauss.   
Further, it should be noted that in this work, the possible impact of the magnetic field on the composition of NS matter (populations of neutrons, protons, and electrons) has not been considered.
%%%%%%%%%%%%%%%%%%%%%%%%%%%%%%%%%  

\subsection{Equation of State}
\label{eos}
We have employed three EoSs, namely APR~\cite{Gusakov_2005,PhysRevC.100.025803,PhysRevC.58.1804}, FPS~\cite{Flowers:1976ux}, and SLY~\cite{Broderick_2000} EoSs to study the thermodynamic behavior of the interiors of NSs. The Akmal-Pandharipande-Ravenhall (APR) EoSs describe the interaction potential by isospin asymmetry and baryon density. This EoS depicts the transition from the lower-density phase to the higher-density phase. In principle, it has also been observed that phase transition speeds up the shrink rate of NSs. The FPS EoS shows a unified description of the liquid core and inner crust of the NSs. In FPS EoSs, the liquid-crust core transition usually takes place at density  $\rho_ {edge}$ = $1.6 \times 10^{14}$ gm/cm$^{3}$. All the existing phase transitions are first-order, with relative density jumps less than $1\%$. Considering different nuclear Hamiltonian, FPS EoS is more suitable than any other EoS. The SLY EoS is based on the effective nuclear interaction model in which the nuclei in the ground state remain spherically downward towards the bottom edge of the inner crust of NSs. In this EoS, the transition towards the ionized matter plasma happens at the density $\rho=1.3\times 10^{14}$ gm/cm$^{3}$. The above EoSs in detail are described in our previous work~\cite{yadav2024thermal}.
\begin{center}
\begin{table}[h]
\label{table:tablemass}
\begin{tabular}{ c | c | c } 
 \hline
 \hline
 $EoS$ \ & Maximum Mass ($M_\odot$) \ & Radius (km) (for 1.4$M_\odot$)  \\ [0.5ex] 
 \hline
  APR \ & $2.13$\ & $10.37$ \\ 

 FPS \ & $1.79$\ & $9.68$ \\

 SLY \ & $1.99$\ & $10.43$\\ [1ex]
 \hline\hline 
\end{tabular}
\caption {Table showing the maximum mass ($M_\odot$) and radius (km) corresponding to the mass $M = 1.4 M_\odot$ of NSs for three EoS.} 
\end{table}
\end{center}
{The maximum mass for the three EoS is obtained using~\cite{bocquet1995rotating,10.1093/mnras/stu2706}
(corresponding to the ratio of the magnetic and fluid pressure close to unity) is shown in Table I.
We have also tabulated the radius of the compact object for the three EoS for a mass of 1.4 $M_{\odot}$.

\subsection{Neutron star cooling}
\label{nsc}
To analyze the state of highly dense matter and the emission properties of the strongly magnetized NSs, we have modified the publicly available NSCool code~\cite{2016ascl.soft09009P}. A Fortran-based cooling computational code that solves the equations related to energy balance and heat transport. This package also contains various pre-built stars, different EoSs, and a TOV integrator to form the stars. Considering the spherically symmetric NSs with interiors as isothermal and solving its energy balance equation, one can reasonably understand the various cooling simulations of such compact objects. The equation for the conservation of energy of a star as per the Newtonian formulation is given by~\cite{2006NuPhA777497P}:
\begin{equation}
L_{a}^\infty=-C_{v}\frac{\mathrm{dT_b^{\infty}} }{\mathrm{d} t}-L_{\nu }^\infty-L_{\gamma }^\infty(T_{s}).
\end{equation}
Here
$C_{v}$ specific heat,
$L_{\nu }^\infty$ is the energy sinks are the total neutrino luminosity, 
$L_{a}^\infty$ is the luminosity of axions, 
$L_{\gamma }^\infty$ is luminosity of photons, 
%$H$ includes all possible heating mechanisms,
$T_{s}$ is the surface temperature and 
$T_b^{\infty}$ internal temperature of the NS.
$L_{\gamma }^\infty$= $4\pi\sigma R^{2}(T_{s}^{\infty})^4$, where $\sigma$ is Stefan-Boltzmann's constant and $R$ is the radius of the NS. $T_{s}^{\infty}$ = $T_{s}\sqrt{1-2GM/c^{2}R}$, here infinity superscript means that the external observer is located at the infinity and measures these terms on the redshifted scale. We have not included any heating mechanisms, $H=0$, in the above equation.
Specifically, for non-rotating NSs of mass $M \sim 1.4M_{\odot}$, $T_{s}^{\infty}/T_{s}$ $~\sim$ $0.7$.
The NS cooling mechanisms~\cite{doi:10.1146/annurev.astro.42.053102.134013,PhysRevLett.106.081101} is the crucial ingredient to study the hypothesis related to the axionic DM.
The structure of heat blanketing envelopes (HBEs) may be affected by neutrino cooling mechanisms~\cite{refId01,potekhin2007heat,potekhin2005magnetic,Potekhin_2003,beznogov2016diffusive,10.1093/mnras/stu1102}. The boundary of the HBEs is $\epsilon_b$ ($\rho_{b} \sim 10^{10}$ gm/cm$^3$) and magnetic field at HBEs is taken as $10^{13}$ Gauss. The HBEs comprise iron or successive iron, carbon, hydrogen, and helium layers. 
The formula that determines the radiative opacity for a completely ionized, non-degenerate plasma is made up of electrons and ions and gets modified in the presence of a magnetic field \cite{beznogov2016diffusive,YAKOVLEV2004523}. Also, the heat blanketing layer of NSs will strongly affect the electron thermal conductivity. In the presence of a magnetic field, the thermal conductivity of electrons considered for current calculations is given in~\cite{beznogov2016diffusive,doi:10.1146/annurev.astro.42.053102.134013}. A detailed explanation is provided in our previous work~\cite{yadav2024thermal}. 

\subsection{Cooper pair Breaking and Formation process and N-N Bremsstrahlung process}
\label{cooper}
The composition of the inner core of NSs is somewhat uncertain, and it depends on the considered EoSs, which is still an open area of research~\cite{Reddy:2021rln}. The concept of pairing has opened up new areas for the dominant neutrino emissions. The photon and neutrino emissions are controlled by the stellar outer layers' structure and the state of superdense nuclear matter, respectively. 
There is a continuous formation and breaking of $n-n$ and $p-p$ cooper pairs, leading to the formation of the superfluid material. This happens lower than the superfluid critical temperature $T_{c}$. During the breaking and formation process, axions may be emitted. The energy liberated in this process is carried out by a $\nu\bar{\nu}$ pairs~\cite{PhysRevD.93.065044}.
\begin{align} 
X+X\to \left [ XX \right ] + \nu +\overline{\nu}\nonumber\\
\end{align}
Here, $X$ can be protons or neutrons. 

\subsubsection*{Axion emission rate by Cooper pair Breaking and Formation process (PBF)}
This section briefly describes the axion emission rate from the Cooper pair breaking and formation (PBF) process~\cite{Keller_2013,PhysRevD.93.065044} inside the NS core. To calculate the axion production rates in the cores, the prerequisites are the temperature profiles (from modified TOV equations), metric, critical temperature profiles, and neutron and proton Fermi's momenta profile (which depends on the EoSs). 
The axion emission rate due to the neutron S-wave pairing is given by~\cite{PhysRevD.93.065044}:

\begin{equation}
\epsilon_{a}^s= \frac{8}{3\pi f_{a}^{2}}\,v_{n}(0)\,v_F(n)^{2}\,T^{5}\,I_{a}^s.
\end{equation}

The integral $I_{a}^s$ is given by:
\begin{equation}
I_{a}^s= z^5_{n}\left ( \int_{1}^{\infty} \frac{y^3}{\sqrt{y^2-1}}\left [ f_{F}\left ( z_{n}y \right ) \right ]^{2} dy\right),
\end{equation}
where,
$\epsilon_{a}^s$ is emissivity of axions,
$f_{a}$ is axion decay constant,
$v_{n}(0)$ density of states at fermi surface and
$v_F(n)$ is the fermi velocity of the neutron.

\begin{eqnarray}
v_{n}\left ( 0 \right )=\frac{m_{n}\,p_{F}\left ( n \right )} {\pi^2} \\
z=\frac{\Delta(T)}{T}, \\
f_{F}\left ( x \right )= \left [ e^x+1 \right ]^{-1},
\end{eqnarray}
with $x=\frac{\omega}{2T}$, where $\omega$ is the axion energy. \\

The ratio between axion and neutrino emissivity is given by: 
\begin{equation}
{\epsilon _{a}^s} = \left ( \frac{59.2}{f_{a}^2G_{F}^2[\Delta\left ( T \right)]^2}r(z) \right )\epsilon_{\nu}^s
\end{equation}
where, 
\begin{align}
\Delta(T)\simeq 3.06\,T_{cn}\sqrt{1-\frac{T}{T_{cn}}}
\end{align}
The numerical values of the $r(z)=z^2\, I_a^s/I_{\nu}^s$ are related to integrals corresponding to axion and neutrino emissivity for the different values of $z$.

\begin{equation}
f_{a}> 5.92\times 10^9 GeV\left [ \frac{0.1 MeV}{\Delta\left (T\right)} \right ].
\end{equation}
Here $f_{a}$ is an axion model-dependent variable. The axion mass $m_{a}$ is related to $f_{a}$ given by the equation:
\begin{equation}
m_{a}= 0.60\; \text{eV}\times \frac{10^{7}\text{GeV}}{f_{a}}.
\end{equation}
Here, the axion emissivity due to proton S-wave pairing is the same as neutron S-wave pairing provided $C_n=C_p$ and $T_{cn}=T_{cp}$, where $T_{cn}$ and $T_{cp}$ are critical temperatures for neutron superfluid and proton superfluid, respectively. We have assumed $T_{cp}=T_{cn}=10^9$ K.

The axion emissivity due to the P-wave paired neutron superfluid is given by~\cite{PhysRevD.93.065044,Page:2005fq}:\\
\begin{equation}
\epsilon_{a}^{P}=\frac{2\,C_{n}^{2}} {3\pi f_{a}^{2}}\,v_{n}(0)\,T^{5}\,I_{a}^p,
\end{equation}
where $C_n$ is a constant that depends on the axion model. 
The integral $I_a^p$ is expressed as:
\begin{equation}
I_{a}^{p}=z_{n}^{5}\left ( \int_{1}^{\propto }\frac{y^{3}}{\sqrt{y^{2}-1}} [f_{F}\, (z_n y)]^2 dy\right ),
\end{equation}
where  $z_{n}=\Delta^P(T,\theta)/T$ where $\theta$ is the angle between the quantization axis and momentum of the neutron. There exist two states (A and B) of P-wave superfluid pairing corresponding to  $m_{J} = 0$ and $2$ (representing the projection of the angular momentum of the Cooper-pair on the z-axis). The superfluid gaps are given by~\cite{PhysRevD.93.065044,Buschmann:2019pfp} $
\Delta^{A}=\Delta _{0}^{A}\sqrt{1+3\,\cos^{2}\theta}$ and $
\Delta^{B}=\Delta _{0}^{B}\,\sin\theta.
$.

\subsubsection*{Axion emission rate by N-N Bremsstrahlung process}
The axions, along with neutrinos, are expected to emit during the nucleon collisions by the bremsstrahlung process~\cite{PhysRevLett.53.1198,PhysRevD.100.083005,Paul2018NeutronSC,PhysRevD.93.065044,raffelt1996stars}. To calculate the production rates in the core, the prerequisites are the processes involving protons, neutrons, and electrons, the purely nucleonic EOS at all possible densities. In the current analysis, the axion production occurs in the fully degenerate nucleon-nucleon bremsstrahlung process, $N + N \rightarrow N+N+a$, where $N$ can be either a neutron or a proton, and a is the emitted axion. The current process involves the emission of radiation (in this case, axions) during the interaction of charged particles (neutrons and protons in this context)~\cite{PhysRevLett.53.1198,raffelt1996stars}. The expression for axion emission rate is given by:
\begin{equation}
\epsilon_{a}^{brem}=\frac{\alpha_{a}n_{B}\Gamma_{\sigma}T^{3} }{4 \pi m_{N}^{2}}I_{a}^{brem}
\end{equation}
The axion integral is expressed as:
\begin{equation}
I_{a}^{brem}=\int_{0}^{\infty} s(x) x^{2} e^{-x}~dx,
\end{equation}
where,
$\Gamma_{\sigma}$ means the spin rate (nucleon) changes under collisions with the other nucleon. For the degenerate limit:
\begin{equation}
\Gamma_{\sigma}= \frac{4\alpha_{\pi}^{2}p_{F}T^{3}}{3\pi n_{B}}
\end{equation}
\begin{equation}
s(x)=\frac{\left ( x^{2}+4 \pi^2 \right )\left| x \right|}{4\pi^{2}\left ( 1-e^{-\left|x \right|} \right )}.
\end{equation}
Here $p_{F}$ is the fermi's momentum of the nucleon.
We note that we have not included the effect of the magnetic field on the mentioned cooling mechanisms.
\subsection{Energy Spectrum of Axions}
\label{cooper}
\subsubsection*{Cooper Breaking and Formation process (PBF)}
As mentioned in the previous work~\cite{Buschmann:2019pfp}, the local energy spectrum of axions emitted from the core by the Cooper Breaking and Formation process follows the modified thermal distribution given as:
\begin{equation}
\frac{dF}{dE} \propto  \frac{(E/T)^3~(E/T)^2+4\pi^2 }{e^{E/T}-1},
\end{equation}
where, $E$ is local axion energy and $F$ is the axion-converted-photon flux.
Both spin$-0$ S-wave and spin$-1$ P-wave nucleon superfluids could be possible inside the core of NSs. The energy spectrum of axions due to the spin$-0$ S-wave process is given as:
\begin{equation}
{J_{a}^{s,PBF}}= \frac{d\epsilon_{a}^{s,PBF}}{d\omega_{a}},
\end{equation}
\begin{equation}
{J_{a}^{s,PBF}}= \frac{N_{a}^{s,PBF}}{2\Delta T}\frac{\left ( \frac{\omega_{a} }{2\Delta T} \right )^3}{\sqrt{\left (\frac{\omega_{a} }{2\Delta T} \right )^2-1}}\left [ f_{F}(\frac{\omega_{a}}{2T})\right ]^2,
\end{equation}
where,
$$
{N_{a}^{s,PBF}} = \epsilon_{a}^{s}~z_{n}^{5}/I_{a}^{s}
$$
is the normalization constant derived from the expression $\int_{2\Delta T}^{\infty }J_{a}^{s,PBF}d\omega_{a} =\epsilon_{a}^{s,PBF}$ and $2y\Delta T$ is an axion energy. $T$ corresponds to a locally measurable quantity at some radius within the interiors of NSs.The energy spectrum of axions due to the P-wave pairing is given by~\cite{Buschmann:2019pfp}:
\begin{equation}
{J_{a}^{p,PBF}}= \frac{d\epsilon_{a}^{p,PBF}}{d\omega_{a}},
\end{equation}

%\begin{equation}\

\begin{multline*}
{J_{a}^{p,PBF}} =  \int_{-1}^{1}dcos\theta\frac{1}{4}\Delta^{P}(T,\theta)^4 {N_{a}^{p,PBF}}\\ \times \frac{\left ( \frac{\omega_{a} }{2\Delta^{P} (T,\theta)} \right )^3}{\sqrt{\left (\frac{\omega_{a} }{2\Delta^{P} (T,\theta)} \right )^2-1}}\left [ f_{F}(\frac{\omega_{a}}{2T})\right ]^2,
\end{multline*}
$$
{N_{a}^{p,PBF}} = \epsilon_{a}^{p}/T^{5}~I_{a}^{p}
$$
is the normalisation constant which is defined as $\int_{2\Delta^{P}(T,\theta)}^{\infty }J_{a}^{p,PBF}d\omega_{a} =\epsilon_{a}^{p,PBF}$and 2y$\Delta^{P}(T,\theta)$ is the axion energy $\omega_{a}$. 
\subsubsection*{N-N Bremsstrahlung process}
\label{bremmms}
The energy spectrum of axions due to the N-N bremsstrahlung process is given by:
\begin{equation}
{J_{a}^{brem}}=\frac{d \epsilon_{a}^{brem}}{d\omega_{a}} = \frac {N_{a}^{brem}\omega_{a}^{5}} {e^{x} T^6},
\end{equation}
where, 
$N_{a}^{brem}= \frac{\epsilon_{a}^{brem}}{4 \pi^{2} I_{a}^{brem}}$
is the normalization constant derived from the expression $\int_{0}^{\infty }J_{a}^{brem}d\omega_{a} =\epsilon_{a}^{brem}$ and $x \times T$ is an axion energy.
Here, $x=\frac{\omega_{a}}{T}$ is the dimensionless quantity and $\omega_{a}$ (equivalently $x\times T$) is the axion energy. 

\subsection{Conversion Probability and Axion-converted-Photon Flux}
Axions produced in the cores escape the stars because of their weak interaction with the ordinary matter and further get converted to the photons in an extended magnetosphere of strongly magnetized NSs~\cite{PhysRevD.37.1237,PhysRevLett.123.061104}. The axions mixing with photons is based on the axion-photon coupling term given by the Lagrangian:
\begin{equation}
{\mathcal L}_{conv} = -\frac{1}{4}g_{a\gamma \gamma }aF\widetilde{F}.
\end{equation}
Here, F $\&$ $\widetilde{F}$ corresponds to an electromagnetic field strength tensor and corresponding dual tensor, respectively, $g_{a\gamma \gamma }$ is axion-photon coupling parameter given as~\cite{Buschmann:2019pfp}:
\begin{equation}
g_{a\gamma \gamma }=\frac{C_{\gamma}\alpha}{2\pi f_{a}},
\end{equation}
where, $C_{\gamma}$ is a axion model dependent factor and $\alpha=1/137$ is QED fine structure constant. In the magnetosphere of strongly magnetized NSs, the axion-photon mixing term may rotate the initial state of an axion to an electromagnetic wave that is polarised exactly in the same direction as the external magnetic fields. The following expression provides an approximate relationship for conversion probability~\cite{PhysRevD.37.1237,PhysRevLett.123.061104,fortin2018constraining,Buschmann:2019pfp}:
\begin{equation}
\begin{split}
P_{a\to\gamma }\approx 1.5\times 10^{-4}\left(\frac{g_{a\gamma\gamma}}{10^{-11}\,GeV^{-1}}\right)^2\left(\frac{1\,keV}{\omega}\right)^{0.8} \\
\times\left(\frac{B_{0}}{10^{13}\,G}\right)^{0.4}\left(\frac{R_{NS}}{10\, km}\right )^{1.2}\sin^{0.4}\theta,
\end{split}
\end{equation}
where, $\omega$ is the axion energy and $R_{NS}$ is radius of NS (in km). $B_{0}$ is the strength of the surface magnetic field at the pole, and $\theta$ is the magnetic axis polar angle. Finally, we have multiplied the axion-photon conversion probability with the axion energy spectrum to obtain the final axion-converted-photon flux at the different values of axion energy. 

%-----------------------------------------------------
\section{Results and Discussions}
\label{result}
\label{result}
\begin{figure*}[htp!]
\begin{tabular}{cc}
\includegraphics[width=7.75cm]{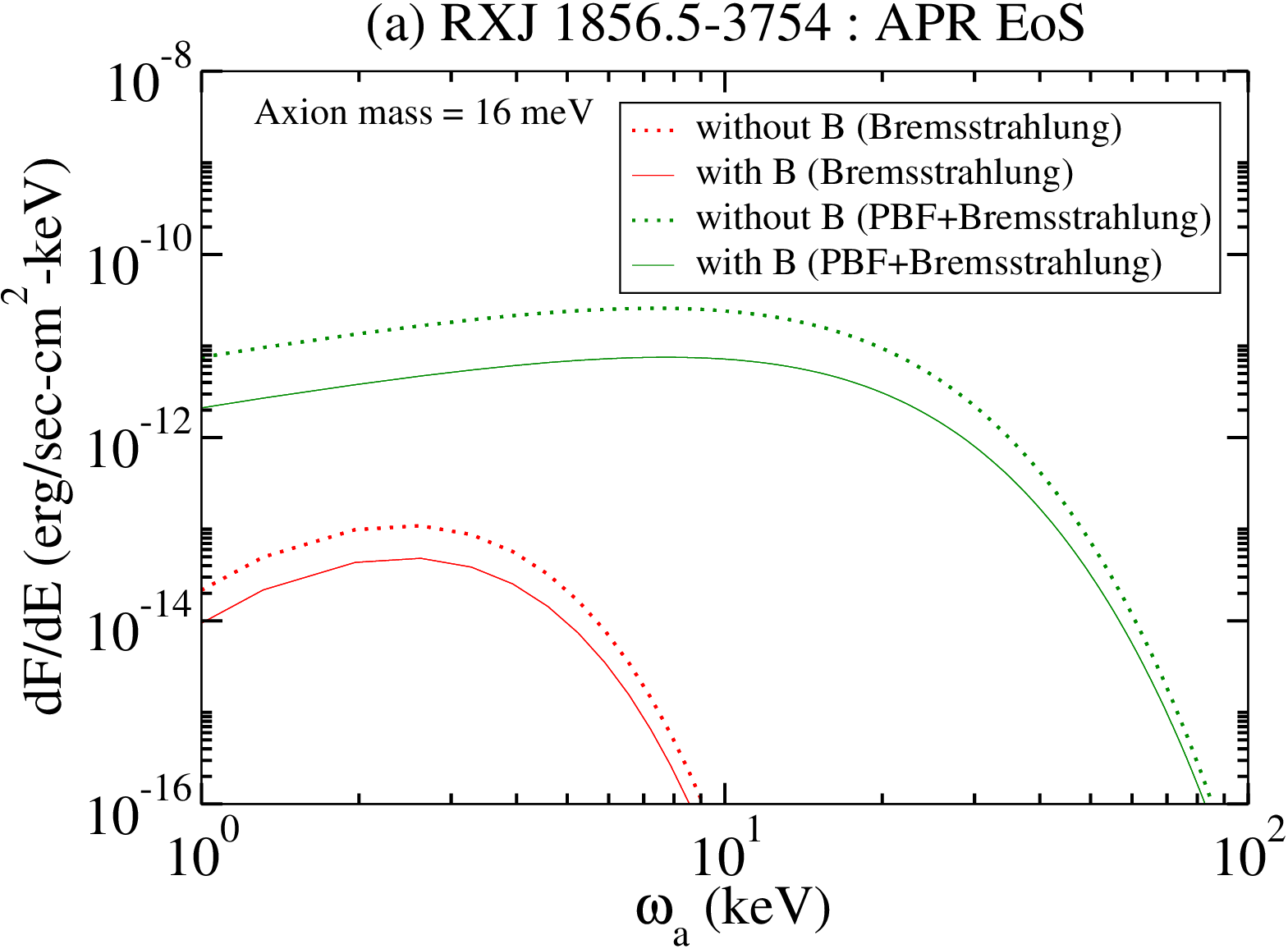}\ & \includegraphics[width=7.75cm]{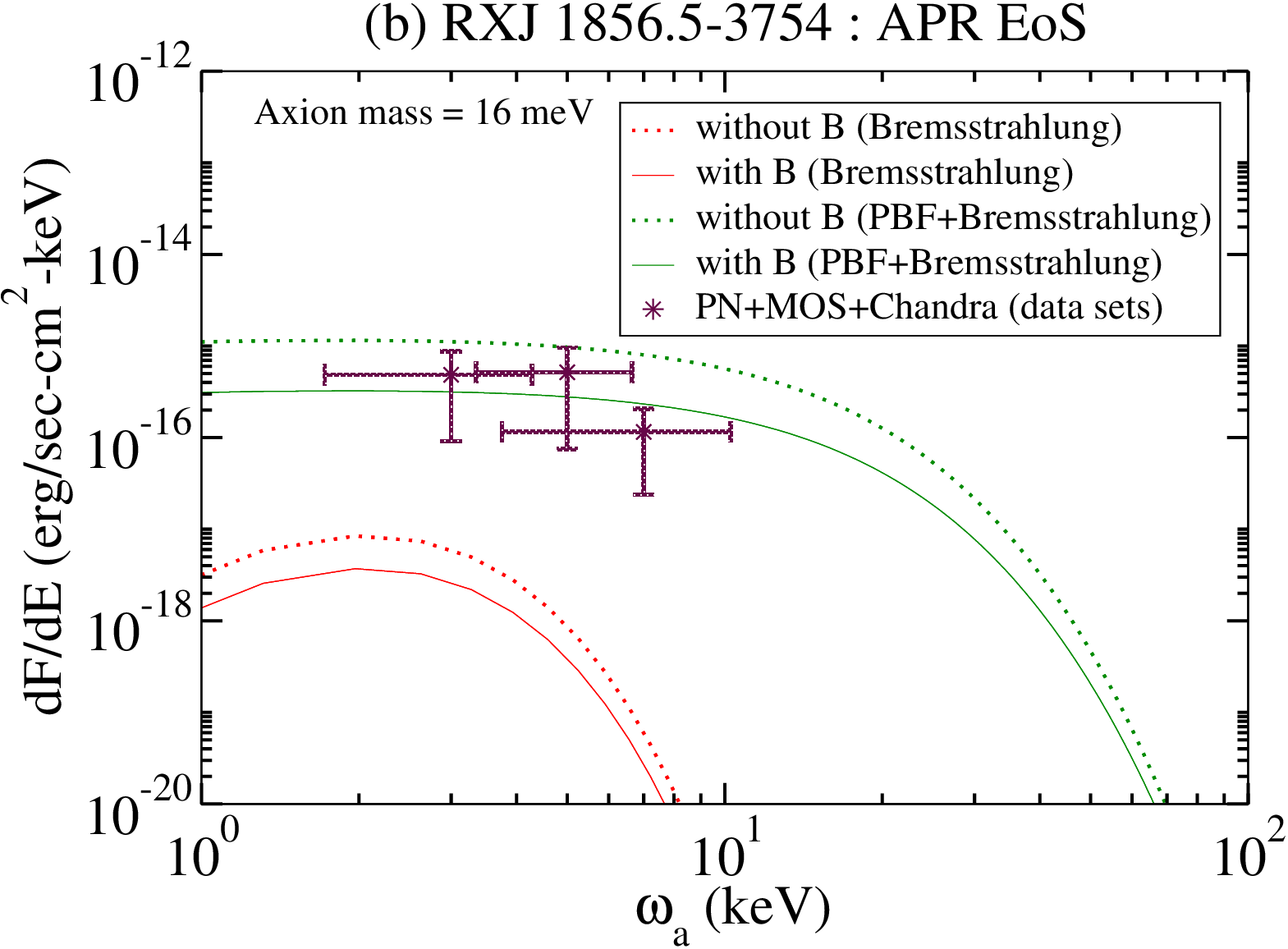} 
\end{tabular}
\caption{The variation of the energy spectrum of axions (left) with axion energies by Bremsstrahlung plus PBF and only Bremsstrahlung process for RXJ 1856.5-3754 (M7) NS at an axion mass 16 meV for APR EoS in the presence of magnetic field.} 
\label{fig:nsener1} 
\end{figure*}
\begin{figure*}[htp!]
\begin{tabular}{cc}
\includegraphics[width=7.75cm]{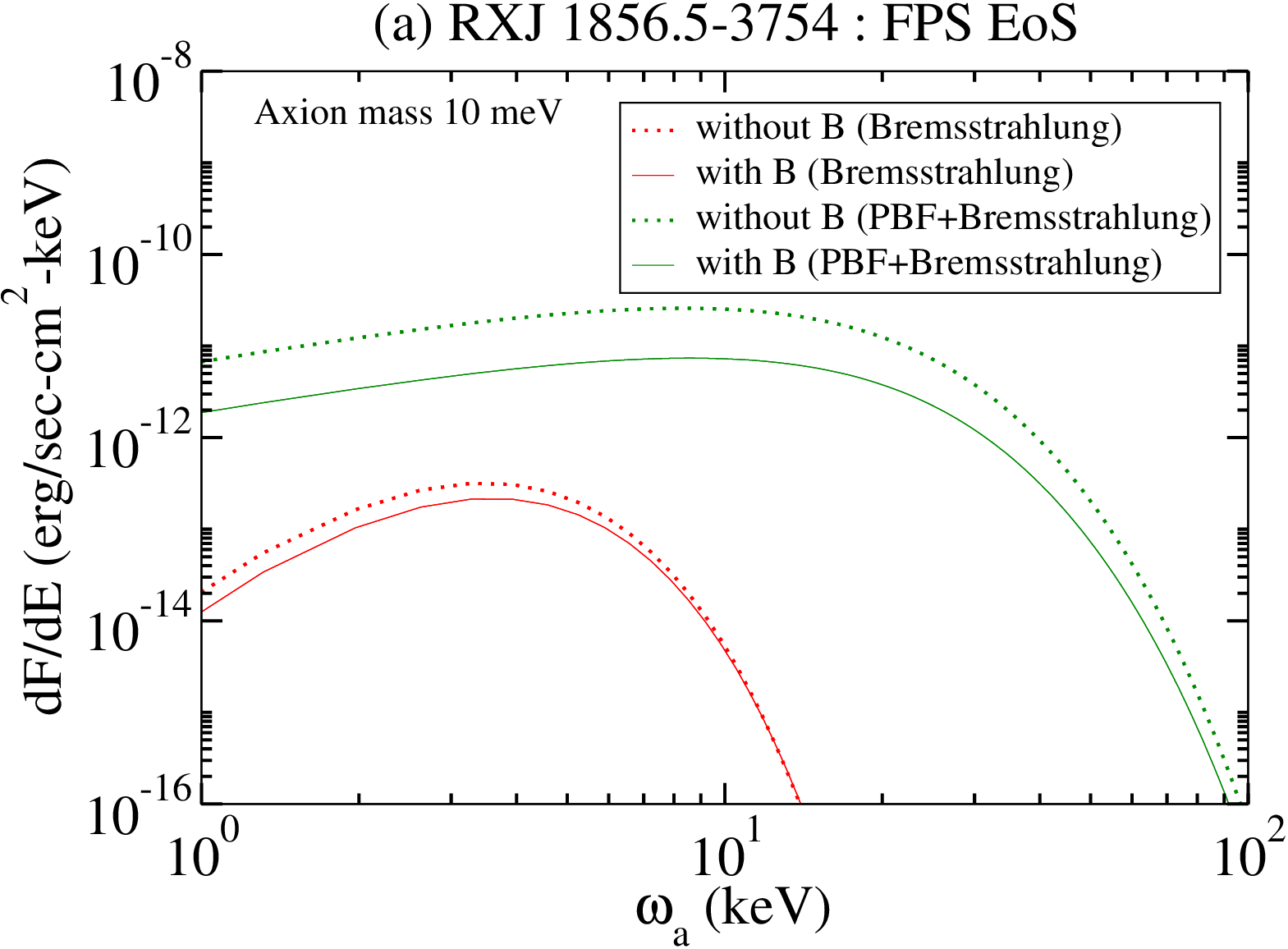}\ & \includegraphics[width=7.75cm]{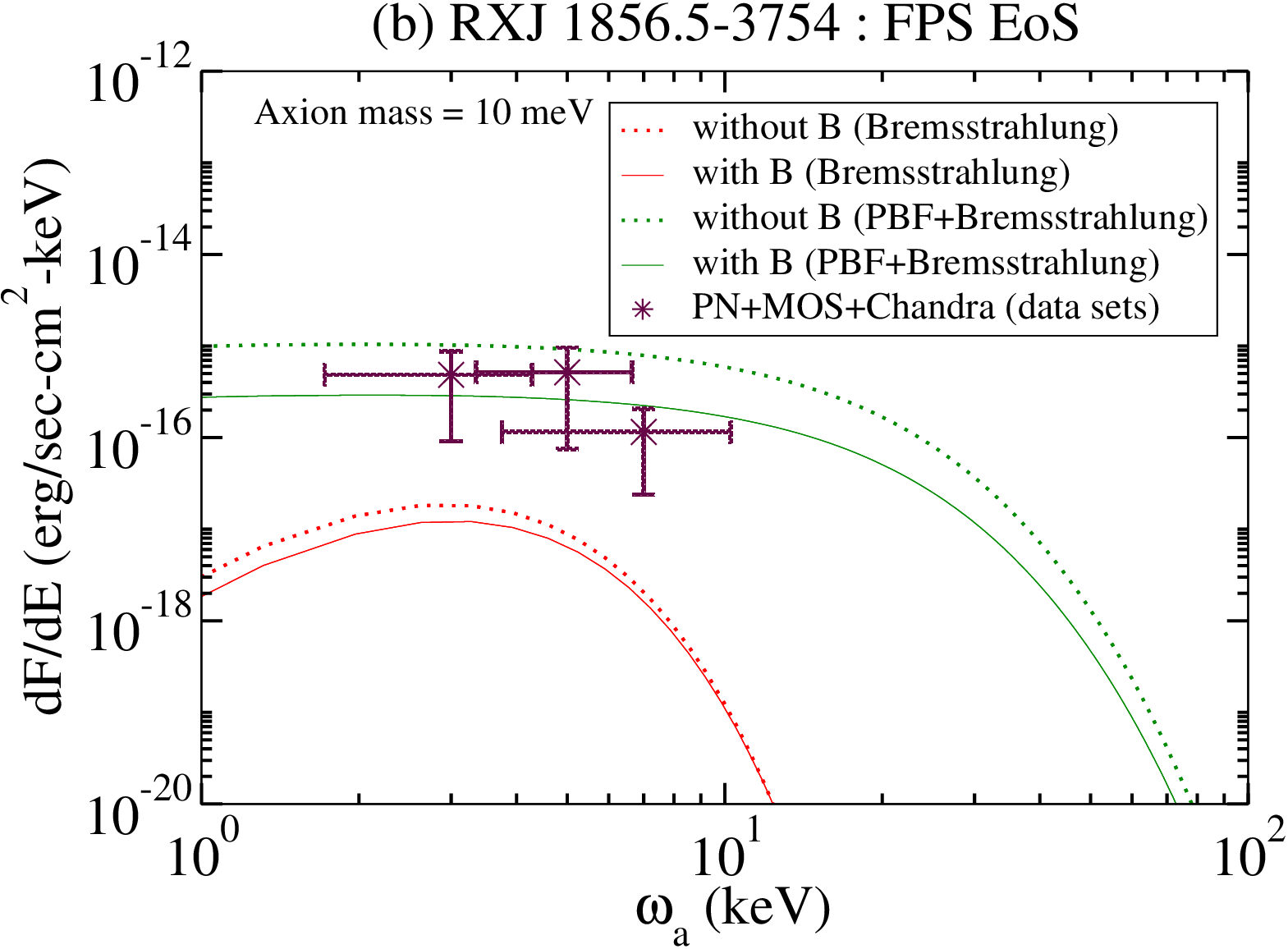}
\end{tabular} 
\caption{The variation of the energy spectrum of axions (left) and axion-converted-photon flux (right) with axion energies by Bremsstrahlung plus PBF and only Bremsstrahlung process for RXJ 1856.5-3754 (M7) NS at an axion mass 10 meV for FPS EoS in the presence and absence of magnetic field.} 
\label{fig:nsener2}  
\end{figure*}
\begin{figure*}[htp!]
\begin{tabular}{cc}
\includegraphics[width=7.75cm]{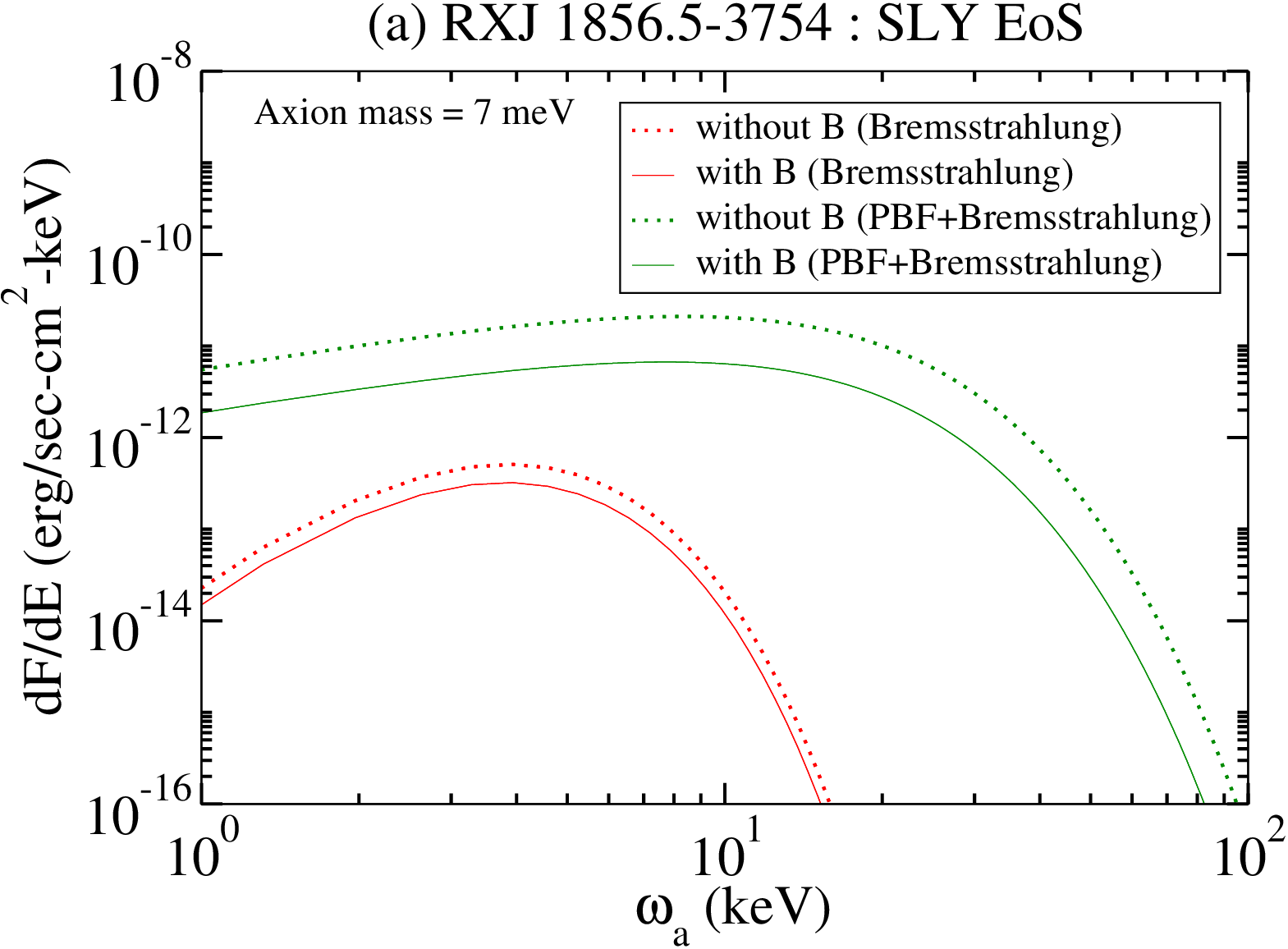}\ & \includegraphics[width=7.75cm]{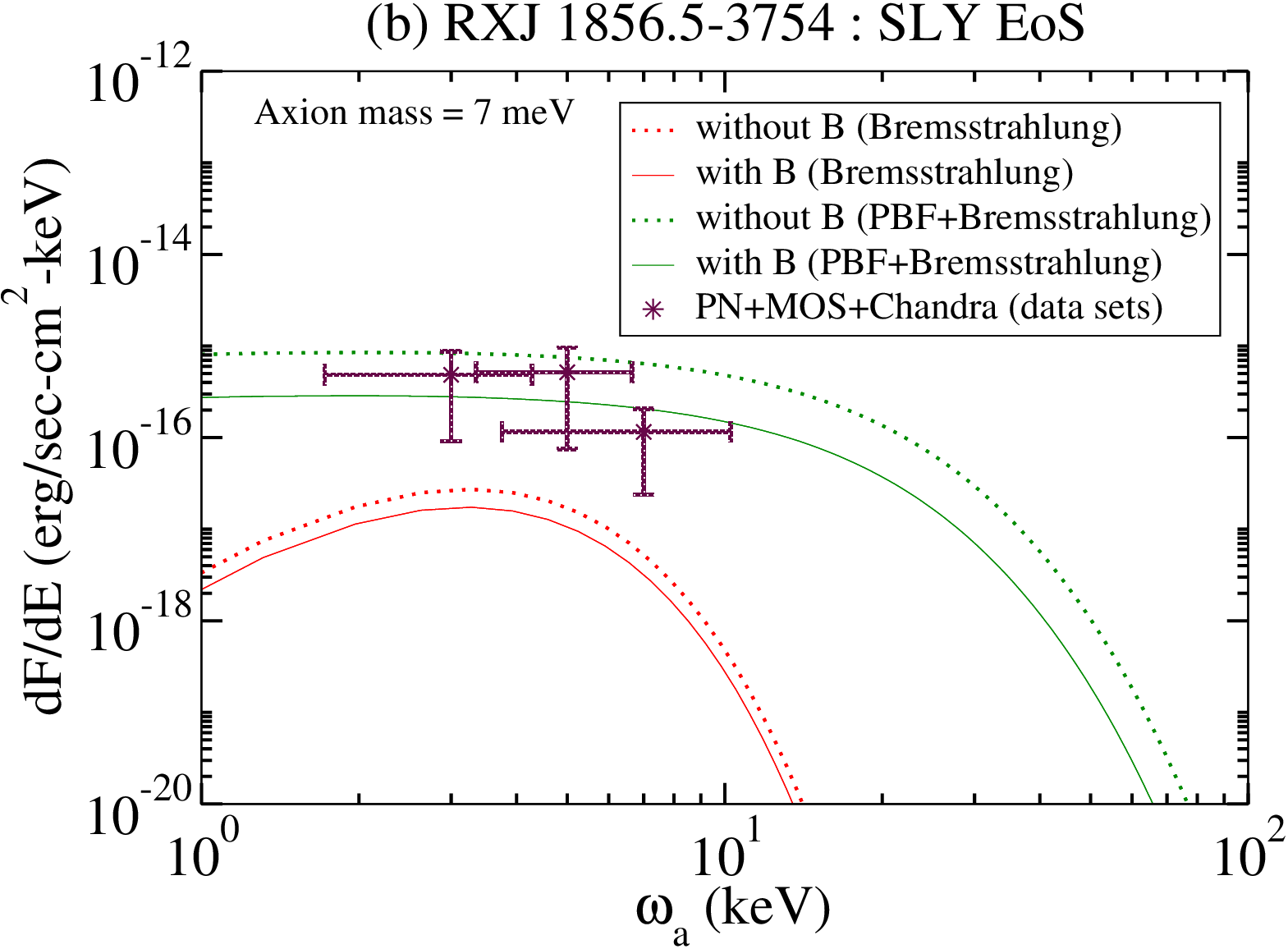} 
\end{tabular}
\caption{The variation of the energy spectrum of axions (left) and axion-converted-photon flux (right) with axion energies by Bremsstrahlung plus PBF and only Bremsstrahlung process for RXJ 1856.5-3754 (M7) NS at an axion mass 7 meV for SLY EoS in the presence and absence of magnetic field.} 
\label{fig:nsener3}  
\end{figure*}
 
Figure (\ref{fig:nsener1}a) shows the energy spectrum of axions as a function of axion energy for APR EoS. The local energy spectrum of axions emitted by the Bremsstrahlung plus PBF process remains dominant over those emitted only by the Bremsstrahlung process for the entire range of the axion energies. As the axion energies increase, for a no-magnetic field case, the energy spectrum of axions remains higher than when there is a magnetic field at all values of axion energies. In the Bremsstrahlung plus PBF process, the separation in the energy spectrum of axions is almost constant up to $10$ keV axion energy, but after that, separation decreases, and both the curves overlap around $10^{2}$ keV energy. 
However, if we consider only the Bremsstrahlung process, a significant difference can be observed in the energy spectrum of axions in both the presence and absence of a magnetic field. The magnitude of the axion energy spectrum and axion-converted-photon flux is quite sensitive to the axion mass, and a significant departure is observed in the presence of the magnetic field.
%-------------------------------------------------------
In figure (\ref{fig:nsener1}b), we have presented the axion-converted-photon flux as a function of axion energies for APR EoS. The axion-converted-photon flux from the Bremsstrahlung plus PBF process dominates from the only Bremsstrahlung process for all ranges of axion energies. Including a magnetic field does not change the qualitative behavior, although a significant difference is observed. As we focus on the Bremsstrahlung plus PBF process, a significant difference in the axion energies spectrum can be observed upto axion energy of $\sim 10$ keV. After that, the separation decreases and the curves overlap beyond $\sim 70$ keV. We also show the data points from X-ray spectra at an axion mass $16$ meV for RXJ 1856.5-3754 (M7) star by combining all the available archival data sets from PN+MOS+Chandra~\cite{Dessert_2020,Buschmann:2019pfp}. The axion mass bound is around $16$ meV and is obtained by fitting the various axion mass spectral models corresponding to available data sets. On comparing with the observational data, our predicted results show a better agreement with the magnetic field case.           
%-----------------------------------------------------------------
The figure (\ref{fig:nsener2}a) shows the variation of the energy spectrum of axions versus axion energies for FPS EoS. Here, the local energy spectrum of axions emitted by the Bremsstrahlung plus PBF processes dominates over the only Bremsstrahlung process for the entire range of axions energies. The separation between (with/without) magnetic fields in the energy spectrum of axions is constant up to $\sim 10$ keV axion energies for the Bremsstrahlung plus PBF processes. 
If we include both processes, we find that in the presence of a magnetic field, the energy spectrum of axions in the low axion energies region is lesser than without a magnetic field. The gap becomes negligible at higher values of axion energies, $\omega_{a}>95\,$ keV.
However, for only the Bremsstrahlung process, a constant difference can be seen in the energy spectrum of axions in the presence/absence of magnetic field for axion energy $\sim 4$ keV. Beyond this energy, this difference decreases until both curves overlap between $10-20$ keV axion energies.
%-----------------------------------------------------------------
The figure (\ref{fig:nsener2}b) depicts the variation of axion-converted-photon flux as a function of axion energies for FPS EoS. After including the magnetic field, the axion-converted-photon flux when both the Bremsstrahlung and PBF processes are included is more significant than the Bremsstrahlung process. This is true for all values of the axion energies. The separation (in the presence/absence of a magnetic field) for the axion-converted-photon flux emitted by the Bremsstrahlung plus PBF process is almost similar as compared to other mentioned EoSs. The separation decreases from $\sim 30$ keV axion energy, and finally, both the curves (with/without magnetic field) overlap beyond $\sim 90$ keV axion energies. If we consider axion-converted-photon flux from the Bremsstrahlung process only. In this case, the separation remains constant up to $\sim 3$ keV, and after that, curves (with/without magnetic field) overlap beyond the axion energy of $\sim 10$ keV. We also show the data points from X-ray spectra for RXJ 1856.5-3754 (M7) star from all available archival data sets from PN+MOS+Chandra~\cite{Dessert_2020,Buschmann:2019pfp}. Again, by fitting the several axion mass spectral models corresponding to the available data sets, we obtain the axion mass bound of $10$ meV in the case of FPS EoS. Compared to the available experimental data, our calculated results demonstrate better agreement with the magnetic field. For FPS EoS, the magnitude of the axion's energy spectrum and axion-converted-photon flux is quite sensitive to the mass of an axion. One can also observe a significant departure in the presence of the magnetic field.
%-----------------------------------------------------------------
In the figure (\ref{fig:nsener3}a), we have shown a variation of the energy spectrum of axions versus axion energies for SLY EoS. In the presence of a magnetic field, at all values of the axion energies, the local energy spectrum of axions emitted by the Bremsstrahlung plus PBF process dominates over the case when only the Bremsstrahlung process is included. This behavior is similar to other mentioned EOSs. Considering the Bremsstrahlung plus PBF processes, the separation (with/without magnetic field) is almost similar to the other EoSs but for lesser values of the axion's energy only. One can observe a significant difference in the energy spectrum for the entire range of axion energy. The imprint of the strong magnetic field on the energy spectrum of axions from the Bremsstrahlung plus PBF processes can be seen in this EoS but overall qualitative features remain the same. For the Bremsstrahlung process, a significant departure in the energy spectrum of axions is observed at lower energies before the spectrum declines quickly at the higher values of axion energies. 
%-----------------------------------------------------------------
In the figure (\ref{fig:nsener3}b), we have presented the variation of axion-converted-photon flux as a function of axion energies for SLY EoS. For the Bremsstrahlung plus PBF process, in the presence of a magnetic field, the axion-converted-photon flux consistently deviates as compared to the case of a no-magnetic field. Still, it does not behave qualitatively differently from the other mentioned EoS. Additionally, the axion-converted-photon flux is always higher in the no-magnetic field scenario than in the case of the magnetic field, regardless of all axion energies. After considering both processes, a significant difference can be seen for the entire range of axion energies ($\sim 1-100$ keV). Specifically, the axion-converted-photon flux is steeply peaked at low axion energies for the Bremsstrahlung process. This rapidly decreases at higher values of axion energies because of energy conservation. Here again, we have presented the X-ray spectra of the RXJ 1856.5-3754 (M7) star using the data sets from PN+MOS+Chandra by employing the SLY EoS~\cite{Dessert_2020,Buschmann:2019pfp}. After fitting the various axion mass spectral models with the available data sets, we report an axion mass bound around $7$ meV for the SLY EoS. Our anticipated results, on comparsion with the observational data, show better agreements with the magnetic field case than the no-magnetic field situation.  
%-------------------------------------------------------------

%-------------------------------------------------------------------
The strong magnetic fields and the corresponding high value of magnetic field energy lead to a drop in the star's thermal energy while the contributions from the overall gravitational energy stay unchanged. This explains the correlation between the luminosity suppression and the magnetic field strength~\cite{10.1093/mnras/sty776}. 
This work has not considered the effect of magnetic fields on the mentioned EoS. It is anticipated that the axion mass bounds would exhibit degeneracies with the mentioned EoS. However, the current analysis suggests that the EoS influences the macroscopic properties (mass and pressure profiles) by modified TOV equations, thereby influencing the bounds on axion mass.
%----------------------------------------------------------------------------------

%-------------------------------------------------------------------------------------
\begin{figure*}[htp!]
\begin{tabular}{cc}
\includegraphics[width=7.75cm]{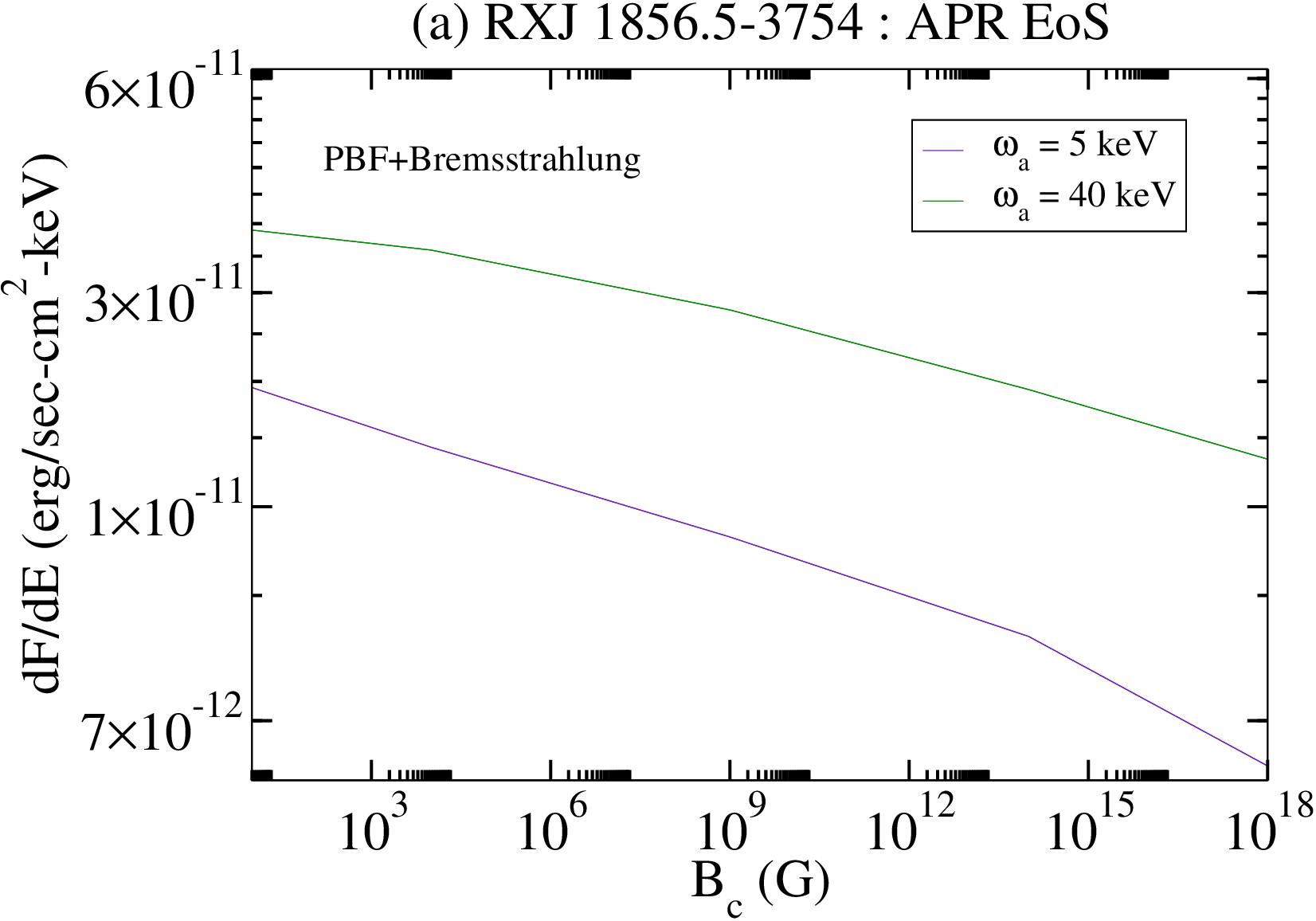}\ & \includegraphics[width=7.75cm]{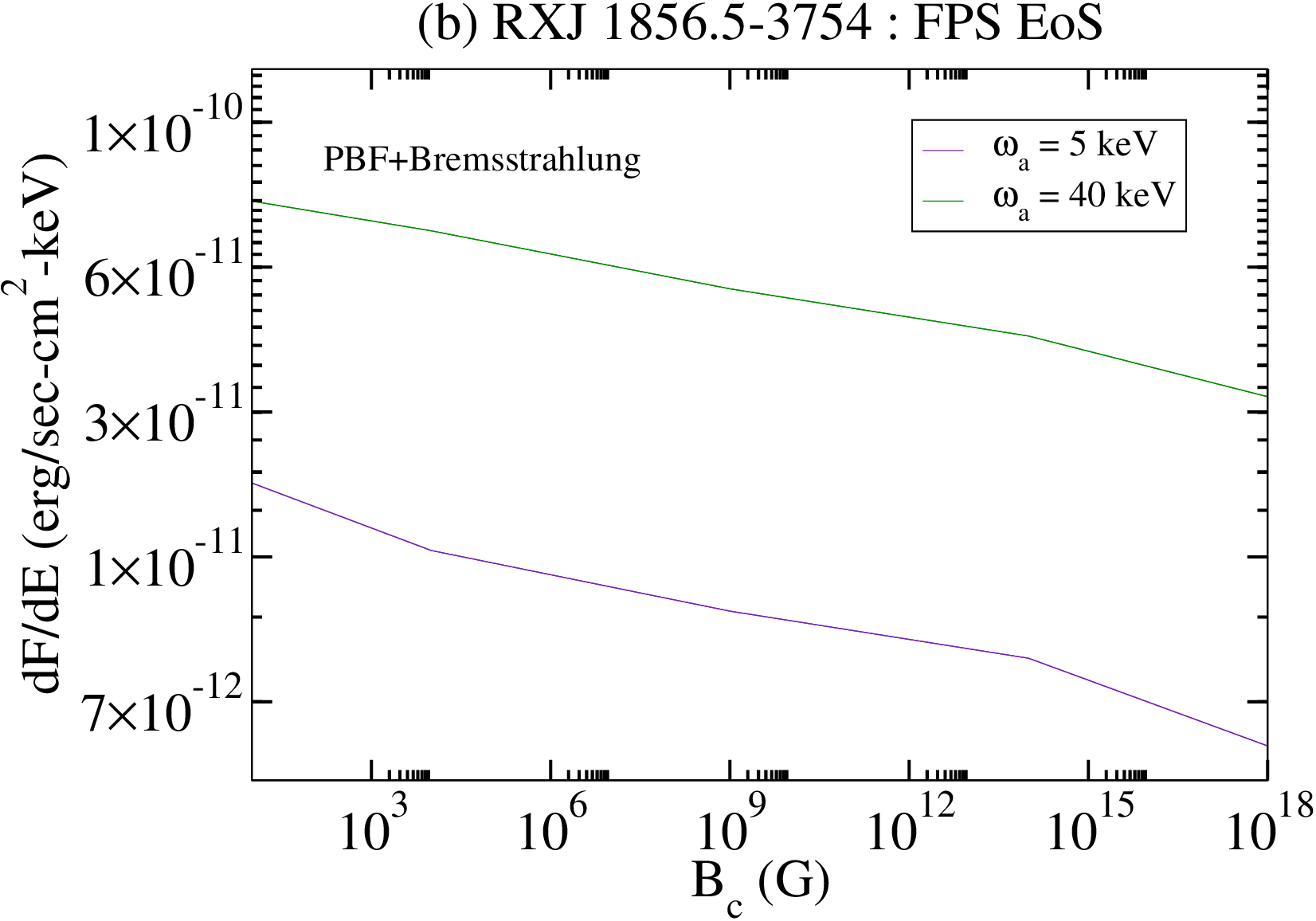} 
\end{tabular}
\end{figure*}
\begin{figure*}[htp!]
\begin{center}
\includegraphics[width=7.75cm]{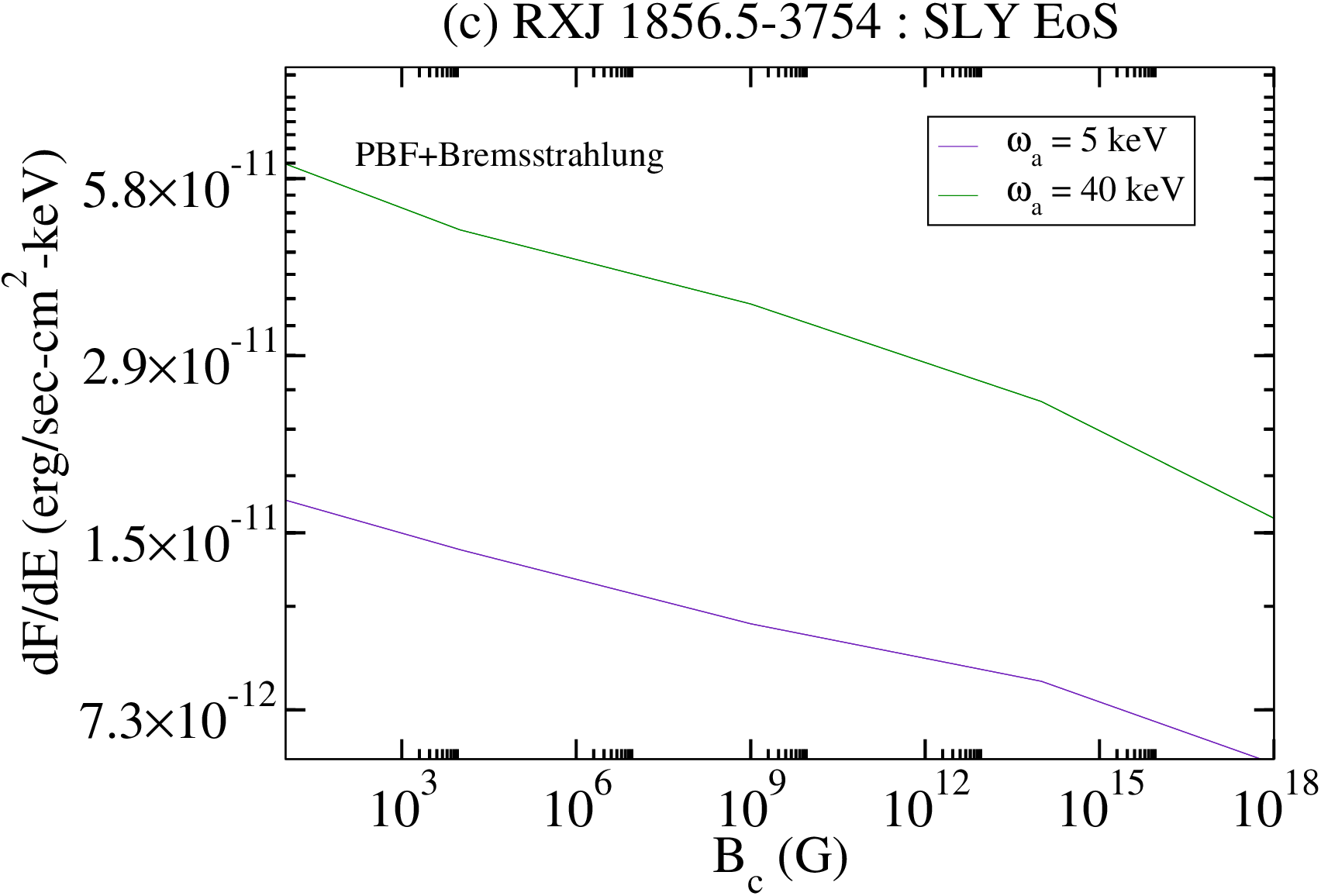}
\end{center}
\caption{The variation of the energy spectrum of axions with different values of central magnetic field $B_{c}$ at axion energies $\omega_{a}= 5$ keV and $\omega_{a} = 40$ keV from PBF plus Bremsstrahlung process for RXJ 1856.5-3754 (M7) NS and for three EoS. The results for energy spectrum of axions corresponding to axion energy $\omega_{a} = 40$ keV are multiplied by a factor of $100$ to compare with the results of $\omega_{a} = 5$ keV.}  
\label{fig:ener_bc}
\end{figure*}

%-------------------------------------------------------------------------------------
In the Bremsstrahlung plus PBF process, the energy spectrum of axions for the APR, FPS, and SLY EoSs show an $\sim$ $18.9$\%, $18.1$\%, $16.5$\% departure from without the magnetic field at an axion energy $\sim 5$ keV.
Similarly, for the Bremsstrahlung process, the energy spectrum of axions for the APR, FPS, and SLY EoSs show an $\sim$ $6.8$\%, $4.7$\%, $5.9$\% departure from without the magnetic field at an axion energy $\sim 5$ keV.

In addition to this, Table II summarises the percentage changes of axion-converted-photon flux for APR, FPS, and SLY EoS at an axion energy of $5$ keV in the presence and absence of a magnetic field for RXJ 1856.5-3754 (M7) NSs.
%------------------------------------------------------------------------
\begin{center}
\begin{table}[h!]
\begin{tabular}{ c | c |  c  } 
 \hline \hline
EoSs \ & Bremsstrahlung plus PBF \ & Bremsstrahlung \\ [1ex] 
 \hline
 APR\ & $18.2\%$\ & $6.3\%$ \\ 

 FPS \ & $17.8\%$\ & $4.4\%$ \\
 
 SLY \ & $16.1\%$\ & $5.2\%$\\ 
 \hline\hline 
\end{tabular}
\caption{Table shows the percentage changes of axion-converted-photon flux by Bremsstrahlung plus PBF and only Bremsstrahlung process of RXJ 1856.5-3754 (M7) NSs for three EoS at an axion energy of $5$ keV.}
\label{tab:tabtwo}
\end{table}
\end{center}

%--------------------------------------------------------------------------------------
The effect of the magnetic field can be visualized by studying the variation of energy spectrum at a fixed energy with the central magnetic field. Figure (\ref{fig:ener_bc}a), (\ref{fig:ener_bc}b) and (\ref{fig:ener_bc}c) shows the variations of the energy spectrum of axions at axion energies $\omega_{a}= 5$ keV and $\omega_{a} = 40$ keV with varying central magnetic field for three EoSs. The rate of decrease in the energy spectrum is slightly lesser at a higher axion energy of 40 keV compared to the axion energy of 5 keV from $B=0$ G to $B=10^{18}$ G. However, the qualitative behavior remains the same for all three EoSs. 
\section{Summary and Conclusions}
\label{conc}
In the present work, we study how the presence of a strong magnetic field affects the energy spectrum of axions and axion-converted-photon flux. Our current work is based on the hypothesis that the axions are produced in the core of NSs by the Cooper pair breaking and formation process and the Bremsstrahlung process, then converted into X-rays due to a strong magnetic field in NS magnetospheres. We have calculated the axion mass bound for the APR, FPS, and SLY EoS. We have also explored the effect of the magnetic field by comparing our numerically computed results with the available archival data of PN+MOS+Chandra~\cite{Dessert_2020,Buschmann:2019pfp} for Magnificient seven (M7) star RXJ 1856.5-3754 (M7). Additionally, the summary of our findings is as follows:

\begin{itemize}

 \item We have not included the influence of magnetic field on the employed EoSs and mechanism related to the production of axions. Initially, the modified Tolman-Oppenheimer-Volkoff (TOV) system of equations was solved for a given EoSs to study the macroscopic properties (mass/pressure) profiles, which are then used to compute luminosities of axions and surface temperature as a function of time using the NSCool code. 
 
 \item Our work encompassed the energy spectrum of axions and axion-converted-photon flux versus axion energies in the presence and absence of magnetic fields. We find the average best-fit axion mass bound for the mentioned three EOSs; namely APR, FPS, and SLY, as $16$ meV, $10$ meV, and $7$ meV, respectively. These mass bounds lie within the limits of the recently reported axion mass bound of $10$ meV and $16$ meV in the case of a no-magnetic field. The results have been calculated with a fixed value of the star's mean radius $\overline{r}= 13$ km and a high value of the central magnetic field $10^{18}$ Gauss.
 \item We emphasize that the axion mass bound slightly varies for APR, FPS, and SLY EoSs. On comparing our results with all available archival data sets of PN+MOS+Chandra, we find that the results for axion-converted-photon flux closely matches with the magnetic field, especially for APR and FPS EOSs.

 \item Our investigation reports that the influence of a strong magnetic field on the observables depends on the employed EoS. The uncertainties are more pronounced, especially with low temperatures and high-density conditions. Thus, the study of axion-photon conversion remains a subject of active research for these highly magnetized NSs. 
\end{itemize}
 
\vskip 1cm
\section*{Acknowledgments}
The authors thank the anonymous referee for insightful feedback that helped us to improve the manuscript. We thank Dany Page, Malte Buschmann, and T. Opferkuch for clearing our problems related to the NSCool code. S. Yadav acknowledges Birla Institute of Technology and Science, Pilani, Pilani Campus, Rajasthan, for research facilities and financial assistance.  
\bibliography{MSTGA}  

\begin{thebibliography}{141}
\expandafter\ifx\csname natexlab\endcsname\relax\def\natexlab#1{#1}\fi
\expandafter\ifx\csname bibnamefont\endcsname\relax
  \def\bibnamefont#1{#1}\fi
\expandafter\ifx\csname bibfnamefont\endcsname\relax
  \def\bibfnamefont#1{#1}\fi
\expandafter\ifx\csname citenamefont\endcsname\relax
  \def\citenamefont#1{#1}\fi
\expandafter\ifx\csname url\endcsname\relax
  \def\url#1{\texttt{#1}}\fi
\expandafter\ifx\csname urlprefix\endcsname\relax\def\urlprefix{URL }\fi
\providecommand{\bibinfo}[2]{#2}
\providecommand{\eprint}[2][]{\url{#2}}

\bibitem[{\citenamefont{Potekhin et~al.}(2019)\citenamefont{Potekhin, Chugunov,
  and Chabrier}}]{potekhin2019thermal}
\bibinfo{author}{\bibfnamefont{A.}~\bibnamefont{Potekhin}},
  \bibinfo{author}{\bibfnamefont{A.}~\bibnamefont{Chugunov}}, \bibnamefont{and}
  \bibinfo{author}{\bibfnamefont{G.}~\bibnamefont{Chabrier}},
  \bibinfo{journal}{Astronomy \& Astrophysics} \textbf{\bibinfo{volume}{629}},
  \bibinfo{pages}{A88} (\bibinfo{year}{2019}).

\bibitem[{\citenamefont{Potekhin and
  Chabrier}(2012)}]{potekhin2012thermonuclear}
\bibinfo{author}{\bibfnamefont{A.}~\bibnamefont{Potekhin}} \bibnamefont{and}
  \bibinfo{author}{\bibfnamefont{G.}~\bibnamefont{Chabrier}},
  \bibinfo{journal}{Astronomy \& Astrophysics} \textbf{\bibinfo{volume}{538}},
  \bibinfo{pages}{A115} (\bibinfo{year}{2012}).

\bibitem[{\citenamefont{Beznogov et~al.}(2016)\citenamefont{Beznogov, Potekhin,
  and Yakovlev}}]{beznogov2016diffusive}
\bibinfo{author}{\bibfnamefont{M.}~\bibnamefont{Beznogov}},
  \bibinfo{author}{\bibfnamefont{A.}~\bibnamefont{Potekhin}}, \bibnamefont{and}
  \bibinfo{author}{\bibfnamefont{D.}~\bibnamefont{Yakovlev}},
  \bibinfo{journal}{Monthly Notices of the Royal Astronomical Society}
  \textbf{\bibinfo{volume}{459}}, \bibinfo{pages}{1569} (\bibinfo{year}{2016}).

\bibitem[{\citenamefont{Potekhin}(2014)}]{potekhin2014atmospheres}
\bibinfo{author}{\bibfnamefont{A.~Y.} \bibnamefont{Potekhin}},
  \bibinfo{journal}{Physics-Uspekhi} \textbf{\bibinfo{volume}{57}},
  \bibinfo{pages}{735} (\bibinfo{year}{2014}).

\bibitem[{\citenamefont{Page}(1998)}]{1998nspt.conf..183P}
\bibinfo{author}{\bibfnamefont{D.}~\bibnamefont{Page}}, p. \bibinfo{pages}{183}
  (\bibinfo{year}{1998}).

\bibitem[{\citenamefont{Buschmann et~al.}(2022)\citenamefont{Buschmann,
  Dessert, Foster, Long, and Safdi}}]{Buschmann:2021juv}
\bibinfo{author}{\bibfnamefont{M.}~\bibnamefont{Buschmann}},
  \bibinfo{author}{\bibfnamefont{C.}~\bibnamefont{Dessert}},
  \bibinfo{author}{\bibfnamefont{J.~W.} \bibnamefont{Foster}},
  \bibinfo{author}{\bibfnamefont{A.~J.} \bibnamefont{Long}}, \bibnamefont{and}
  \bibinfo{author}{\bibfnamefont{B.~R.} \bibnamefont{Safdi}},
  \bibinfo{journal}{Phys. Rev. Lett.} \textbf{\bibinfo{volume}{128}},
  \bibinfo{pages}{091102} (\bibinfo{year}{2022}).

\bibitem[{\citenamefont{Buschmann et~al.}(2020)\citenamefont{Buschmann, Foster,
  and Safdi}}]{Buschmann:2019icd}
\bibinfo{author}{\bibfnamefont{M.}~\bibnamefont{Buschmann}},
  \bibinfo{author}{\bibfnamefont{J.~W.} \bibnamefont{Foster}},
  \bibnamefont{and} \bibinfo{author}{\bibfnamefont{B.~R.} \bibnamefont{Safdi}},
  \bibinfo{journal}{Phys. Rev. Lett.} \textbf{\bibinfo{volume}{124}},
  \bibinfo{pages}{161103} (\bibinfo{year}{2020}).

\bibitem[{\citenamefont{Chadha-Day et~al.}(2022)\citenamefont{Chadha-Day,
  Ellis, and Marsh}}]{chadha2022axion}
\bibinfo{author}{\bibfnamefont{F.}~\bibnamefont{Chadha-Day}},
  \bibinfo{author}{\bibfnamefont{J.}~\bibnamefont{Ellis}}, \bibnamefont{and}
  \bibinfo{author}{\bibfnamefont{D.~J.} \bibnamefont{Marsh}},
  \bibinfo{journal}{Science advances} \textbf{\bibinfo{volume}{8}},
  \bibinfo{pages}{eabj3618} (\bibinfo{year}{2022}).

\bibitem[{\citenamefont{Dine and Fischler}(1983)}]{DINE1983137}
\bibinfo{author}{\bibfnamefont{M.}~\bibnamefont{Dine}} \bibnamefont{and}
  \bibinfo{author}{\bibfnamefont{W.}~\bibnamefont{Fischler}},
  \bibinfo{journal}{Physics Letters B} \textbf{\bibinfo{volume}{120}},
  \bibinfo{pages}{137} (\bibinfo{year}{1983}).

\bibitem[{\citenamefont{Hartle and Thorne}(1968)}]{1968ApJ...153..807H}
\bibinfo{author}{\bibfnamefont{J.~B.} \bibnamefont{Hartle}} \bibnamefont{and}
  \bibinfo{author}{\bibfnamefont{K.~S.} \bibnamefont{Thorne}},
  \bibinfo{journal}{\apj} \textbf{\bibinfo{volume}{153}}, \bibinfo{pages}{807}
  (\bibinfo{year}{1968}).

\bibitem[{\citenamefont{Pattersons and Sulaksono}(2021)}]{2021EPJC...81..698P}
\bibinfo{author}{\bibfnamefont{M.~L.} \bibnamefont{Pattersons}}
  \bibnamefont{and}
  \bibinfo{author}{\bibfnamefont{A.}~\bibnamefont{Sulaksono}},
  \bibinfo{journal}{European Physical Journal C} \textbf{\bibinfo{volume}{81}},
  \bibinfo{eid}{698} (\bibinfo{year}{2021}).

\bibitem[{\citenamefont{Psaltis and Özel}(2014)}]{Psaltis_2014}
\bibinfo{author}{\bibfnamefont{D.}~\bibnamefont{Psaltis}} \bibnamefont{and}
  \bibinfo{author}{\bibfnamefont{F.}~\bibnamefont{Özel}},
  \bibinfo{journal}{The Astrophysical Journal} \textbf{\bibinfo{volume}{792}},
  \bibinfo{pages}{87} (\bibinfo{year}{2014}).

\bibitem[{\citenamefont{Sedrakian}(2007)}]{Sedrakian:2006mq}
\bibinfo{author}{\bibfnamefont{A.}~\bibnamefont{Sedrakian}},
  \bibinfo{journal}{Prog. Part. Nucl. Phys.} \textbf{\bibinfo{volume}{58}},
  \bibinfo{pages}{168} (\bibinfo{year}{2007}), \eprint{nucl-th/0601086}.

\bibitem[{\citenamefont{Gudmundsson et~al.}(1983)\citenamefont{Gudmundsson,
  Pethick, and Epstein}}]{gudmundsson1983structure}
\bibinfo{author}{\bibfnamefont{E.~H.} \bibnamefont{Gudmundsson}},
  \bibinfo{author}{\bibfnamefont{C.}~\bibnamefont{Pethick}}, \bibnamefont{and}
  \bibinfo{author}{\bibfnamefont{R.~I.} \bibnamefont{Epstein}},
  \bibinfo{journal}{Astrophysical Journal, Part 1 (ISSN 0004-637X), vol. 272,
  Sept. 1, 1983, p. 286-300.} \textbf{\bibinfo{volume}{272}},
  \bibinfo{pages}{286} (\bibinfo{year}{1983}).

\bibitem[{\citenamefont{Zavlin}(2007)}]{Zavlin_2007}
\bibinfo{author}{\bibfnamefont{V.~E.} \bibnamefont{Zavlin}},
  \bibinfo{journal}{The Astrophysical Journal} \textbf{\bibinfo{volume}{665}},
  \bibinfo{pages}{L143} (\bibinfo{year}{2007}).

\bibitem[{\citenamefont{Beznogov et~al.}(2021)\citenamefont{Beznogov, Potekhin,
  and Yakovlev}}]{BEZNOGOV20211}
\bibinfo{author}{\bibfnamefont{M.}~\bibnamefont{Beznogov}},
  \bibinfo{author}{\bibfnamefont{A.}~\bibnamefont{Potekhin}}, \bibnamefont{and}
  \bibinfo{author}{\bibfnamefont{D.}~\bibnamefont{Yakovlev}},
  \bibinfo{journal}{Physics Reports} \textbf{\bibinfo{volume}{919}},
  \bibinfo{pages}{1} (\bibinfo{year}{2021}), \bibinfo{note}{heat blanketing
  envelopes of neutron stars}.

\bibitem[{\citenamefont{Buschmann et~al.}(2021)\citenamefont{Buschmann, Co,
  Dessert, and Safdi}}]{Buschmann:2019pfp}
\bibinfo{author}{\bibfnamefont{M.}~\bibnamefont{Buschmann}},
  \bibinfo{author}{\bibfnamefont{R.~T.} \bibnamefont{Co}},
  \bibinfo{author}{\bibfnamefont{C.}~\bibnamefont{Dessert}}, \bibnamefont{and}
  \bibinfo{author}{\bibfnamefont{B.~R.} \bibnamefont{Safdi}},
  \bibinfo{journal}{Phys. Rev. Lett.} \textbf{\bibinfo{volume}{126}},
  \bibinfo{pages}{021102} (\bibinfo{year}{2021}).

\bibitem[{\citenamefont{Dexheimer
  et~al.}(2017{\natexlab{a}})\citenamefont{Dexheimer, Franzon, Gomes, Farias,
  Avancini, and Schramm}}]{DEXHEIMER2017487}
\bibinfo{author}{\bibfnamefont{V.}~\bibnamefont{Dexheimer}},
  \bibinfo{author}{\bibfnamefont{B.}~\bibnamefont{Franzon}},
  \bibinfo{author}{\bibfnamefont{R.}~\bibnamefont{Gomes}},
  \bibinfo{author}{\bibfnamefont{R.}~\bibnamefont{Farias}},
  \bibinfo{author}{\bibfnamefont{S.}~\bibnamefont{Avancini}}, \bibnamefont{and}
  \bibinfo{author}{\bibfnamefont{S.}~\bibnamefont{Schramm}},
  \bibinfo{journal}{Physics Letters B} \textbf{\bibinfo{volume}{773}},
  \bibinfo{pages}{487} (\bibinfo{year}{2017}{\natexlab{a}}).

\bibitem[{\citenamefont{Sinha et~al.}(2013)\citenamefont{Sinha, Mukhopadhyay,
  and Sedrakian}}]{SINHA201343}
\bibinfo{author}{\bibfnamefont{M.}~\bibnamefont{Sinha}},
  \bibinfo{author}{\bibfnamefont{B.}~\bibnamefont{Mukhopadhyay}},
  \bibnamefont{and}
  \bibinfo{author}{\bibfnamefont{A.}~\bibnamefont{Sedrakian}},
  \bibinfo{journal}{Nuclear Physics A} \textbf{\bibinfo{volume}{898}},
  \bibinfo{pages}{43} (\bibinfo{year}{2013}).

\bibitem[{\citenamefont{Dexheimer
  et~al.}(2017{\natexlab{b}})\citenamefont{Dexheimer, Franzon, Gomes, Farias,
  Avancini, and Schramm}}]{Dexheimer:2017fhy}
\bibinfo{author}{\bibfnamefont{V.}~\bibnamefont{Dexheimer}},
  \bibinfo{author}{\bibfnamefont{B.}~\bibnamefont{Franzon}},
  \bibinfo{author}{\bibfnamefont{R.~O.} \bibnamefont{Gomes}},
  \bibinfo{author}{\bibfnamefont{R.~L.~S.} \bibnamefont{Farias}},
  \bibinfo{author}{\bibfnamefont{S.~S.} \bibnamefont{Avancini}},
  \bibnamefont{and} \bibinfo{author}{\bibfnamefont{S.}~\bibnamefont{Schramm}},
  \bibinfo{journal}{Astron. Nachr.} \textbf{\bibinfo{volume}{338}},
  \bibinfo{pages}{1052} (\bibinfo{year}{2017}{\natexlab{b}}).

\bibitem[{\citenamefont{Gomes and Vasconcellos}(2013)}]{2013arXiv1307.7450G}
\bibinfo{author}{\bibfnamefont{V.}~\bibnamefont{Gomes},
  \bibfnamefont{R.~O.~Dexheimer}} \bibnamefont{and}
  \bibinfo{author}{\bibfnamefont{C.~A.~Z.} \bibnamefont{Vasconcellos}},
  \bibinfo{journal}{arXiv e-prints}  (\bibinfo{year}{2013}).

\bibitem[{\citenamefont{Chatterjee et~al.}(2015)\citenamefont{Chatterjee,
  Elghozi, Novak, and Oertel}}]{10.1093/mnras/stu2706}
\bibinfo{author}{\bibfnamefont{D.}~\bibnamefont{Chatterjee}},
  \bibinfo{author}{\bibfnamefont{T.}~\bibnamefont{Elghozi}},
  \bibinfo{author}{\bibfnamefont{J.}~\bibnamefont{Novak}}, \bibnamefont{and}
  \bibinfo{author}{\bibfnamefont{M.}~\bibnamefont{Oertel}},
  \bibinfo{journal}{Monthly Notices of the Royal Astronomical Society}
  \textbf{\bibinfo{volume}{447}}, \bibinfo{pages}{3785} (\bibinfo{year}{2015}).

\bibitem[{\citenamefont{Lattimer and Prakash}(2001)}]{Lattimer_2001}
\bibinfo{author}{\bibfnamefont{J.~M.} \bibnamefont{Lattimer}} \bibnamefont{and}
  \bibinfo{author}{\bibfnamefont{M.}~\bibnamefont{Prakash}},
  \bibinfo{journal}{The Astrophysical Journal} \textbf{\bibinfo{volume}{550}},
  \bibinfo{pages}{426} (\bibinfo{year}{2001}).

\bibitem[{\citenamefont{Davis}(1985)}]{PhysRevD.32.3172}
\bibinfo{author}{\bibfnamefont{R.~L.} \bibnamefont{Davis}},
  \bibinfo{journal}{Phys. Rev. D} \textbf{\bibinfo{volume}{32}},
  \bibinfo{pages}{3172} (\bibinfo{year}{1985}).

\bibitem[{\citenamefont{Dexheimer et~al.}(2012)\citenamefont{Dexheimer,
  Negreiros, and Schramm}}]{dexheimer2012hybrid}
\bibinfo{author}{\bibfnamefont{V.}~\bibnamefont{Dexheimer}},
  \bibinfo{author}{\bibfnamefont{R.}~\bibnamefont{Negreiros}},
  \bibnamefont{and} \bibinfo{author}{\bibfnamefont{S.}~\bibnamefont{Schramm}},
  \bibinfo{journal}{The European Physical Journal A}
  \textbf{\bibinfo{volume}{48}}, \bibinfo{pages}{1} (\bibinfo{year}{2012}).

\bibitem[{\citenamefont{Potekhin et~al.}(2007)\citenamefont{Potekhin, Chabrier,
  and Yakovlev}}]{potekhin2007heat}
\bibinfo{author}{\bibfnamefont{A.~Y.} \bibnamefont{Potekhin}},
  \bibinfo{author}{\bibfnamefont{G.}~\bibnamefont{Chabrier}}, \bibnamefont{and}
  \bibinfo{author}{\bibfnamefont{D.~G.} \bibnamefont{Yakovlev}}, pp.
  \bibinfo{pages}{353--361} (\bibinfo{year}{2007}).

\bibitem[{\citenamefont{Potekhin et~al.}(2005)\citenamefont{Potekhin, Urpin,
  and Chabrier}}]{potekhin2005magnetic}
\bibinfo{author}{\bibfnamefont{A.~Y.} \bibnamefont{Potekhin}},
  \bibinfo{author}{\bibfnamefont{V.}~\bibnamefont{Urpin}}, \bibnamefont{and}
  \bibinfo{author}{\bibfnamefont{G.}~\bibnamefont{Chabrier}},
  \bibinfo{journal}{Astronomy \& Astrophysics} \textbf{\bibinfo{volume}{443}},
  \bibinfo{pages}{1025} (\bibinfo{year}{2005}).

\bibitem[{\citenamefont{Potekhin et~al.}(2015)\citenamefont{Potekhin, Pons, and
  Page}}]{potekhin2015neutron}
\bibinfo{author}{\bibfnamefont{A.~Y.} \bibnamefont{Potekhin}},
  \bibinfo{author}{\bibfnamefont{J.~A.} \bibnamefont{Pons}}, \bibnamefont{and}
  \bibinfo{author}{\bibfnamefont{D.}~\bibnamefont{Page}},
  \bibinfo{journal}{Space Science Reviews} \textbf{\bibinfo{volume}{191}},
  \bibinfo{pages}{239} (\bibinfo{year}{2015}).

\bibitem[{\citenamefont{Lopes and Menezes}(2015)}]{Lopes_2015}
\bibinfo{author}{\bibfnamefont{L.}~\bibnamefont{Lopes}} \bibnamefont{and}
  \bibinfo{author}{\bibfnamefont{D.}~\bibnamefont{Menezes}},
  \bibinfo{journal}{Journal of Cosmology and Astroparticle Physics}
  \textbf{\bibinfo{volume}{2015}}, \bibinfo{pages}{002} (\bibinfo{year}{2015}).

\bibitem[{\citenamefont{Geppert et~al.}(2004)\citenamefont{Geppert, Kuker, and
  Page}}]{refId0}
\bibinfo{author}{\bibfnamefont{U.}~\bibnamefont{Geppert}},
  \bibinfo{author}{\bibfnamefont{M.}~\bibnamefont{Kuker}}, \bibnamefont{and}
  \bibinfo{author}{\bibfnamefont{D.}~\bibnamefont{Page}},
  \bibinfo{journal}{A\&A} \textbf{\bibinfo{volume}{426}}, \bibinfo{pages}{267}
  (\bibinfo{year}{2004}).

\bibitem[{\citenamefont{Kaminker et~al.}(2006)\citenamefont{Kaminker, Yakovlev,
  Potekhin, Shibazaki, Shternin, and Gnedin}}]{Kaminker_2006}
\bibinfo{author}{\bibfnamefont{A.~D.} \bibnamefont{Kaminker}},
  \bibinfo{author}{\bibfnamefont{D.~G.} \bibnamefont{Yakovlev}},
  \bibinfo{author}{\bibfnamefont{A.~Y.} \bibnamefont{Potekhin}},
  \bibinfo{author}{\bibfnamefont{N.}~\bibnamefont{Shibazaki}},
  \bibinfo{author}{\bibfnamefont{P.~S.} \bibnamefont{Shternin}},
  \bibnamefont{and} \bibinfo{author}{\bibfnamefont{O.~Y.}
  \bibnamefont{Gnedin}}, \bibinfo{journal}{Monthly Notices of the Royal
  Astronomical Society} \textbf{\bibinfo{volume}{371}}, \bibinfo{pages}{477}
  (\bibinfo{year}{2006}).

\bibitem[{\citenamefont{Valyavin et~al.}(2014)\citenamefont{Valyavin, Shulyak,
  Wade, Antonyuk, Zharikov, Galazutdinov, Plachinda, Bagnulo, Machado, 'in
  lvarez et~al.}}]{Valyavin2014SuppressionOC}
\bibinfo{author}{\bibfnamefont{G.}~\bibnamefont{Valyavin}},
  \bibinfo{author}{\bibfnamefont{D.}~\bibnamefont{Shulyak}},
  \bibinfo{author}{\bibfnamefont{G.~A.} \bibnamefont{Wade}},
  \bibinfo{author}{\bibfnamefont{K.~A.} \bibnamefont{Antonyuk}},
  \bibinfo{author}{\bibfnamefont{S.}~\bibnamefont{Zharikov}},
  \bibinfo{author}{\bibfnamefont{G.~A.} \bibnamefont{Galazutdinov}},
  \bibinfo{author}{\bibfnamefont{S.~I.} \bibnamefont{Plachinda}},
  \bibinfo{author}{\bibfnamefont{S.}~\bibnamefont{Bagnulo}},
  \bibinfo{author}{\bibfnamefont{L.~F.} \bibnamefont{Machado}},
  \bibinfo{author}{\bibfnamefont{M.~B.} \bibnamefont{'in lvarez}},
  \bibnamefont{et~al.}, \bibinfo{journal}{Nature}
  \textbf{\bibinfo{volume}{515}}, \bibinfo{pages}{88} (\bibinfo{year}{2014}).

\bibitem[{\citenamefont{Kaspi and
  Beloborodov}(2017)}]{doi:10.1146/annurev-astro-081915-023329}
\bibinfo{author}{\bibfnamefont{V.~M.} \bibnamefont{Kaspi}} \bibnamefont{and}
  \bibinfo{author}{\bibfnamefont{A.~M.} \bibnamefont{Beloborodov}},
  \bibinfo{journal}{Annual Review of Astronomy and Astrophysics}
  \textbf{\bibinfo{volume}{55}}, \bibinfo{pages}{261} (\bibinfo{year}{2017}).

\bibitem[{\citenamefont{Gomes et~al.}(2017)\citenamefont{Gomes, Franzon,
  Dexheimer, and Schramm}}]{Gomes_2017}
\bibinfo{author}{\bibfnamefont{R.~O.} \bibnamefont{Gomes}},
  \bibinfo{author}{\bibfnamefont{B.}~\bibnamefont{Franzon}},
  \bibinfo{author}{\bibfnamefont{V.}~\bibnamefont{Dexheimer}},
  \bibnamefont{and} \bibinfo{author}{\bibfnamefont{S.}~\bibnamefont{Schramm}},
  \bibinfo{journal}{The Astrophysical Journal} \textbf{\bibinfo{volume}{850}},
  \bibinfo{pages}{20} (\bibinfo{year}{2017}).

\bibitem[{\citenamefont{Braithwaite and Spruit}(2004)}]{braithwaite2004fossil}
\bibinfo{author}{\bibfnamefont{J.}~\bibnamefont{Braithwaite}} \bibnamefont{and}
  \bibinfo{author}{\bibfnamefont{H.~C.} \bibnamefont{Spruit}},
  \bibinfo{journal}{nature} \textbf{\bibinfo{volume}{431}},
  \bibinfo{pages}{819} (\bibinfo{year}{2004}).

\bibitem[{\citenamefont{Bocquet et~al.}(1995)\citenamefont{Bocquet, Bonazzola,
  Gourgoulhon, and Novak}}]{bocquet1995rotating}
\bibinfo{author}{\bibfnamefont{M.}~\bibnamefont{Bocquet}},
  \bibinfo{author}{\bibfnamefont{S.}~\bibnamefont{Bonazzola}},
  \bibinfo{author}{\bibfnamefont{E.}~\bibnamefont{Gourgoulhon}},
  \bibnamefont{and} \bibinfo{author}{\bibfnamefont{J.}~\bibnamefont{Novak}},
  \bibinfo{journal}{arXiv preprint gr-qc/9503044}  (\bibinfo{year}{1995}).

\bibitem[{\citenamefont{Braithwaite}(2009)}]{10.1111/j.1365-2966.2008.14034.x}
\bibinfo{author}{\bibfnamefont{J.}~\bibnamefont{Braithwaite}},
  \bibinfo{journal}{Monthly Notices of the Royal Astronomical Society}
  \textbf{\bibinfo{volume}{397}}, \bibinfo{pages}{763} (\bibinfo{year}{2009}).

\bibitem[{\citenamefont{Braithwaite and
  Nordlund}(2006)}]{braithwaite2006stable}
\bibinfo{author}{\bibfnamefont{J.}~\bibnamefont{Braithwaite}} \bibnamefont{and}
  \bibinfo{author}{\bibfnamefont{{\AA}.}~\bibnamefont{Nordlund}},
  \bibinfo{journal}{Astronomy \& Astrophysics} \textbf{\bibinfo{volume}{450}},
  \bibinfo{pages}{1077} (\bibinfo{year}{2006}).

\bibitem[{\citenamefont{Potekhin et~al.}(2003)\citenamefont{Potekhin, Yakovlev,
  Chabrier, and Gnedin}}]{Potekhin_2003}
\bibinfo{author}{\bibfnamefont{A.~Y.} \bibnamefont{Potekhin}},
  \bibinfo{author}{\bibfnamefont{D.~G.} \bibnamefont{Yakovlev}},
  \bibinfo{author}{\bibfnamefont{G.}~\bibnamefont{Chabrier}}, \bibnamefont{and}
  \bibinfo{author}{\bibfnamefont{O.~Y.} \bibnamefont{Gnedin}},
  \bibinfo{journal}{The Astrophysical Journal} \textbf{\bibinfo{volume}{594}},
  \bibinfo{pages}{404} (\bibinfo{year}{2003}).

\bibitem[{\citenamefont{Kolomeitsev and Voskresensky}(2008)}]{Kolomeitsev_2008}
\bibinfo{author}{\bibfnamefont{E.~E.} \bibnamefont{Kolomeitsev}}
  \bibnamefont{and} \bibinfo{author}{\bibfnamefont{D.~N.}
  \bibnamefont{Voskresensky}}, \bibinfo{journal}{Physical Review C}
  \textbf{\bibinfo{volume}{77}} (\bibinfo{year}{2008}).

\bibitem[{\citenamefont{Leinson}(2000)}]{Leinson_2000}
\bibinfo{author}{\bibfnamefont{L.}~\bibnamefont{Leinson}},
  \bibinfo{journal}{Physics Letters B} \textbf{\bibinfo{volume}{473}},
  \bibinfo{pages}{318} (\bibinfo{year}{2000}).

\bibitem[{\citenamefont{Bhattacharya et~al.}(2018)\citenamefont{Bhattacharya,
  Mukhopadhyay, and Mukerjee}}]{10.1093/mnras/sty776}
\bibinfo{author}{\bibfnamefont{M.}~\bibnamefont{Bhattacharya}},
  \bibinfo{author}{\bibfnamefont{B.}~\bibnamefont{Mukhopadhyay}},
  \bibnamefont{and} \bibinfo{author}{\bibfnamefont{S.}~\bibnamefont{Mukerjee}},
  \bibinfo{journal}{Monthly Notices of the Royal Astronomical Society}
  \textbf{\bibinfo{volume}{477}}, \bibinfo{pages}{2705} (\bibinfo{year}{2018}).

\bibitem[{\citenamefont{Iwamoto et~al.}(1995)\citenamefont{Iwamoto, Qin,
  Fukugita, and Tsuruta}}]{PhysRevD.51.348}
\bibinfo{author}{\bibfnamefont{N.}~\bibnamefont{Iwamoto}},
  \bibinfo{author}{\bibfnamefont{L.}~\bibnamefont{Qin}},
  \bibinfo{author}{\bibfnamefont{M.}~\bibnamefont{Fukugita}}, \bibnamefont{and}
  \bibinfo{author}{\bibfnamefont{S.}~\bibnamefont{Tsuruta}},
  \bibinfo{journal}{Phys. Rev. D} \textbf{\bibinfo{volume}{51}},
  \bibinfo{pages}{348} (\bibinfo{year}{1995}).

\bibitem[{\citenamefont{Yakovlev et~al.}(2001)\citenamefont{Yakovlev, Gnedin,
  and Haensel}}]{2001PhR...354....1Y}
\bibinfo{author}{\bibfnamefont{A.}~\bibnamefont{Yakovlev},
  \bibfnamefont{D.G.~Kaminker}},
  \bibinfo{author}{\bibfnamefont{O.}~\bibnamefont{Gnedin}}, \bibnamefont{and}
  \bibinfo{author}{\bibfnamefont{P.}~\bibnamefont{Haensel}},
  \bibinfo{journal}{Phys. Rep.} \textbf{\bibinfo{volume}{354}},
  \bibinfo{pages}{1} (\bibinfo{year}{2001}).

\bibitem[{\citenamefont{Morris}(1986)}]{PhysRevD.34.843}
\bibinfo{author}{\bibfnamefont{D.~E.} \bibnamefont{Morris}},
  \bibinfo{journal}{Phys. Rev. D} \textbf{\bibinfo{volume}{34}},
  \bibinfo{pages}{843} (\bibinfo{year}{1986}).

\bibitem[{\citenamefont{Nomoto and Tsuruta}(1987)}]{nomoto1987cooling}
\bibinfo{author}{\bibfnamefont{K.}~\bibnamefont{Nomoto}} \bibnamefont{and}
  \bibinfo{author}{\bibfnamefont{S.}~\bibnamefont{Tsuruta}},
  \bibinfo{journal}{Astrophysical Journal, Part 1 (ISSN 0004-637X), vol. 312,
  Jan. 15, 1987, p. 711-726.} \textbf{\bibinfo{volume}{312}},
  \bibinfo{pages}{711} (\bibinfo{year}{1987}).

\bibitem[{\citenamefont{Peccei and Quinn}(1977{\natexlab{a}})}]{Peccei:1977hh}
\bibinfo{author}{\bibfnamefont{R.~D.} \bibnamefont{Peccei}} \bibnamefont{and}
  \bibinfo{author}{\bibfnamefont{H.~R.} \bibnamefont{Quinn}},
  \bibinfo{journal}{Phys. Rev. Lett.} \textbf{\bibinfo{volume}{38}},
  \bibinfo{pages}{1440} (\bibinfo{year}{1977}{\natexlab{a}}).

\bibitem[{\citenamefont{Peccei and Quinn}(1977{\natexlab{b}})}]{Peccei:1977ur}
\bibinfo{author}{\bibfnamefont{R.~D.} \bibnamefont{Peccei}} \bibnamefont{and}
  \bibinfo{author}{\bibfnamefont{H.~R.} \bibnamefont{Quinn}},
  \bibinfo{journal}{Phys. Rev. D} \textbf{\bibinfo{volume}{16}},
  \bibinfo{pages}{1791} (\bibinfo{year}{1977}{\natexlab{b}}).

\bibitem[{\citenamefont{Umeda et~al.}(1998)\citenamefont{Umeda, Iwamoto,
  Tsuruta, Qin, and Nomoto}}]{umeda1998axion}
\bibinfo{author}{\bibfnamefont{H.}~\bibnamefont{Umeda}},
  \bibinfo{author}{\bibfnamefont{N.}~\bibnamefont{Iwamoto}},
  \bibinfo{author}{\bibfnamefont{S.}~\bibnamefont{Tsuruta}},
  \bibinfo{author}{\bibfnamefont{L.}~\bibnamefont{Qin}}, \bibnamefont{and}
  \bibinfo{author}{\bibfnamefont{K.}~\bibnamefont{Nomoto}},
  \bibinfo{journal}{arXiv preprint astro-ph/9806337}  (\bibinfo{year}{1998}).

\bibitem[{\citenamefont{Kim}(1979)}]{PhysRevLett.43.103}
\bibinfo{author}{\bibfnamefont{J.~E.} \bibnamefont{Kim}},
  \bibinfo{journal}{Phys. Rev. Lett.} \textbf{\bibinfo{volume}{43}},
  \bibinfo{pages}{103} (\bibinfo{year}{1979}).

\bibitem[{\citenamefont{Shifman et~al.}(1980)\citenamefont{Shifman, Vainshtein,
  and Zakharov}}]{SHIFMAN1980493}
\bibinfo{author}{\bibfnamefont{M.}~\bibnamefont{Shifman}},
  \bibinfo{author}{\bibfnamefont{A.}~\bibnamefont{Vainshtein}},
  \bibnamefont{and} \bibinfo{author}{\bibfnamefont{V.}~\bibnamefont{Zakharov}},
  \bibinfo{journal}{Nuclear Physics B} \textbf{\bibinfo{volume}{166}},
  \bibinfo{pages}{493} (\bibinfo{year}{1980}), ISSN \bibinfo{issn}{0550-3213}.

\bibitem[{\citenamefont{Weinberg}(1978)}]{Weinberg:1977ma}
\bibinfo{author}{\bibfnamefont{S.}~\bibnamefont{Weinberg}},
  \bibinfo{journal}{Phys. Rev. Lett.} \textbf{\bibinfo{volume}{40}},
  \bibinfo{pages}{223} (\bibinfo{year}{1978}).

\bibitem[{\citenamefont{Dine et~al.}(1981)\citenamefont{Dine, Fischler, and
  Srednicki}}]{DINE1981199}
\bibinfo{author}{\bibfnamefont{M.}~\bibnamefont{Dine}},
  \bibinfo{author}{\bibfnamefont{W.}~\bibnamefont{Fischler}}, \bibnamefont{and}
  \bibinfo{author}{\bibfnamefont{M.}~\bibnamefont{Srednicki}},
  \bibinfo{journal}{Physics Letters B} \textbf{\bibinfo{volume}{104}},
  \bibinfo{pages}{199} (\bibinfo{year}{1981}), ISSN \bibinfo{issn}{0370-2693}.

\bibitem[{\citenamefont{Wilczek}(1978)}]{PhysRevLett.40.279}
\bibinfo{author}{\bibfnamefont{F.}~\bibnamefont{Wilczek}},
  \bibinfo{journal}{Phys. Rev. Lett.} \textbf{\bibinfo{volume}{40}},
  \bibinfo{pages}{279} (\bibinfo{year}{1978}).

\bibitem[{\citenamefont{Bogorad et~al.}(2019)\citenamefont{Bogorad, Hook, Kahn,
  and Soreq}}]{PhysRevLett.123.021801}
\bibinfo{author}{\bibfnamefont{Z.}~\bibnamefont{Bogorad}},
  \bibinfo{author}{\bibfnamefont{A.}~\bibnamefont{Hook}},
  \bibinfo{author}{\bibfnamefont{Y.}~\bibnamefont{Kahn}}, \bibnamefont{and}
  \bibinfo{author}{\bibfnamefont{Y.}~\bibnamefont{Soreq}},
  \bibinfo{journal}{Phys. Rev. Lett.} \textbf{\bibinfo{volume}{123}},
  \bibinfo{pages}{021801} (\bibinfo{year}{2019}).

\bibitem[{\citenamefont{Dessert et~al.}(2022)\citenamefont{Dessert, Long, and
  Safdi}}]{Dessert_2022}
\bibinfo{author}{\bibfnamefont{C.}~\bibnamefont{Dessert}},
  \bibinfo{author}{\bibfnamefont{A.~J.} \bibnamefont{Long}}, \bibnamefont{and}
  \bibinfo{author}{\bibfnamefont{B.~R.} \bibnamefont{Safdi}},
  \bibinfo{journal}{Physical Review Letters} \textbf{\bibinfo{volume}{128}}
  (\bibinfo{year}{2022}).

\bibitem[{\citenamefont{Duffy and van Bibber}(2009)}]{Duffy_2009}
\bibinfo{author}{\bibfnamefont{L.~D.} \bibnamefont{Duffy}} \bibnamefont{and}
  \bibinfo{author}{\bibfnamefont{K.}~\bibnamefont{van Bibber}},
  \bibinfo{journal}{New Journal of Physics} \textbf{\bibinfo{volume}{11}},
  \bibinfo{pages}{105008} (\bibinfo{year}{2009}).

\bibitem[{\citenamefont{Abbott and Sikivie}(1983)}]{ABBOTT1983133}
\bibinfo{author}{\bibfnamefont{L.}~\bibnamefont{Abbott}} \bibnamefont{and}
  \bibinfo{author}{\bibfnamefont{P.}~\bibnamefont{Sikivie}},
  \bibinfo{journal}{Physics Letters B} \textbf{\bibinfo{volume}{120}},
  \bibinfo{pages}{133} (\bibinfo{year}{1983}).

\bibitem[{\citenamefont{Preskill et~al.}(1983)\citenamefont{Preskill, Wise, and
  Wilczek}}]{PRESKILL1983127}
\bibinfo{author}{\bibfnamefont{J.}~\bibnamefont{Preskill}},
  \bibinfo{author}{\bibfnamefont{M.~B.} \bibnamefont{Wise}}, \bibnamefont{and}
  \bibinfo{author}{\bibfnamefont{F.}~\bibnamefont{Wilczek}},
  \bibinfo{journal}{Physics Letters B} \textbf{\bibinfo{volume}{120}},
  \bibinfo{pages}{127} (\bibinfo{year}{1983}).

\bibitem[{\citenamefont{Kahn et~al.}(2016)\citenamefont{Kahn, Safdi, and
  Thaler}}]{PhysRevLett.117.141801}
\bibinfo{author}{\bibfnamefont{Y.}~\bibnamefont{Kahn}},
  \bibinfo{author}{\bibfnamefont{B.~R.} \bibnamefont{Safdi}}, \bibnamefont{and}
  \bibinfo{author}{\bibfnamefont{J.}~\bibnamefont{Thaler}},
  \bibinfo{journal}{Phys. Rev. Lett.} \textbf{\bibinfo{volume}{117}},
  \bibinfo{pages}{141801} (\bibinfo{year}{2016}).

\bibitem[{\citenamefont{Foster et~al.}(2018)\citenamefont{Foster, Rodd, and
  Safdi}}]{PhysRevD.97.123006}
\bibinfo{author}{\bibfnamefont{J.~W.} \bibnamefont{Foster}},
  \bibinfo{author}{\bibfnamefont{N.~L.} \bibnamefont{Rodd}}, \bibnamefont{and}
  \bibinfo{author}{\bibfnamefont{B.~R.} \bibnamefont{Safdi}},
  \bibinfo{journal}{Phys. Rev. D} \textbf{\bibinfo{volume}{97}},
  \bibinfo{pages}{123006} (\bibinfo{year}{2018}).

\bibitem[{\citenamefont{Marsh et~al.}(2017)\citenamefont{Marsh, Russell,
  Fabian, McNamara, Nulsen, and Reynolds}}]{Marsh_2017}
\bibinfo{author}{\bibfnamefont{M.~D.} \bibnamefont{Marsh}},
  \bibinfo{author}{\bibfnamefont{H.~R.} \bibnamefont{Russell}},
  \bibinfo{author}{\bibfnamefont{A.~C.} \bibnamefont{Fabian}},
  \bibinfo{author}{\bibfnamefont{B.~R.} \bibnamefont{McNamara}},
  \bibinfo{author}{\bibfnamefont{P.}~\bibnamefont{Nulsen}}, \bibnamefont{and}
  \bibinfo{author}{\bibfnamefont{C.~S.} \bibnamefont{Reynolds}},
  \bibinfo{journal}{Journal of Cosmology and Astroparticle Physics}
  \textbf{\bibinfo{volume}{2017}}, \bibinfo{pages}{036} (\bibinfo{year}{2017}).

\bibitem[{\citenamefont{Beznogov et~al.}(2018)\citenamefont{Beznogov, Rrapaj,
  Page, and Reddy}}]{PhysRevC.98.035802}
\bibinfo{author}{\bibfnamefont{M.~V.} \bibnamefont{Beznogov}},
  \bibinfo{author}{\bibfnamefont{E.}~\bibnamefont{Rrapaj}},
  \bibinfo{author}{\bibfnamefont{D.}~\bibnamefont{Page}}, \bibnamefont{and}
  \bibinfo{author}{\bibfnamefont{S.}~\bibnamefont{Reddy}},
  \bibinfo{journal}{Phys. Rev. C} \textbf{\bibinfo{volume}{98}},
  \bibinfo{pages}{035802} (\bibinfo{year}{2018}).

\bibitem[{\citenamefont{Beznogov et~al.}(2020)\citenamefont{Beznogov, Page, and
  Ramirez-Ruiz}}]{beznogov2020thermal}
\bibinfo{author}{\bibfnamefont{M.~V.} \bibnamefont{Beznogov}},
  \bibinfo{author}{\bibfnamefont{D.}~\bibnamefont{Page}}, \bibnamefont{and}
  \bibinfo{author}{\bibfnamefont{E.}~\bibnamefont{Ramirez-Ruiz}},
  \bibinfo{journal}{The Astrophysical Journal} \textbf{\bibinfo{volume}{888}},
  \bibinfo{pages}{97} (\bibinfo{year}{2020}).

\bibitem[{\citenamefont{Yakovlev et~al.}(2003)\citenamefont{Yakovlev,
  Levenfish, and Haensel}}]{yakovlev2003thermal}
\bibinfo{author}{\bibfnamefont{D.}~\bibnamefont{Yakovlev}},
  \bibinfo{author}{\bibfnamefont{K.}~\bibnamefont{Levenfish}},
  \bibnamefont{and} \bibinfo{author}{\bibfnamefont{P.}~\bibnamefont{Haensel}},
  \bibinfo{journal}{Astronomy \& Astrophysics} \textbf{\bibinfo{volume}{407}},
  \bibinfo{pages}{265} (\bibinfo{year}{2003}).

\bibitem[{\citenamefont{Adams et~al.}(2022)\citenamefont{Adams, Aggarwal,
  Agrawal, Balafendiev, Bartram, Baryakhtar, Bekker, Belov, Berggren, Berlin
  et~al.}}]{adams2022axion}
\bibinfo{author}{\bibfnamefont{C.}~\bibnamefont{Adams}},
  \bibinfo{author}{\bibfnamefont{N.}~\bibnamefont{Aggarwal}},
  \bibinfo{author}{\bibfnamefont{A.}~\bibnamefont{Agrawal}},
  \bibinfo{author}{\bibfnamefont{R.}~\bibnamefont{Balafendiev}},
  \bibinfo{author}{\bibfnamefont{C.}~\bibnamefont{Bartram}},
  \bibinfo{author}{\bibfnamefont{M.}~\bibnamefont{Baryakhtar}},
  \bibinfo{author}{\bibfnamefont{H.}~\bibnamefont{Bekker}},
  \bibinfo{author}{\bibfnamefont{P.}~\bibnamefont{Belov}},
  \bibinfo{author}{\bibfnamefont{K.}~\bibnamefont{Berggren}},
  \bibinfo{author}{\bibfnamefont{A.}~\bibnamefont{Berlin}},
  \bibnamefont{et~al.}, \bibinfo{journal}{arXiv preprint arXiv:2203.14923}
  (\bibinfo{year}{2022}).

\bibitem[{\citenamefont{Backes et~al.}(2021)\citenamefont{Backes, Palken,
  Kenany, Brubaker, Cahn, Droster, Hilton, Ghosh, Jackson, Lamoreaux
  et~al.}}]{backes2021quantum}
\bibinfo{author}{\bibfnamefont{K.~M.} \bibnamefont{Backes}},
  \bibinfo{author}{\bibfnamefont{D.~A.} \bibnamefont{Palken}},
  \bibinfo{author}{\bibfnamefont{S.~A.} \bibnamefont{Kenany}},
  \bibinfo{author}{\bibfnamefont{B.~M.} \bibnamefont{Brubaker}},
  \bibinfo{author}{\bibfnamefont{S.}~\bibnamefont{Cahn}},
  \bibinfo{author}{\bibfnamefont{A.}~\bibnamefont{Droster}},
  \bibinfo{author}{\bibfnamefont{G.~C.} \bibnamefont{Hilton}},
  \bibinfo{author}{\bibfnamefont{S.}~\bibnamefont{Ghosh}},
  \bibinfo{author}{\bibfnamefont{H.}~\bibnamefont{Jackson}},
  \bibinfo{author}{\bibfnamefont{S.~K.} \bibnamefont{Lamoreaux}},
  \bibnamefont{et~al.}, \bibinfo{journal}{Nature}
  \textbf{\bibinfo{volume}{590}}, \bibinfo{pages}{238} (\bibinfo{year}{2021}).

\bibitem[{\citenamefont{Chen et~al.}(2022)\citenamefont{Chen, Kobakhidze,
  O’Hare, Picker, and Pierobon}}]{chen2022phenomenology}
\bibinfo{author}{\bibfnamefont{Z.}~\bibnamefont{Chen}},
  \bibinfo{author}{\bibfnamefont{A.}~\bibnamefont{Kobakhidze}},
  \bibinfo{author}{\bibfnamefont{C.~A.} \bibnamefont{O’Hare}},
  \bibinfo{author}{\bibfnamefont{Z.~S.} \bibnamefont{Picker}},
  \bibnamefont{and} \bibinfo{author}{\bibfnamefont{G.}~\bibnamefont{Pierobon}},
  \bibinfo{journal}{The European Physical Journal C}
  \textbf{\bibinfo{volume}{82}}, \bibinfo{pages}{1} (\bibinfo{year}{2022}).

\bibitem[{\citenamefont{Arza}(2019)}]{arza2019photon}
\bibinfo{author}{\bibfnamefont{A.}~\bibnamefont{Arza}}, \bibinfo{journal}{The
  European Physical Journal C} \textbf{\bibinfo{volume}{79}},
  \bibinfo{pages}{1} (\bibinfo{year}{2019}).

\bibitem[{\citenamefont{Song et~al.}(2024)\citenamefont{Song, Su, and
  Wu}}]{song2024polarization}
\bibinfo{author}{\bibfnamefont{N.}~\bibnamefont{Song}},
  \bibinfo{author}{\bibfnamefont{L.}~\bibnamefont{Su}}, \bibnamefont{and}
  \bibinfo{author}{\bibfnamefont{L.}~\bibnamefont{Wu}}, \bibinfo{journal}{arXiv
  preprint arXiv:2402.15144}  (\bibinfo{year}{2024}).

\bibitem[{\citenamefont{Tjemsland et~al.}(2024)\citenamefont{Tjemsland,
  McDonald, and Witte}}]{tjemsland2024adiabatic}
\bibinfo{author}{\bibfnamefont{J.}~\bibnamefont{Tjemsland}},
  \bibinfo{author}{\bibfnamefont{J.}~\bibnamefont{McDonald}}, \bibnamefont{and}
  \bibinfo{author}{\bibfnamefont{S.~J.} \bibnamefont{Witte}},
  \bibinfo{journal}{Physical Review D} \textbf{\bibinfo{volume}{109}},
  \bibinfo{pages}{023015} (\bibinfo{year}{2024}).

\bibitem[{\citenamefont{Du et~al.}(2018)\citenamefont{Du, Force, Khatiwada,
  Lentz, Ottens, Rosenberg, Rybka, Carosi, Woollett, Bowring
  et~al.}}]{PhysRevLett.120.151301}
\bibinfo{author}{\bibfnamefont{N.}~\bibnamefont{Du}},
  \bibinfo{author}{\bibfnamefont{N.}~\bibnamefont{Force}},
  \bibinfo{author}{\bibfnamefont{R.}~\bibnamefont{Khatiwada}},
  \bibinfo{author}{\bibfnamefont{E.}~\bibnamefont{Lentz}},
  \bibinfo{author}{\bibfnamefont{R.}~\bibnamefont{Ottens}},
  \bibinfo{author}{\bibfnamefont{L.~J.} \bibnamefont{Rosenberg}},
  \bibinfo{author}{\bibfnamefont{G.}~\bibnamefont{Rybka}},
  \bibinfo{author}{\bibfnamefont{G.}~\bibnamefont{Carosi}},
  \bibinfo{author}{\bibfnamefont{N.}~\bibnamefont{Woollett}},
  \bibinfo{author}{\bibfnamefont{D.}~\bibnamefont{Bowring}},
  \bibnamefont{et~al.} (\bibinfo{collaboration}{ADMX Collaboration}),
  \bibinfo{journal}{Phys. Rev. Lett.} \textbf{\bibinfo{volume}{120}},
  \bibinfo{pages}{151301} (\bibinfo{year}{2018}).

\bibitem[{\citenamefont{Braine et~al.}(2020)\citenamefont{Braine, Cervantes,
  Crisosto, Du, Kimes, Rosenberg, Rybka, Yang, Bowring, Chou
  et~al.}}]{PhysRevLett.124.101303}
\bibinfo{author}{\bibfnamefont{T.}~\bibnamefont{Braine}},
  \bibinfo{author}{\bibfnamefont{R.}~\bibnamefont{Cervantes}},
  \bibinfo{author}{\bibfnamefont{N.}~\bibnamefont{Crisosto}},
  \bibinfo{author}{\bibfnamefont{N.}~\bibnamefont{Du}},
  \bibinfo{author}{\bibfnamefont{S.}~\bibnamefont{Kimes}},
  \bibinfo{author}{\bibfnamefont{L.~J.} \bibnamefont{Rosenberg}},
  \bibinfo{author}{\bibfnamefont{G.}~\bibnamefont{Rybka}},
  \bibinfo{author}{\bibfnamefont{J.}~\bibnamefont{Yang}},
  \bibinfo{author}{\bibfnamefont{D.}~\bibnamefont{Bowring}},
  \bibinfo{author}{\bibfnamefont{A.~S.} \bibnamefont{Chou}},
  \bibnamefont{et~al.} (\bibinfo{collaboration}{ADMX Collaboration}),
  \bibinfo{journal}{Phys. Rev. Lett.} \textbf{\bibinfo{volume}{124}},
  \bibinfo{pages}{101303} (\bibinfo{year}{2020}).

\bibitem[{\citenamefont{Bartram et~al.}(2021)\citenamefont{Bartram, Braine,
  Burns, Cervantes, Crisosto, Du, Korandla, Leum, Mohapatra, Nitta
  et~al.}}]{PhysRevLett.127.261803}
\bibinfo{author}{\bibfnamefont{C.}~\bibnamefont{Bartram}},
  \bibinfo{author}{\bibfnamefont{T.}~\bibnamefont{Braine}},
  \bibinfo{author}{\bibfnamefont{E.}~\bibnamefont{Burns}},
  \bibinfo{author}{\bibfnamefont{R.}~\bibnamefont{Cervantes}},
  \bibinfo{author}{\bibfnamefont{N.}~\bibnamefont{Crisosto}},
  \bibinfo{author}{\bibfnamefont{N.}~\bibnamefont{Du}},
  \bibinfo{author}{\bibfnamefont{H.}~\bibnamefont{Korandla}},
  \bibinfo{author}{\bibfnamefont{G.}~\bibnamefont{Leum}},
  \bibinfo{author}{\bibfnamefont{P.}~\bibnamefont{Mohapatra}},
  \bibinfo{author}{\bibfnamefont{T.}~\bibnamefont{Nitta}}, \bibnamefont{et~al.}
  (\bibinfo{collaboration}{ADMX Collaboration}), \bibinfo{journal}{Phys. Rev.
  Lett.} \textbf{\bibinfo{volume}{127}}, \bibinfo{pages}{261803}
  (\bibinfo{year}{2021}).

\bibitem[{\citenamefont{Asztalos et~al.}(2001)\citenamefont{Asztalos, Daw,
  Peng, Rosenberg, Hagmann, Kinion, Stoeffl, van Bibber, Sikivie, Sullivan
  et~al.}}]{PhysRevD.64.092003}
\bibinfo{author}{\bibfnamefont{S.}~\bibnamefont{Asztalos}},
  \bibinfo{author}{\bibfnamefont{E.}~\bibnamefont{Daw}},
  \bibinfo{author}{\bibfnamefont{H.}~\bibnamefont{Peng}},
  \bibinfo{author}{\bibfnamefont{L.~J.} \bibnamefont{Rosenberg}},
  \bibinfo{author}{\bibfnamefont{C.}~\bibnamefont{Hagmann}},
  \bibinfo{author}{\bibfnamefont{D.}~\bibnamefont{Kinion}},
  \bibinfo{author}{\bibfnamefont{W.}~\bibnamefont{Stoeffl}},
  \bibinfo{author}{\bibfnamefont{K.}~\bibnamefont{van Bibber}},
  \bibinfo{author}{\bibfnamefont{P.}~\bibnamefont{Sikivie}},
  \bibinfo{author}{\bibfnamefont{N.~S.} \bibnamefont{Sullivan}},
  \bibnamefont{et~al.}, \bibinfo{journal}{Phys. Rev. D}
  \textbf{\bibinfo{volume}{64}}, \bibinfo{pages}{092003}
  (\bibinfo{year}{2001}).

\bibitem[{\citenamefont{Leinson}(2019)}]{Leinson_2019}
\bibinfo{author}{\bibfnamefont{L.~B.} \bibnamefont{Leinson}},
  \bibinfo{journal}{Journal of Cosmology and Astroparticle Physics}
  \textbf{\bibinfo{volume}{2019}}, \bibinfo{pages}{031} (\bibinfo{year}{2019}).

\bibitem[{\citenamefont{Zhitnitsky}(1980)}]{Zhitnitsky:1980tq}
\bibinfo{author}{\bibfnamefont{A.~R.} \bibnamefont{Zhitnitsky}},
  \bibinfo{journal}{Sov. J. Nucl. Phys.} \textbf{\bibinfo{volume}{31}},
  \bibinfo{pages}{260} (\bibinfo{year}{1980}).

\bibitem[{\citenamefont{Brun et~al.}(2019)\citenamefont{Brun, Caldwell,
  Chevalier, Dvali, Freire, Garutti, Heyminck, Jochum, Knirck, Kramer
  et~al.}}]{brun2019new}
\bibinfo{author}{\bibfnamefont{P.}~\bibnamefont{Brun}},
  \bibinfo{author}{\bibfnamefont{A.}~\bibnamefont{Caldwell}},
  \bibinfo{author}{\bibfnamefont{L.}~\bibnamefont{Chevalier}},
  \bibinfo{author}{\bibfnamefont{G.}~\bibnamefont{Dvali}},
  \bibinfo{author}{\bibfnamefont{P.}~\bibnamefont{Freire}},
  \bibinfo{author}{\bibfnamefont{E.}~\bibnamefont{Garutti}},
  \bibinfo{author}{\bibfnamefont{S.}~\bibnamefont{Heyminck}},
  \bibinfo{author}{\bibfnamefont{J.}~\bibnamefont{Jochum}},
  \bibinfo{author}{\bibfnamefont{S.}~\bibnamefont{Knirck}},
  \bibinfo{author}{\bibfnamefont{M.}~\bibnamefont{Kramer}},
  \bibnamefont{et~al.}, \bibinfo{journal}{The European Physical Journal C}
  \textbf{\bibinfo{volume}{79}}, \bibinfo{pages}{1} (\bibinfo{year}{2019}).

\bibitem[{\citenamefont{Lawson et~al.}(2019)\citenamefont{Lawson, Millar,
  Pancaldi, Vitagliano, and Wilczek}}]{PhysRevLett.123.141802}
\bibinfo{author}{\bibfnamefont{M.}~\bibnamefont{Lawson}},
  \bibinfo{author}{\bibfnamefont{A.~J.} \bibnamefont{Millar}},
  \bibinfo{author}{\bibfnamefont{M.}~\bibnamefont{Pancaldi}},
  \bibinfo{author}{\bibfnamefont{E.}~\bibnamefont{Vitagliano}},
  \bibnamefont{and} \bibinfo{author}{\bibfnamefont{F.}~\bibnamefont{Wilczek}},
  \bibinfo{journal}{Phys. Rev. Lett.} \textbf{\bibinfo{volume}{123}},
  \bibinfo{pages}{141802} (\bibinfo{year}{2019}).

\bibitem[{\citenamefont{Budker et~al.}(2014)\citenamefont{Budker, Graham,
  Ledbetter, Rajendran, and Sushkov}}]{PhysRevX.4.021030}
\bibinfo{author}{\bibfnamefont{D.}~\bibnamefont{Budker}},
  \bibinfo{author}{\bibfnamefont{P.~W.} \bibnamefont{Graham}},
  \bibinfo{author}{\bibfnamefont{M.}~\bibnamefont{Ledbetter}},
  \bibinfo{author}{\bibfnamefont{S.}~\bibnamefont{Rajendran}},
  \bibnamefont{and} \bibinfo{author}{\bibfnamefont{A.~O.}
  \bibnamefont{Sushkov}}, \bibinfo{journal}{Phys. Rev. X}
  \textbf{\bibinfo{volume}{4}}, \bibinfo{pages}{021030} (\bibinfo{year}{2014}).

\bibitem[{\citenamefont{Aybas et~al.}(2021)\citenamefont{Aybas, Adam,
  Blumenthal, Gramolin, Johnson, Kleyheeg, Afach, Blanchard, Centers, Garcon
  et~al.}}]{PhysRevLett.126.141802}
\bibinfo{author}{\bibfnamefont{D.}~\bibnamefont{Aybas}},
  \bibinfo{author}{\bibfnamefont{J.}~\bibnamefont{Adam}},
  \bibinfo{author}{\bibfnamefont{E.}~\bibnamefont{Blumenthal}},
  \bibinfo{author}{\bibfnamefont{A.~V.} \bibnamefont{Gramolin}},
  \bibinfo{author}{\bibfnamefont{D.}~\bibnamefont{Johnson}},
  \bibinfo{author}{\bibfnamefont{A.}~\bibnamefont{Kleyheeg}},
  \bibinfo{author}{\bibfnamefont{S.}~\bibnamefont{Afach}},
  \bibinfo{author}{\bibfnamefont{J.~W.} \bibnamefont{Blanchard}},
  \bibinfo{author}{\bibfnamefont{G.~P.} \bibnamefont{Centers}},
  \bibinfo{author}{\bibfnamefont{A.}~\bibnamefont{Garcon}},
  \bibnamefont{et~al.}, \bibinfo{journal}{Phys. Rev. Lett.}
  \textbf{\bibinfo{volume}{126}}, \bibinfo{pages}{141802}
  (\bibinfo{year}{2021}).

\bibitem[{\citenamefont{Garcon et~al.}(2017)\citenamefont{Garcon, Aybas,
  Blanchard, Centers, Figueroa, Graham, Kimball, Rajendran, Sendra, Sushkov
  et~al.}}]{Garcon_2018}
\bibinfo{author}{\bibfnamefont{A.}~\bibnamefont{Garcon}},
  \bibinfo{author}{\bibfnamefont{D.}~\bibnamefont{Aybas}},
  \bibinfo{author}{\bibfnamefont{J.~W.} \bibnamefont{Blanchard}},
  \bibinfo{author}{\bibfnamefont{G.}~\bibnamefont{Centers}},
  \bibinfo{author}{\bibfnamefont{N.~L.} \bibnamefont{Figueroa}},
  \bibinfo{author}{\bibfnamefont{P.~W.} \bibnamefont{Graham}},
  \bibinfo{author}{\bibfnamefont{D.~F.~J.} \bibnamefont{Kimball}},
  \bibinfo{author}{\bibfnamefont{S.}~\bibnamefont{Rajendran}},
  \bibinfo{author}{\bibfnamefont{M.~G.} \bibnamefont{Sendra}},
  \bibinfo{author}{\bibfnamefont{A.~O.} \bibnamefont{Sushkov}},
  \bibnamefont{et~al.}, \bibinfo{journal}{Quantum Science and Technology}
  \textbf{\bibinfo{volume}{3}}, \bibinfo{pages}{014008} (\bibinfo{year}{2017}).

\bibitem[{\citenamefont{Silva-Feaver et~al.}(2017)\citenamefont{Silva-Feaver,
  Chaudhuri, Cho, Dawson, Graham, Irwin, Kuenstner, Li, Mardon, Moseley
  et~al.}}]{7750582}
\bibinfo{author}{\bibfnamefont{M.}~\bibnamefont{Silva-Feaver}},
  \bibinfo{author}{\bibfnamefont{S.}~\bibnamefont{Chaudhuri}},
  \bibinfo{author}{\bibfnamefont{H.-M.} \bibnamefont{Cho}},
  \bibinfo{author}{\bibfnamefont{C.}~\bibnamefont{Dawson}},
  \bibinfo{author}{\bibfnamefont{P.}~\bibnamefont{Graham}},
  \bibinfo{author}{\bibfnamefont{K.}~\bibnamefont{Irwin}},
  \bibinfo{author}{\bibfnamefont{S.}~\bibnamefont{Kuenstner}},
  \bibinfo{author}{\bibfnamefont{D.}~\bibnamefont{Li}},
  \bibinfo{author}{\bibfnamefont{J.}~\bibnamefont{Mardon}},
  \bibinfo{author}{\bibfnamefont{H.}~\bibnamefont{Moseley}},
  \bibnamefont{et~al.}, \bibinfo{journal}{IEEE Transactions on Applied
  Superconductivity} \textbf{\bibinfo{volume}{27}}, \bibinfo{pages}{1}
  (\bibinfo{year}{2017}).

\bibitem[{\citenamefont{Ouellet et~al.}(2019)\citenamefont{Ouellet, Salemi,
  Foster, Henning, Bogorad, Conrad, Formaggio, Kahn, Minervini, Radovinsky
  et~al.}}]{PhysRevLett.122.121802}
\bibinfo{author}{\bibfnamefont{J.~L.} \bibnamefont{Ouellet}},
  \bibinfo{author}{\bibfnamefont{C.~P.} \bibnamefont{Salemi}},
  \bibinfo{author}{\bibfnamefont{J.~W.} \bibnamefont{Foster}},
  \bibinfo{author}{\bibfnamefont{R.}~\bibnamefont{Henning}},
  \bibinfo{author}{\bibfnamefont{Z.}~\bibnamefont{Bogorad}},
  \bibinfo{author}{\bibfnamefont{J.~M.} \bibnamefont{Conrad}},
  \bibinfo{author}{\bibfnamefont{J.~A.} \bibnamefont{Formaggio}},
  \bibinfo{author}{\bibfnamefont{Y.}~\bibnamefont{Kahn}},
  \bibinfo{author}{\bibfnamefont{J.}~\bibnamefont{Minervini}},
  \bibinfo{author}{\bibfnamefont{A.}~\bibnamefont{Radovinsky}},
  \bibnamefont{et~al.}, \bibinfo{journal}{Phys. Rev. Lett.}
  \textbf{\bibinfo{volume}{122}}, \bibinfo{pages}{121802}
  (\bibinfo{year}{2019}).

\bibitem[{\citenamefont{Salemi et~al.}(2021)\citenamefont{Salemi, Foster,
  Ouellet, Gavin, Pappas, Cheng, Richardson, Henning, Kahn, Nguyen
  et~al.}}]{PhysRevLett.127.081801}
\bibinfo{author}{\bibfnamefont{C.~P.} \bibnamefont{Salemi}},
  \bibinfo{author}{\bibfnamefont{J.~W.} \bibnamefont{Foster}},
  \bibinfo{author}{\bibfnamefont{J.~L.} \bibnamefont{Ouellet}},
  \bibinfo{author}{\bibfnamefont{A.}~\bibnamefont{Gavin}},
  \bibinfo{author}{\bibfnamefont{K.~M.~W.} \bibnamefont{Pappas}},
  \bibinfo{author}{\bibfnamefont{S.}~\bibnamefont{Cheng}},
  \bibinfo{author}{\bibfnamefont{K.~A.} \bibnamefont{Richardson}},
  \bibinfo{author}{\bibfnamefont{R.}~\bibnamefont{Henning}},
  \bibinfo{author}{\bibfnamefont{Y.}~\bibnamefont{Kahn}},
  \bibinfo{author}{\bibfnamefont{R.}~\bibnamefont{Nguyen}},
  \bibnamefont{et~al.}, \bibinfo{journal}{Phys. Rev. Lett.}
  \textbf{\bibinfo{volume}{127}}, \bibinfo{pages}{081801}
  (\bibinfo{year}{2021}).

\bibitem[{\citenamefont{Arvanitaki et~al.}(2010)\citenamefont{Arvanitaki,
  Dimopoulos, Dubovsky, Kaloper, and March-Russell}}]{PhysRevD.81.123530}
\bibinfo{author}{\bibfnamefont{A.}~\bibnamefont{Arvanitaki}},
  \bibinfo{author}{\bibfnamefont{S.}~\bibnamefont{Dimopoulos}},
  \bibinfo{author}{\bibfnamefont{S.}~\bibnamefont{Dubovsky}},
  \bibinfo{author}{\bibfnamefont{N.}~\bibnamefont{Kaloper}}, \bibnamefont{and}
  \bibinfo{author}{\bibfnamefont{J.}~\bibnamefont{March-Russell}},
  \bibinfo{journal}{Phys. Rev. D} \textbf{\bibinfo{volume}{81}},
  \bibinfo{pages}{123530} (\bibinfo{year}{2010}).

\bibitem[{\citenamefont{Arvanitaki and Dubovsky}(2011)}]{PhysRevD.83.044026}
\bibinfo{author}{\bibfnamefont{A.}~\bibnamefont{Arvanitaki}} \bibnamefont{and}
  \bibinfo{author}{\bibfnamefont{S.}~\bibnamefont{Dubovsky}},
  \bibinfo{journal}{Phys. Rev. D} \textbf{\bibinfo{volume}{83}},
  \bibinfo{pages}{044026} (\bibinfo{year}{2011}).

\bibitem[{\citenamefont{Armengaud et~al.}(2019)\citenamefont{Armengaud, Attié,
  Basso, Brun, Bykovskiy, Carmona, Castel, Cebrián, Cicoli, Civitani
  et~al.}}]{Armengaud_2019}
\bibinfo{author}{\bibfnamefont{E.}~\bibnamefont{Armengaud}},
  \bibinfo{author}{\bibfnamefont{D.}~\bibnamefont{Attié}},
  \bibinfo{author}{\bibfnamefont{S.}~\bibnamefont{Basso}},
  \bibinfo{author}{\bibfnamefont{P.}~\bibnamefont{Brun}},
  \bibinfo{author}{\bibfnamefont{N.}~\bibnamefont{Bykovskiy}},
  \bibinfo{author}{\bibfnamefont{J.}~\bibnamefont{Carmona}},
  \bibinfo{author}{\bibfnamefont{J.}~\bibnamefont{Castel}},
  \bibinfo{author}{\bibfnamefont{S.}~\bibnamefont{Cebrián}},
  \bibinfo{author}{\bibfnamefont{M.}~\bibnamefont{Cicoli}},
  \bibinfo{author}{\bibfnamefont{M.}~\bibnamefont{Civitani}},
  \bibnamefont{et~al.}, \bibinfo{journal}{Journal of Cosmology and
  Astroparticle Physics} \textbf{\bibinfo{volume}{2019}}, \bibinfo{pages}{047}
  (\bibinfo{year}{2019}).

\bibitem[{\citenamefont{Liu et~al.}(2022)\citenamefont{Liu, Dona, Hoshino,
  Knirck, Kurinsky, Malaker, Miller, Sonnenschein, Awida, Barry
  et~al.}}]{Liu_2022}
\bibinfo{author}{\bibfnamefont{J.}~\bibnamefont{Liu}},
  \bibinfo{author}{\bibfnamefont{K.}~\bibnamefont{Dona}},
  \bibinfo{author}{\bibfnamefont{G.}~\bibnamefont{Hoshino}},
  \bibinfo{author}{\bibfnamefont{S.}~\bibnamefont{Knirck}},
  \bibinfo{author}{\bibfnamefont{N.}~\bibnamefont{Kurinsky}},
  \bibinfo{author}{\bibfnamefont{M.}~\bibnamefont{Malaker}},
  \bibinfo{author}{\bibfnamefont{D.~W.} \bibnamefont{Miller}},
  \bibinfo{author}{\bibfnamefont{A.}~\bibnamefont{Sonnenschein}},
  \bibinfo{author}{\bibfnamefont{M.~H.} \bibnamefont{Awida}},
  \bibinfo{author}{\bibfnamefont{P.~S.} \bibnamefont{Barry}},
  \bibnamefont{et~al.}, \bibinfo{journal}{Physical Review Letters}
  \textbf{\bibinfo{volume}{128}} (\bibinfo{year}{2022}), ISSN
  \bibinfo{issn}{1079-7114}.

\bibitem[{\citenamefont{Co et~al.}(2020)\citenamefont{Co, Hall, Harigaya,
  Olive, and Verner}}]{Co_2020}
\bibinfo{author}{\bibfnamefont{R.~T.} \bibnamefont{Co}},
  \bibinfo{author}{\bibfnamefont{L.~J.} \bibnamefont{Hall}},
  \bibinfo{author}{\bibfnamefont{K.}~\bibnamefont{Harigaya}},
  \bibinfo{author}{\bibfnamefont{K.~A.} \bibnamefont{Olive}}, \bibnamefont{and}
  \bibinfo{author}{\bibfnamefont{S.}~\bibnamefont{Verner}},
  \bibinfo{journal}{Journal of Cosmology and Astroparticle Physics}
  \textbf{\bibinfo{volume}{2020}}, \bibinfo{pages}{036} (\bibinfo{year}{2020}).

\bibitem[{\citenamefont{Gorghetto et~al.}(2021)\citenamefont{Gorghetto, Hardy,
  and Villadoro}}]{10.21468/SciPostPhys.10.2.050}
\bibinfo{author}{\bibfnamefont{M.}~\bibnamefont{Gorghetto}},
  \bibinfo{author}{\bibfnamefont{E.}~\bibnamefont{Hardy}}, \bibnamefont{and}
  \bibinfo{author}{\bibfnamefont{G.}~\bibnamefont{Villadoro}},
  \bibinfo{journal}{SciPost Phys.} \textbf{\bibinfo{volume}{10}},
  \bibinfo{pages}{050} (\bibinfo{year}{2021}).

\bibitem[{\citenamefont{Arvanitaki and
  Geraci}(2014)}]{arvanitaki2014resonantly}
\bibinfo{author}{\bibfnamefont{A.}~\bibnamefont{Arvanitaki}} \bibnamefont{and}
  \bibinfo{author}{\bibfnamefont{A.~A.} \bibnamefont{Geraci}},
  \bibinfo{journal}{Physical Review Letters} \textbf{\bibinfo{volume}{113}},
  \bibinfo{pages}{161801} (\bibinfo{year}{2014}).

\bibitem[{\citenamefont{Carosi et~al.}(2020)\citenamefont{Carosi, Rybka, and
  Van~Bibber}}]{carosi2020microwave}
\bibinfo{author}{\bibfnamefont{G.}~\bibnamefont{Carosi}},
  \bibinfo{author}{\bibfnamefont{G.}~\bibnamefont{Rybka}}, \bibnamefont{and}
  \bibinfo{author}{\bibfnamefont{K.}~\bibnamefont{Van~Bibber}},
  \emph{\bibinfo{title}{Microwave cavities and detectors for axion research}}
  (\bibinfo{publisher}{Springer}, \bibinfo{year}{2020}).

\bibitem[{\citenamefont{Lella et~al.}(2023)\citenamefont{Lella, Carenza,
  Lucente, Giannotti, and Mirizzi}}]{PhysRevD.107.103017}
\bibinfo{author}{\bibfnamefont{A.}~\bibnamefont{Lella}},
  \bibinfo{author}{\bibfnamefont{P.}~\bibnamefont{Carenza}},
  \bibinfo{author}{\bibfnamefont{G.}~\bibnamefont{Lucente}},
  \bibinfo{author}{\bibfnamefont{M.}~\bibnamefont{Giannotti}},
  \bibnamefont{and} \bibinfo{author}{\bibfnamefont{A.}~\bibnamefont{Mirizzi}},
  \bibinfo{journal}{Phys. Rev. D} \textbf{\bibinfo{volume}{107}},
  \bibinfo{pages}{103017} (\bibinfo{year}{2023}).

\bibitem[{\citenamefont{Lella et~al.}(2024)\citenamefont{Lella, Carenza, Co',
  Lucente, Giannotti, Mirizzi, and Rauscher}}]{PhysRevD.109.023001}
\bibinfo{author}{\bibfnamefont{A.}~\bibnamefont{Lella}},
  \bibinfo{author}{\bibfnamefont{P.}~\bibnamefont{Carenza}},
  \bibinfo{author}{\bibfnamefont{G.}~\bibnamefont{Co'}},
  \bibinfo{author}{\bibfnamefont{G.}~\bibnamefont{Lucente}},
  \bibinfo{author}{\bibfnamefont{M.}~\bibnamefont{Giannotti}},
  \bibinfo{author}{\bibfnamefont{A.}~\bibnamefont{Mirizzi}}, \bibnamefont{and}
  \bibinfo{author}{\bibfnamefont{T.}~\bibnamefont{Rauscher}},
  \bibinfo{journal}{Phys. Rev. D} \textbf{\bibinfo{volume}{109}},
  \bibinfo{pages}{023001} (\bibinfo{year}{2024}).

\bibitem[{\citenamefont{Yadav et~al.}(2024)\citenamefont{Yadav, Mishra, Sarkar,
  and Singh}}]{yadav2024thermal}
\bibinfo{author}{\bibfnamefont{S.}~\bibnamefont{Yadav}},
  \bibinfo{author}{\bibfnamefont{M.}~\bibnamefont{Mishra}},
  \bibinfo{author}{\bibfnamefont{T.~G.} \bibnamefont{Sarkar}},
  \bibnamefont{and} \bibinfo{author}{\bibfnamefont{C.~R.} \bibnamefont{Singh}},
  \bibinfo{journal}{The European Physical Journal C}
  \textbf{\bibinfo{volume}{84}}, \bibinfo{pages}{225} (\bibinfo{year}{2024}).

\bibitem[{\citenamefont{Hook et~al.}(2018)\citenamefont{Hook, Kahn, Safdi, and
  Sun}}]{hook2018radio}
\bibinfo{author}{\bibfnamefont{A.}~\bibnamefont{Hook}},
  \bibinfo{author}{\bibfnamefont{Y.}~\bibnamefont{Kahn}},
  \bibinfo{author}{\bibfnamefont{B.~R.} \bibnamefont{Safdi}}, \bibnamefont{and}
  \bibinfo{author}{\bibfnamefont{Z.}~\bibnamefont{Sun}},
  \bibinfo{journal}{Phys. Rev. Lett.} \textbf{\bibinfo{volume}{121}},
  \bibinfo{pages}{241102} (\bibinfo{year}{2018}).

\bibitem[{\citenamefont{Raffelt and Stodolsky}(1988)}]{PhysRevD.37.1237}
\bibinfo{author}{\bibfnamefont{G.}~\bibnamefont{Raffelt}} \bibnamefont{and}
  \bibinfo{author}{\bibfnamefont{L.}~\bibnamefont{Stodolsky}},
  \bibinfo{journal}{Phys. Rev. D} \textbf{\bibinfo{volume}{37}},
  \bibinfo{pages}{1237} (\bibinfo{year}{1988}).

\bibitem[{\citenamefont{Brown et~al.}(2018)\citenamefont{Brown, Cumming,
  Fattoyev, Horowitz, Page, and Reddy}}]{PhysRevLett.120.182701}
\bibinfo{author}{\bibfnamefont{E.~F.} \bibnamefont{Brown}},
  \bibinfo{author}{\bibfnamefont{A.}~\bibnamefont{Cumming}},
  \bibinfo{author}{\bibfnamefont{F.~J.} \bibnamefont{Fattoyev}},
  \bibinfo{author}{\bibfnamefont{C.~J.} \bibnamefont{Horowitz}},
  \bibinfo{author}{\bibfnamefont{D.}~\bibnamefont{Page}}, \bibnamefont{and}
  \bibinfo{author}{\bibfnamefont{S.}~\bibnamefont{Reddy}},
  \bibinfo{journal}{Phys. Rev. Lett.} \textbf{\bibinfo{volume}{120}},
  \bibinfo{pages}{182701} (\bibinfo{year}{2018}).

\bibitem[{\citenamefont{Jonker et~al.}(2007)\citenamefont{Jonker, Steeghs,
  Chakrabarty, and Juett}}]{Jonker_2007}
\bibinfo{author}{\bibfnamefont{P.~G.} \bibnamefont{Jonker}},
  \bibinfo{author}{\bibfnamefont{D.}~\bibnamefont{Steeghs}},
  \bibinfo{author}{\bibfnamefont{D.}~\bibnamefont{Chakrabarty}},
  \bibnamefont{and} \bibinfo{author}{\bibfnamefont{A.~M.} \bibnamefont{Juett}},
  \bibinfo{journal}{The Astrophysical Journal} \textbf{\bibinfo{volume}{665}},
  \bibinfo{pages}{L147} (\bibinfo{year}{2007}).

\bibitem[{\citenamefont{Heinke et~al.}(2009)\citenamefont{Heinke, Jonker,
  Wijnands, Deloye, and Taam}}]{Heinke_2009}
\bibinfo{author}{\bibfnamefont{C.~O.} \bibnamefont{Heinke}},
  \bibinfo{author}{\bibfnamefont{P.~G.} \bibnamefont{Jonker}},
  \bibinfo{author}{\bibfnamefont{R.}~\bibnamefont{Wijnands}},
  \bibinfo{author}{\bibfnamefont{C.~J.} \bibnamefont{Deloye}},
  \bibnamefont{and} \bibinfo{author}{\bibfnamefont{R.~E.} \bibnamefont{Taam}},
  \bibinfo{journal}{The Astrophysical Journal} \textbf{\bibinfo{volume}{691}},
  \bibinfo{pages}{1035} (\bibinfo{year}{2009}).

\bibitem[{\citenamefont{Witte et~al.}(2021)\citenamefont{Witte, Noordhuis,
  Edwards, and Weniger}}]{witte2021axion}
\bibinfo{author}{\bibfnamefont{S.~J.} \bibnamefont{Witte}},
  \bibinfo{author}{\bibfnamefont{D.}~\bibnamefont{Noordhuis}},
  \bibinfo{author}{\bibfnamefont{T.~D.} \bibnamefont{Edwards}},
  \bibnamefont{and} \bibinfo{author}{\bibfnamefont{C.}~\bibnamefont{Weniger}},
  \bibinfo{journal}{Physical Review D} \textbf{\bibinfo{volume}{104}},
  \bibinfo{pages}{103030} (\bibinfo{year}{2021}).

\bibitem[{\citenamefont{McDonald et~al.}(2023)\citenamefont{McDonald,
  Garbrecht, and Millington}}]{McDonald_2023}
\bibinfo{author}{\bibfnamefont{J.}~\bibnamefont{McDonald}},
  \bibinfo{author}{\bibfnamefont{B.}~\bibnamefont{Garbrecht}},
  \bibnamefont{and}
  \bibinfo{author}{\bibfnamefont{P.}~\bibnamefont{Millington}},
  \bibinfo{journal}{Journal of Cosmology and Astroparticle Physics}
  \textbf{\bibinfo{volume}{2023}}, \bibinfo{pages}{031} (\bibinfo{year}{2023}).

\bibitem[{\citenamefont{Millar et~al.}(2021)\citenamefont{Millar, Baum, Lawson,
  and Marsh}}]{millar2021axion}
\bibinfo{author}{\bibfnamefont{A.~J.} \bibnamefont{Millar}},
  \bibinfo{author}{\bibfnamefont{S.}~\bibnamefont{Baum}},
  \bibinfo{author}{\bibfnamefont{M.}~\bibnamefont{Lawson}}, \bibnamefont{and}
  \bibinfo{author}{\bibfnamefont{M.~D.} \bibnamefont{Marsh}},
  \bibinfo{journal}{Journal of Cosmology and Astroparticle Physics}
  \textbf{\bibinfo{volume}{2021}}, \bibinfo{pages}{013} (\bibinfo{year}{2021}).

\bibitem[{\citenamefont{Tolman}(1939)}]{PhysRev.55.364}
\bibinfo{author}{\bibfnamefont{R.~C.} \bibnamefont{Tolman}},
  \bibinfo{journal}{Phys. Rev.} \textbf{\bibinfo{volume}{55}},
  \bibinfo{pages}{364} (\bibinfo{year}{1939}).

\bibitem[{\citenamefont{Oppenheimer and Volkoff}(1939)}]{PhysRev.55.374}
\bibinfo{author}{\bibfnamefont{J.~R.} \bibnamefont{Oppenheimer}}
  \bibnamefont{and} \bibinfo{author}{\bibfnamefont{G.~M.}
  \bibnamefont{Volkoff}}, \bibinfo{journal}{Phys. Rev.}
  \textbf{\bibinfo{volume}{55}}, \bibinfo{pages}{374} (\bibinfo{year}{1939}).

\bibitem[{\citenamefont{Mignani et~al.}(2013)\citenamefont{Mignani, Putte,
  Cropper, Turolla, Zane, Pellizza, Bignone, Sartore, and
  Treves}}]{Mignani_2013}
\bibinfo{author}{\bibfnamefont{R.~P.} \bibnamefont{Mignani}},
  \bibinfo{author}{\bibfnamefont{D.~V.} \bibnamefont{Putte}},
  \bibinfo{author}{\bibfnamefont{M.}~\bibnamefont{Cropper}},
  \bibinfo{author}{\bibfnamefont{R.}~\bibnamefont{Turolla}},
  \bibinfo{author}{\bibfnamefont{S.}~\bibnamefont{Zane}},
  \bibinfo{author}{\bibfnamefont{L.~J.} \bibnamefont{Pellizza}},
  \bibinfo{author}{\bibfnamefont{L.~A.} \bibnamefont{Bignone}},
  \bibinfo{author}{\bibfnamefont{N.}~\bibnamefont{Sartore}}, \bibnamefont{and}
  \bibinfo{author}{\bibfnamefont{A.}~\bibnamefont{Treves}},
  \bibinfo{journal}{Monthly Notices of the Royal Astronomical Society}
  \textbf{\bibinfo{volume}{429}}, \bibinfo{pages}{3517} (\bibinfo{year}{2013}).

\bibitem[{\citenamefont{Tetzlaff et~al.}(2011)\citenamefont{Tetzlaff,
  Eisenbeiss, Neuhäuser, and Hohle}}]{10.1111/j.1365-2966.2011.19302.x}
\bibinfo{author}{\bibfnamefont{N.}~\bibnamefont{Tetzlaff}},
  \bibinfo{author}{\bibfnamefont{T.}~\bibnamefont{Eisenbeiss}},
  \bibinfo{author}{\bibfnamefont{R.}~\bibnamefont{Neuhäuser}},
  \bibnamefont{and} \bibinfo{author}{\bibfnamefont{M.~M.} \bibnamefont{Hohle}},
  \bibinfo{journal}{Monthly Notices of the Royal Astronomical Society}
  \textbf{\bibinfo{volume}{417}}, \bibinfo{pages}{617} (\bibinfo{year}{2011}).

\bibitem[{\citenamefont{Dessert et~al.}(2020)\citenamefont{Dessert, Foster, and
  Safdi}}]{Dessert_2020}
\bibinfo{author}{\bibfnamefont{C.}~\bibnamefont{Dessert}},
  \bibinfo{author}{\bibfnamefont{J.~W.} \bibnamefont{Foster}},
  \bibnamefont{and} \bibinfo{author}{\bibfnamefont{B.~R.} \bibnamefont{Safdi}},
  \bibinfo{journal}{The Astrophysical Journal} \textbf{\bibinfo{volume}{904}},
  \bibinfo{pages}{42} (\bibinfo{year}{2020}).

\bibitem[{\citenamefont{Chatterjee et~al.}(2019)\citenamefont{Chatterjee,
  Novak, and Oertel}}]{chatterjee2019magnetic}
\bibinfo{author}{\bibfnamefont{D.}~\bibnamefont{Chatterjee}},
  \bibinfo{author}{\bibfnamefont{J.}~\bibnamefont{Novak}}, \bibnamefont{and}
  \bibinfo{author}{\bibfnamefont{M.}~\bibnamefont{Oertel}},
  \bibinfo{journal}{Physical Review C} \textbf{\bibinfo{volume}{99}},
  \bibinfo{pages}{055811} (\bibinfo{year}{2019}).

\bibitem[{\citenamefont{Dexheimer
  et~al.}(2017{\natexlab{c}})\citenamefont{Dexheimer, Franzon, and
  Schramm}}]{dexheimer2017self}
\bibinfo{author}{\bibfnamefont{V.}~\bibnamefont{Dexheimer}},
  \bibinfo{author}{\bibfnamefont{B.}~\bibnamefont{Franzon}}, \bibnamefont{and}
  \bibinfo{author}{\bibfnamefont{S.}~\bibnamefont{Schramm}}, in
  \emph{\bibinfo{booktitle}{Journal of Physics: Conference Series}}
  (\bibinfo{organization}{IOP Publishing}, \bibinfo{year}{2017}{\natexlab{c}}),
  vol. \bibinfo{volume}{861}, p. \bibinfo{pages}{012012}.

\bibitem[{\citenamefont{Bonazzola et~al.}(1993)\citenamefont{Bonazzola,
  Gourgoulhon, Salgado, and Marck}}]{bonazzola1993axisymmetric}
\bibinfo{author}{\bibfnamefont{S.}~\bibnamefont{Bonazzola}},
  \bibinfo{author}{\bibfnamefont{E.}~\bibnamefont{Gourgoulhon}},
  \bibinfo{author}{\bibfnamefont{M.}~\bibnamefont{Salgado}}, \bibnamefont{and}
  \bibinfo{author}{\bibfnamefont{J.}~\bibnamefont{Marck}},
  \bibinfo{journal}{Astronomy and Astrophysics (ISSN 0004-6361), vol. 278, no.
  2, p. 421-443} \textbf{\bibinfo{volume}{278}}, \bibinfo{pages}{421}
  (\bibinfo{year}{1993}).

\bibitem[{\citenamefont{Konno et~al.}(1999)\citenamefont{Konno, Obata, and
  Kojima}}]{konno1999deformation}
\bibinfo{author}{\bibfnamefont{K.}~\bibnamefont{Konno}},
  \bibinfo{author}{\bibfnamefont{T.}~\bibnamefont{Obata}}, \bibnamefont{and}
  \bibinfo{author}{\bibfnamefont{Y.}~\bibnamefont{Kojima}},
  \bibinfo{journal}{arXiv preprint gr-qc/9910038}  (\bibinfo{year}{1999}).

\bibitem[{\citenamefont{Bowers and Liang}(1974)}]{bowers1974anisotropic}
\bibinfo{author}{\bibfnamefont{R.~L.} \bibnamefont{Bowers}} \bibnamefont{and}
  \bibinfo{author}{\bibfnamefont{E.}~\bibnamefont{Liang}},
  \bibinfo{journal}{Astrophysical Journal, Vol. 188, p. 657 (1974)}
  \textbf{\bibinfo{volume}{188}}, \bibinfo{pages}{657} (\bibinfo{year}{1974}).

\bibitem[{\citenamefont{Prakash et~al.}(1997)\citenamefont{Prakash, Bombaci,
  Prakash, Ellis, Lattimer, and Knorren}}]{PRAKASH19971}
\bibinfo{author}{\bibfnamefont{M.}~\bibnamefont{Prakash}},
  \bibinfo{author}{\bibfnamefont{I.}~\bibnamefont{Bombaci}},
  \bibinfo{author}{\bibfnamefont{M.}~\bibnamefont{Prakash}},
  \bibinfo{author}{\bibfnamefont{P.~J.} \bibnamefont{Ellis}},
  \bibinfo{author}{\bibfnamefont{J.~M.} \bibnamefont{Lattimer}},
  \bibnamefont{and} \bibinfo{author}{\bibfnamefont{R.}~\bibnamefont{Knorren}},
  \bibinfo{journal}{Physics Reports} \textbf{\bibinfo{volume}{280}},
  \bibinfo{pages}{1} (\bibinfo{year}{1997}).

\bibitem[{\citenamefont{Beznogov and Yakovlev}(2015)}]{beznogov2015statistical}
\bibinfo{author}{\bibfnamefont{M.}~\bibnamefont{Beznogov}} \bibnamefont{and}
  \bibinfo{author}{\bibfnamefont{D.}~\bibnamefont{Yakovlev}},
  \bibinfo{journal}{Monthly Notices of the Royal Astronomical Society}
  \textbf{\bibinfo{volume}{447}}, \bibinfo{pages}{1598} (\bibinfo{year}{2015}).

\bibitem[{\citenamefont{Gusakov et~al.}(2005)\citenamefont{Gusakov, Kaminker,
  Yakovlev, and Gnedin}}]{Gusakov_2005}
\bibinfo{author}{\bibfnamefont{M.~E.} \bibnamefont{Gusakov}},
  \bibinfo{author}{\bibfnamefont{A.~D.} \bibnamefont{Kaminker}},
  \bibinfo{author}{\bibfnamefont{D.~G.} \bibnamefont{Yakovlev}},
  \bibnamefont{and} \bibinfo{author}{\bibfnamefont{O.~Y.}
  \bibnamefont{Gnedin}}, \bibinfo{journal}{Monthly Notices of the Royal
  Astronomical Society} \textbf{\bibinfo{volume}{363}}, \bibinfo{pages}{555}
  (\bibinfo{year}{2005}).

\bibitem[{\citenamefont{Schneider et~al.}(2019)\citenamefont{Schneider,
  Constantinou, Muccioli, and Prakash}}]{PhysRevC.100.025803}
\bibinfo{author}{\bibfnamefont{A.~S.} \bibnamefont{Schneider}},
  \bibinfo{author}{\bibfnamefont{C.}~\bibnamefont{Constantinou}},
  \bibinfo{author}{\bibfnamefont{B.}~\bibnamefont{Muccioli}}, \bibnamefont{and}
  \bibinfo{author}{\bibfnamefont{M.}~\bibnamefont{Prakash}},
  \bibinfo{journal}{Phys. Rev. C} \textbf{\bibinfo{volume}{100}},
  \bibinfo{pages}{025803} (\bibinfo{year}{2019}).

\bibitem[{\citenamefont{Akmal et~al.}(1998)\citenamefont{Akmal, Pandharipande,
  and Ravenhall}}]{PhysRevC.58.1804}
\bibinfo{author}{\bibfnamefont{A.}~\bibnamefont{Akmal}},
  \bibinfo{author}{\bibfnamefont{V.~R.} \bibnamefont{Pandharipande}},
  \bibnamefont{and} \bibinfo{author}{\bibfnamefont{D.~G.}
  \bibnamefont{Ravenhall}}, \bibinfo{journal}{Phys. Rev. C}
  \textbf{\bibinfo{volume}{58}}, \bibinfo{pages}{1804} (\bibinfo{year}{1998}).

\bibitem[{\citenamefont{Flowers et~al.}(1976)\citenamefont{Flowers, Ruderman,
  and Sutherland}}]{Flowers:1976ux}
\bibinfo{author}{\bibfnamefont{E.}~\bibnamefont{Flowers}},
  \bibinfo{author}{\bibfnamefont{M.}~\bibnamefont{Ruderman}}, \bibnamefont{and}
  \bibinfo{author}{\bibfnamefont{P.}~\bibnamefont{Sutherland}},
  \bibinfo{journal}{Astrophys. J.} \textbf{\bibinfo{volume}{205}},
  \bibinfo{pages}{541} (\bibinfo{year}{1976}).

\bibitem[{\citenamefont{Broderick et~al.}(2000)\citenamefont{Broderick,
  Prakash, and Lattimer}}]{Broderick_2000}
\bibinfo{author}{\bibfnamefont{A.}~\bibnamefont{Broderick}},
  \bibinfo{author}{\bibfnamefont{M.}~\bibnamefont{Prakash}}, \bibnamefont{and}
  \bibinfo{author}{\bibfnamefont{J.~M.} \bibnamefont{Lattimer}},
  \bibinfo{journal}{The Astrophysical Journal} \textbf{\bibinfo{volume}{537}},
  \bibinfo{pages}{351} (\bibinfo{year}{2000}).

\bibitem[{\citenamefont{Page}(2016)}]{2016ascl.soft09009P}
\bibinfo{author}{\bibfnamefont{D.}~\bibnamefont{Page}},
  \bibinfo{eid}{ascl:1609.009} (\bibinfo{year}{2016}), \eprint{1609.009}.

\bibitem[{\citenamefont{Page et~al.}(2006{\natexlab{a}})\citenamefont{Page,
  Geppert, and Weber}}]{2006NuPhA777497P}
\bibinfo{author}{\bibfnamefont{D.}~\bibnamefont{Page}},
  \bibinfo{author}{\bibfnamefont{U.}~\bibnamefont{Geppert}}, \bibnamefont{and}
  \bibinfo{author}{\bibfnamefont{F.}~\bibnamefont{Weber}},
  \bibinfo{journal}{Nucl. Phys. A} \textbf{\bibinfo{volume}{777}},
  \bibinfo{pages}{497} (\bibinfo{year}{2006}{\natexlab{a}}).

\bibitem[{\citenamefont{Yakovlev and
  Pethick}(2004)}]{doi:10.1146/annurev.astro.42.053102.134013}
\bibinfo{author}{\bibfnamefont{D.}~\bibnamefont{Yakovlev}} \bibnamefont{and}
  \bibinfo{author}{\bibfnamefont{C.}~\bibnamefont{Pethick}},
  \bibinfo{journal}{Annual Review of Astronomy and Astrophysics}
  \textbf{\bibinfo{volume}{42}}, \bibinfo{pages}{169} (\bibinfo{year}{2004}).

\bibitem[{\citenamefont{Page et~al.}(2011)\citenamefont{Page, Prakash,
  Lattimer, and Steiner}}]{PhysRevLett.106.081101}
\bibinfo{author}{\bibfnamefont{D.}~\bibnamefont{Page}},
  \bibinfo{author}{\bibfnamefont{M.}~\bibnamefont{Prakash}},
  \bibinfo{author}{\bibfnamefont{J.~M.} \bibnamefont{Lattimer}},
  \bibnamefont{and} \bibinfo{author}{\bibfnamefont{A.~W.}
  \bibnamefont{Steiner}}, \bibinfo{journal}{Phys. Rev. Lett.}
  \textbf{\bibinfo{volume}{106}}, \bibinfo{pages}{081101}
  (\bibinfo{year}{2011}).

\bibitem[{\citenamefont{Potekhin and Yakovlev}(2001)}]{refId01}
\bibinfo{author}{\bibfnamefont{A.~Y.} \bibnamefont{Potekhin}} \bibnamefont{and}
  \bibinfo{author}{\bibfnamefont{D.~G.} \bibnamefont{Yakovlev}},
  \bibinfo{journal}{A\&A} \textbf{\bibinfo{volume}{374}}, \bibinfo{pages}{213}
  (\bibinfo{year}{2001}).

\bibitem[{\citenamefont{Kaminker et~al.}(2014)\citenamefont{Kaminker, Kaurov,
  Potekhin, and Yakovlev}}]{10.1093/mnras/stu1102}
\bibinfo{author}{\bibfnamefont{A.~D.} \bibnamefont{Kaminker}},
  \bibinfo{author}{\bibfnamefont{A.~A.} \bibnamefont{Kaurov}},
  \bibinfo{author}{\bibfnamefont{A.~Y.} \bibnamefont{Potekhin}},
  \bibnamefont{and} \bibinfo{author}{\bibfnamefont{D.~G.}
  \bibnamefont{Yakovlev}}, \bibinfo{journal}{Monthly Notices of the Royal
  Astronomical Society} \textbf{\bibinfo{volume}{442}}, \bibinfo{pages}{3484}
  (\bibinfo{year}{2014}).

\bibitem[{\citenamefont{Yakovlev et~al.}(2004)\citenamefont{Yakovlev, Gnedin,
  Kaminker, Levenfish, and Potekhin}}]{YAKOVLEV2004523}
\bibinfo{author}{\bibfnamefont{D.}~\bibnamefont{Yakovlev}},
  \bibinfo{author}{\bibfnamefont{O.}~\bibnamefont{Gnedin}},
  \bibinfo{author}{\bibfnamefont{A.}~\bibnamefont{Kaminker}},
  \bibinfo{author}{\bibfnamefont{K.}~\bibnamefont{Levenfish}},
  \bibnamefont{and} \bibinfo{author}{\bibfnamefont{A.}~\bibnamefont{Potekhin}},
  \bibinfo{journal}{Advances in Space Research} \textbf{\bibinfo{volume}{33}},
  \bibinfo{pages}{523} (\bibinfo{year}{2004}).

\bibitem[{\citenamefont{Reddy and Zhou}(2022)}]{Reddy:2021rln}
\bibinfo{author}{\bibfnamefont{S.}~\bibnamefont{Reddy}} \bibnamefont{and}
  \bibinfo{author}{\bibfnamefont{D.}~\bibnamefont{Zhou}},
  \bibinfo{journal}{Phys. Rev. D} \textbf{\bibinfo{volume}{105}},
  \bibinfo{pages}{023026} (\bibinfo{year}{2022}), \eprint{2107.06279}.

\bibitem[{\citenamefont{Sedrakian}(2016)}]{PhysRevD.93.065044}
\bibinfo{author}{\bibfnamefont{A.}~\bibnamefont{Sedrakian}},
  \bibinfo{journal}{Phys. Rev. D} \textbf{\bibinfo{volume}{93}},
  \bibinfo{pages}{065044} (\bibinfo{year}{2016}).

\bibitem[{\citenamefont{Keller and Sedrakian}(2013)}]{Keller_2013}
\bibinfo{author}{\bibfnamefont{J.}~\bibnamefont{Keller}} \bibnamefont{and}
  \bibinfo{author}{\bibfnamefont{A.}~\bibnamefont{Sedrakian}},
  \bibinfo{journal}{Nuclear Physics A} \textbf{\bibinfo{volume}{897}},
  \bibinfo{pages}{62} (\bibinfo{year}{2013}).

\bibitem[{\citenamefont{Page et~al.}(2006{\natexlab{b}})\citenamefont{Page,
  Geppert, and Weber}}]{Page:2005fq}
\bibinfo{author}{\bibfnamefont{D.}~\bibnamefont{Page}},
  \bibinfo{author}{\bibfnamefont{U.}~\bibnamefont{Geppert}}, \bibnamefont{and}
  \bibinfo{author}{\bibfnamefont{F.}~\bibnamefont{Weber}},
  \bibinfo{journal}{Nucl. Phys. A} \textbf{\bibinfo{volume}{777}},
  \bibinfo{pages}{497} (\bibinfo{year}{2006}{\natexlab{b}}),
  \eprint{astro-ph/0508056}.

\bibitem[{\citenamefont{Iwamoto}(1984)}]{PhysRevLett.53.1198}
\bibinfo{author}{\bibfnamefont{N.}~\bibnamefont{Iwamoto}},
  \bibinfo{journal}{Phys. Rev. Lett.} \textbf{\bibinfo{volume}{53}},
  \bibinfo{pages}{1198} (\bibinfo{year}{1984}).

\bibitem[{\citenamefont{Dietrich and Clough}(2019)}]{PhysRevD.100.083005}
\bibinfo{author}{\bibfnamefont{T.}~\bibnamefont{Dietrich}} \bibnamefont{and}
  \bibinfo{author}{\bibfnamefont{K.}~\bibnamefont{Clough}},
  \bibinfo{journal}{Phys. Rev. D} \textbf{\bibinfo{volume}{100}},
  \bibinfo{pages}{083005} (\bibinfo{year}{2019}).

\bibitem[{\citenamefont{Paul et~al.}(2018)\citenamefont{Paul, Majumdar, and
  Modak}}]{Paul2018NeutronSC}
\bibinfo{author}{\bibfnamefont{A.}~\bibnamefont{Paul}},
  \bibinfo{author}{\bibfnamefont{D.}~\bibnamefont{Majumdar}}, \bibnamefont{and}
  \bibinfo{author}{\bibfnamefont{K.~P.} \bibnamefont{Modak}},
  \bibinfo{journal}{Pramana} \textbf{\bibinfo{volume}{92}}, \bibinfo{pages}{1}
  (\bibinfo{year}{2018}).

\bibitem[{\citenamefont{Raffelt}(1996)}]{raffelt1996stars}
\bibinfo{author}{\bibfnamefont{G.~G.} \bibnamefont{Raffelt}},
  \emph{\bibinfo{title}{Stars as laboratories for fundamental physics}}
  (\bibinfo{publisher}{University of Chicago press}, \bibinfo{year}{1996}).

\bibitem[{\citenamefont{Dessert et~al.}(2019)\citenamefont{Dessert, Long, and
  Safdi}}]{PhysRevLett.123.061104}
\bibinfo{author}{\bibfnamefont{C.}~\bibnamefont{Dessert}},
  \bibinfo{author}{\bibfnamefont{A.~J.} \bibnamefont{Long}}, \bibnamefont{and}
  \bibinfo{author}{\bibfnamefont{B.~R.} \bibnamefont{Safdi}},
  \bibinfo{journal}{Phys. Rev. Lett.} \textbf{\bibinfo{volume}{123}},
  \bibinfo{pages}{061104} (\bibinfo{year}{2019}).

\bibitem[{\citenamefont{Fortin and Sinha}(2018)}]{fortin2018constraining}
\bibinfo{author}{\bibfnamefont{J.-F.} \bibnamefont{Fortin}} \bibnamefont{and}
  \bibinfo{author}{\bibfnamefont{K.}~\bibnamefont{Sinha}},
  \bibinfo{journal}{Journal of High Energy Physics}
  \textbf{\bibinfo{volume}{2018}}, \bibinfo{pages}{1} (\bibinfo{year}{2018}).

\bibitem[{\citenamefont{O'Dea et~al.}(2014)\citenamefont{O'Dea, Jenet, Cheng,
  Buu, Beroiz, Asmar, and Armstrong}}]{O'Dea_2014}
\bibinfo{author}{\bibfnamefont{J.~A.} \bibnamefont{O'Dea}},
  \bibinfo{author}{\bibfnamefont{F.~A.} \bibnamefont{Jenet}},
  \bibinfo{author}{\bibfnamefont{T.-H.} \bibnamefont{Cheng}},
  \bibinfo{author}{\bibfnamefont{C.~M.} \bibnamefont{Buu}},
  \bibinfo{author}{\bibfnamefont{M.}~\bibnamefont{Beroiz}},
  \bibinfo{author}{\bibfnamefont{S.~W.} \bibnamefont{Asmar}}, \bibnamefont{and}
  \bibinfo{author}{\bibfnamefont{J.~W.} \bibnamefont{Armstrong}},
  \bibinfo{journal}{The Astronomical Journal} \textbf{\bibinfo{volume}{147}},
  \bibinfo{pages}{100} (\bibinfo{year}{2014}).

\bibitem[{\citenamefont{Xu et~al.}(2021)\citenamefont{Xu, Geng, Wang, Li, and
  Huang}}]{Xu_2021}
\bibinfo{author}{\bibfnamefont{F.}~\bibnamefont{Xu}},
  \bibinfo{author}{\bibfnamefont{J.-J.} \bibnamefont{Geng}},
  \bibinfo{author}{\bibfnamefont{X.}~\bibnamefont{Wang}},
  \bibinfo{author}{\bibfnamefont{L.}~\bibnamefont{Li}}, \bibnamefont{and}
  \bibinfo{author}{\bibfnamefont{Y.-F.} \bibnamefont{Huang}},
  \bibinfo{journal}{Monthly Notices of the Royal Astronomical Society}
  \textbf{\bibinfo{volume}{509}}, \bibinfo{pages}{4916} (\bibinfo{year}{2021}).

\bibitem[{\citenamefont{Ng et~al.}(2007)\citenamefont{Ng, Romani, Brisken,
  Chatterjee, and Kramer}}]{Ng_2007}
\bibinfo{author}{\bibfnamefont{C.-Y.} \bibnamefont{Ng}},
  \bibinfo{author}{\bibfnamefont{R.~W.} \bibnamefont{Romani}},
  \bibinfo{author}{\bibfnamefont{W.~F.} \bibnamefont{Brisken}},
  \bibinfo{author}{\bibfnamefont{S.}~\bibnamefont{Chatterjee}},
  \bibnamefont{and} \bibinfo{author}{\bibfnamefont{M.}~\bibnamefont{Kramer}},
  \bibinfo{journal}{The Astrophysical Journal} \textbf{\bibinfo{volume}{654}},
  \bibinfo{pages}{487} (\bibinfo{year}{2007}).

\end{thebibliography}

\section*{Appendix}
\begin{figure*}[htb!]
\begin{tabular}{cc}
\includegraphics[width=7.5cm]{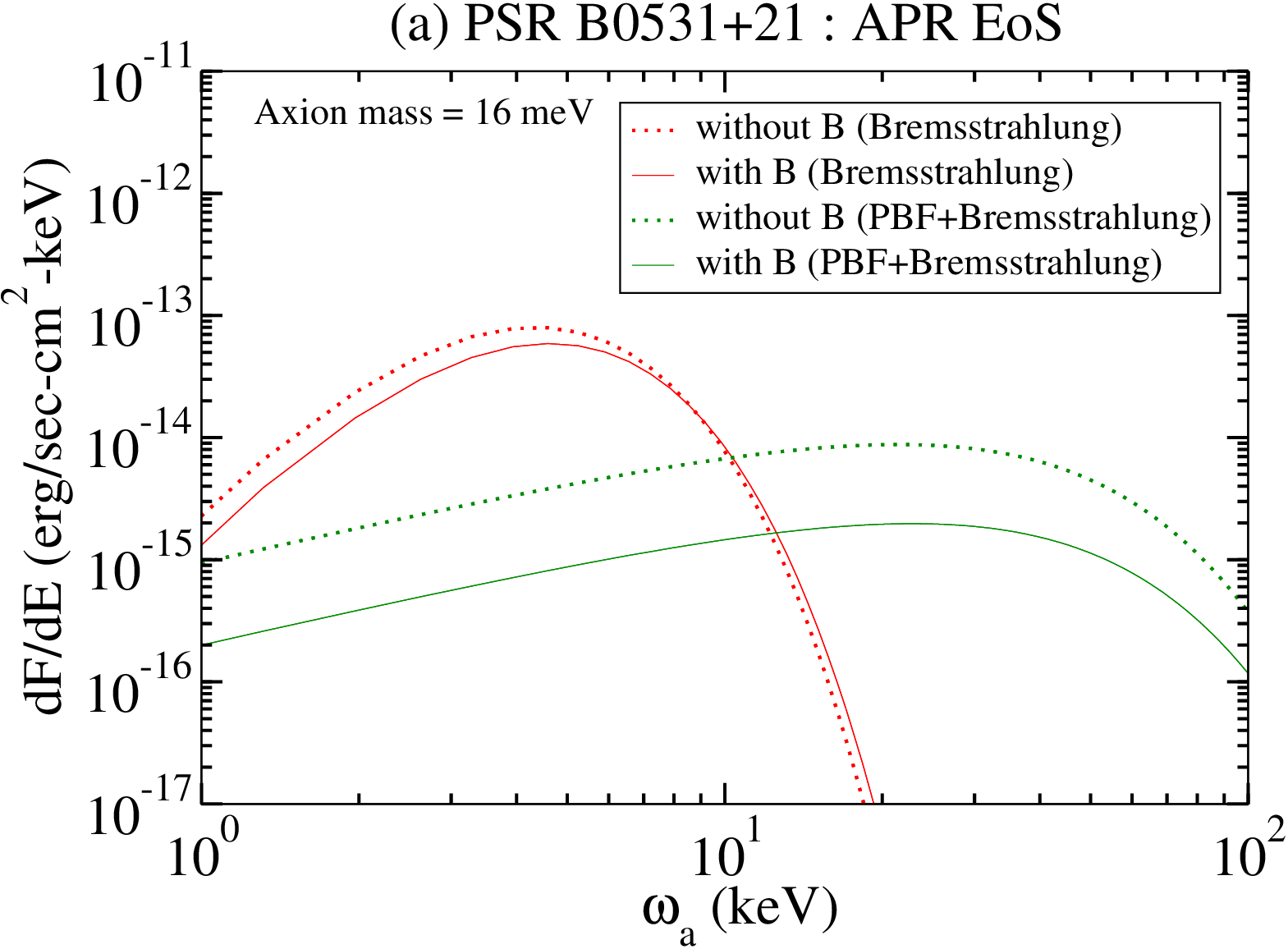}  \includegraphics[width=7.5cm]{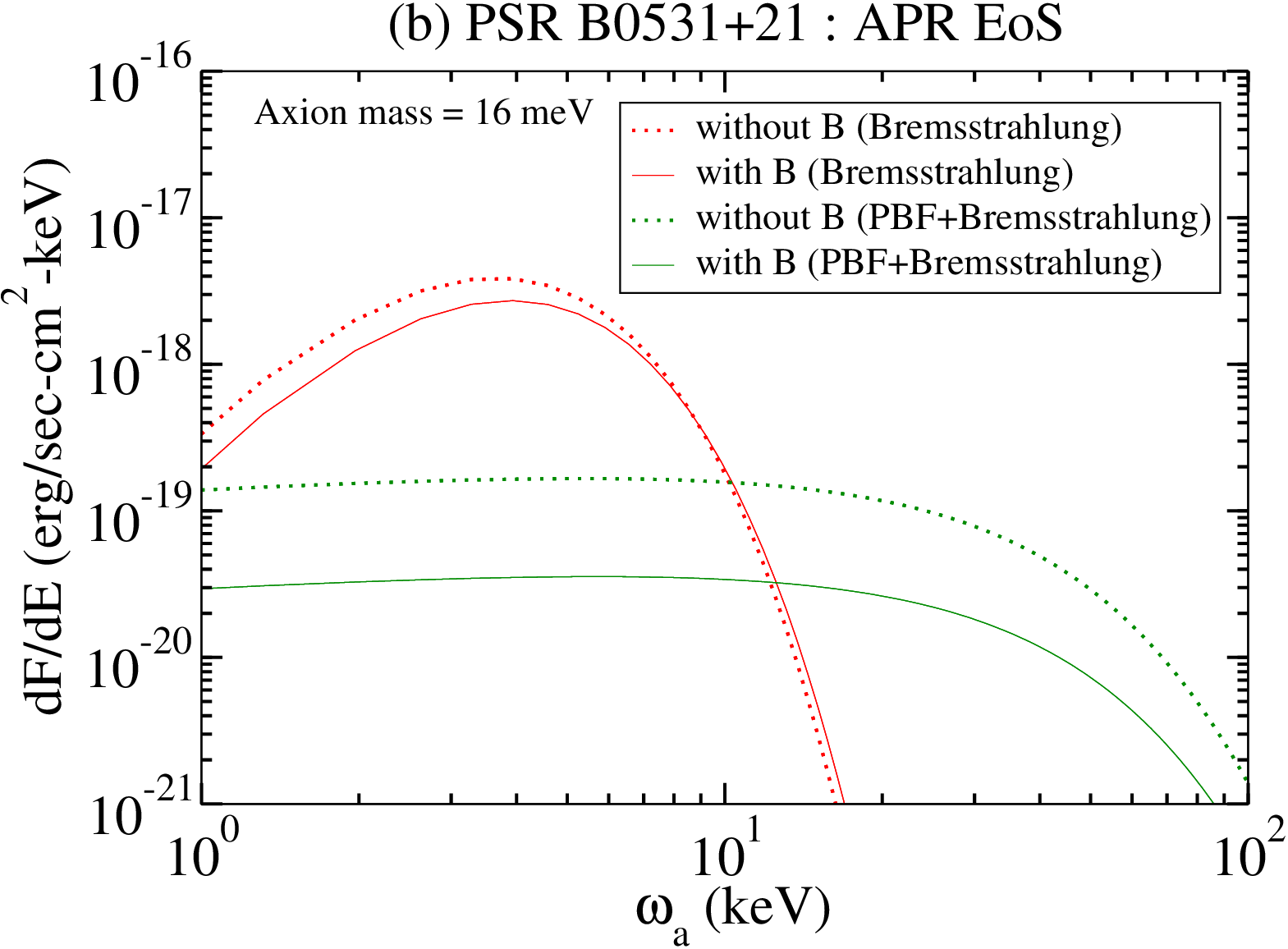}
\end{tabular}
\caption{The variation of the energy spectrum of axions (left) and the axion-converted-photon flux (right) with the axion energies by Bremsstrahlung plus PBF and only Bremsstrahlung process at an axion mass $16$ meV of star PSR B0531+21 for APR EoS with and without including magnetic field.}   
\label{fig:ltimevar5}
\end{figure*}

\begin{figure*}[htb!]
\begin{tabular}{cc}
\includegraphics[width=7.5cm]{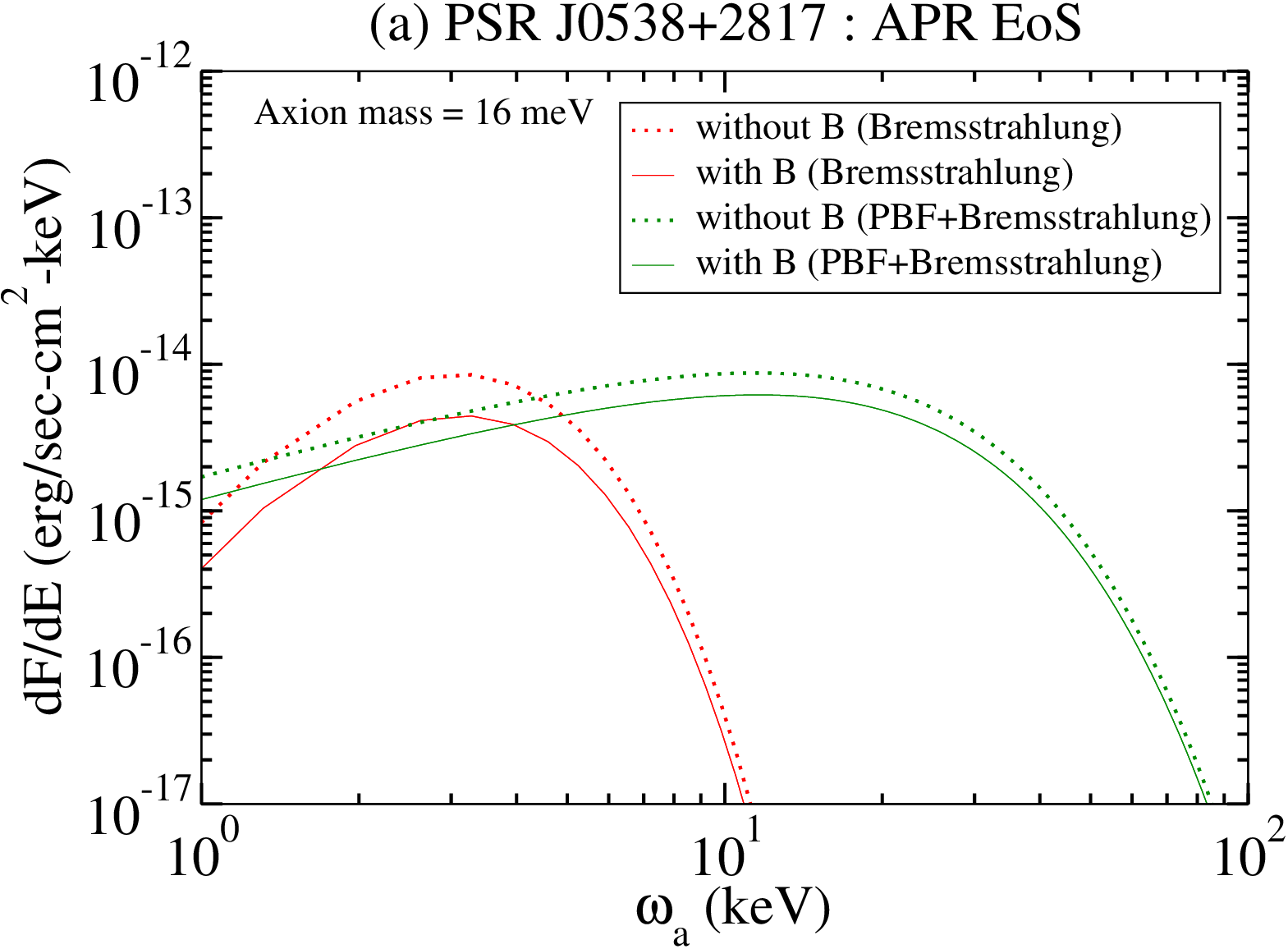} & \includegraphics[width=7.5cm]{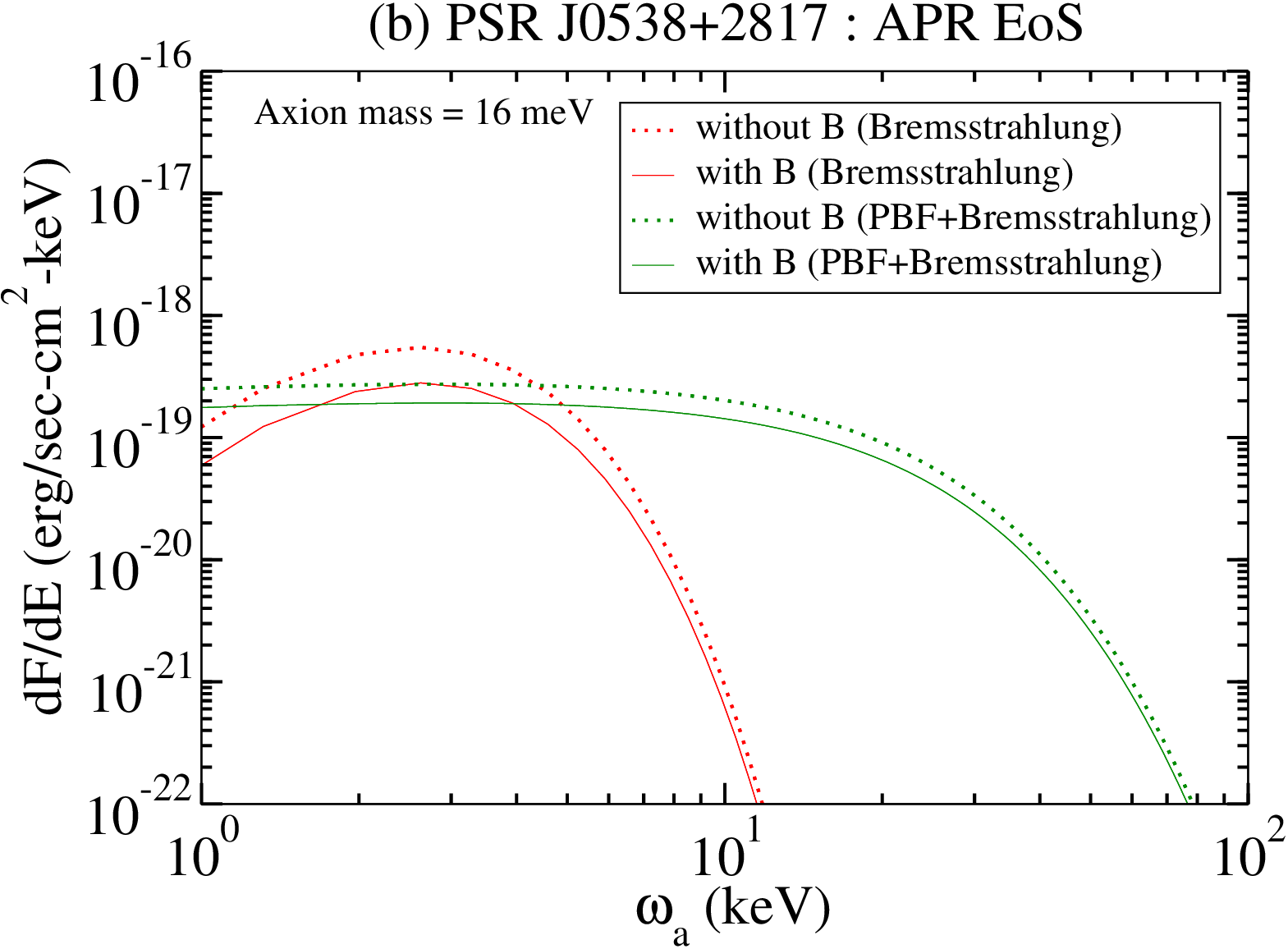}
\end{tabular}
\caption{The variation of the energy spectrum of axions (left) and the axion-converted-photon flux (right) with the axion energies by Bremsstrahlung plus PBF and only Bremsstrahlung process at an axion mass $16$ meV of star PSR J0538+2817 for APR EoS in the presence and absence of magnetic field.}  
\label{fig:ltimevar6}
\end{figure*}

\begin{figure*}[htb!]
\begin{tabular}{cc}
\includegraphics[width=7.5cm]{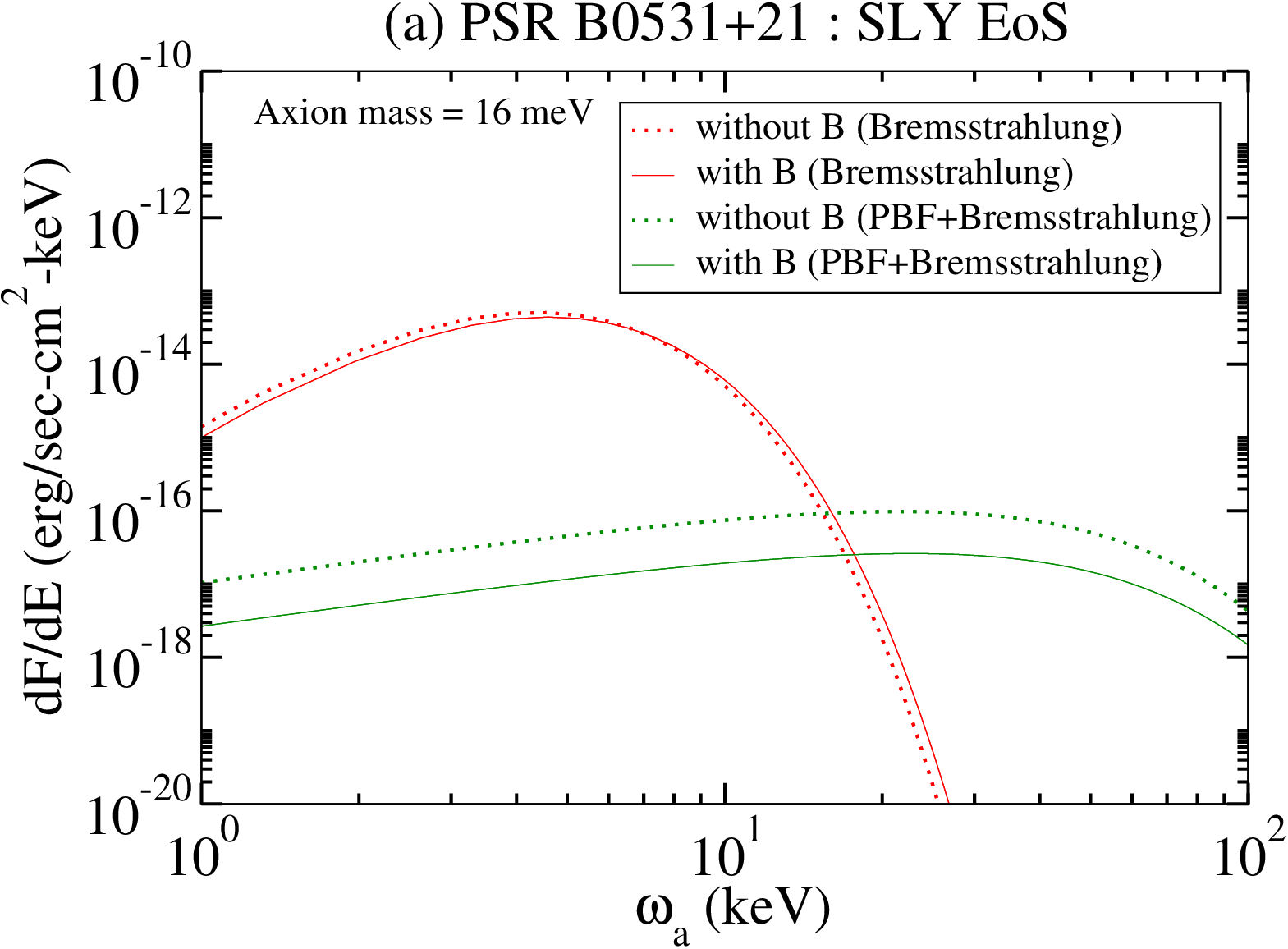} & \includegraphics[width=7.5cm]{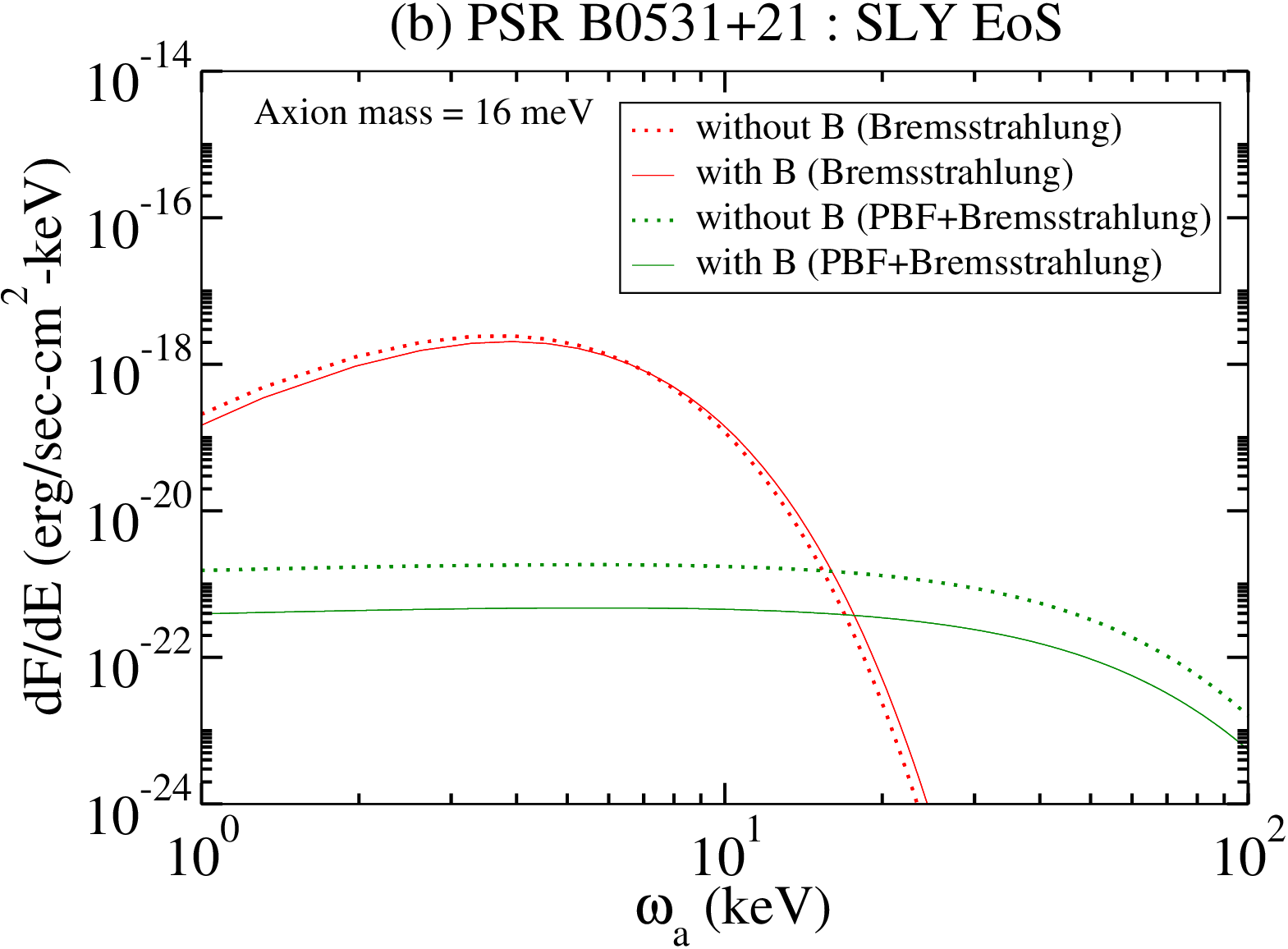}
\end{tabular}
\caption{ The variation of the energy spectrum of axions (left) and the axion-converted-photon flux (right) with the axion energies by Bremsstrahlung plus PBF and only Bremsstrahlung process at an axion mass $16$ meV of star PSR J0531+21 for SLY EoS in the presence and absence of magnetic field.}   
\label{fig:ltimevar7}
\end{figure*}

\begin{figure*}[htb!]
\begin{tabular}{cc}
\includegraphics[width=7.5cm]{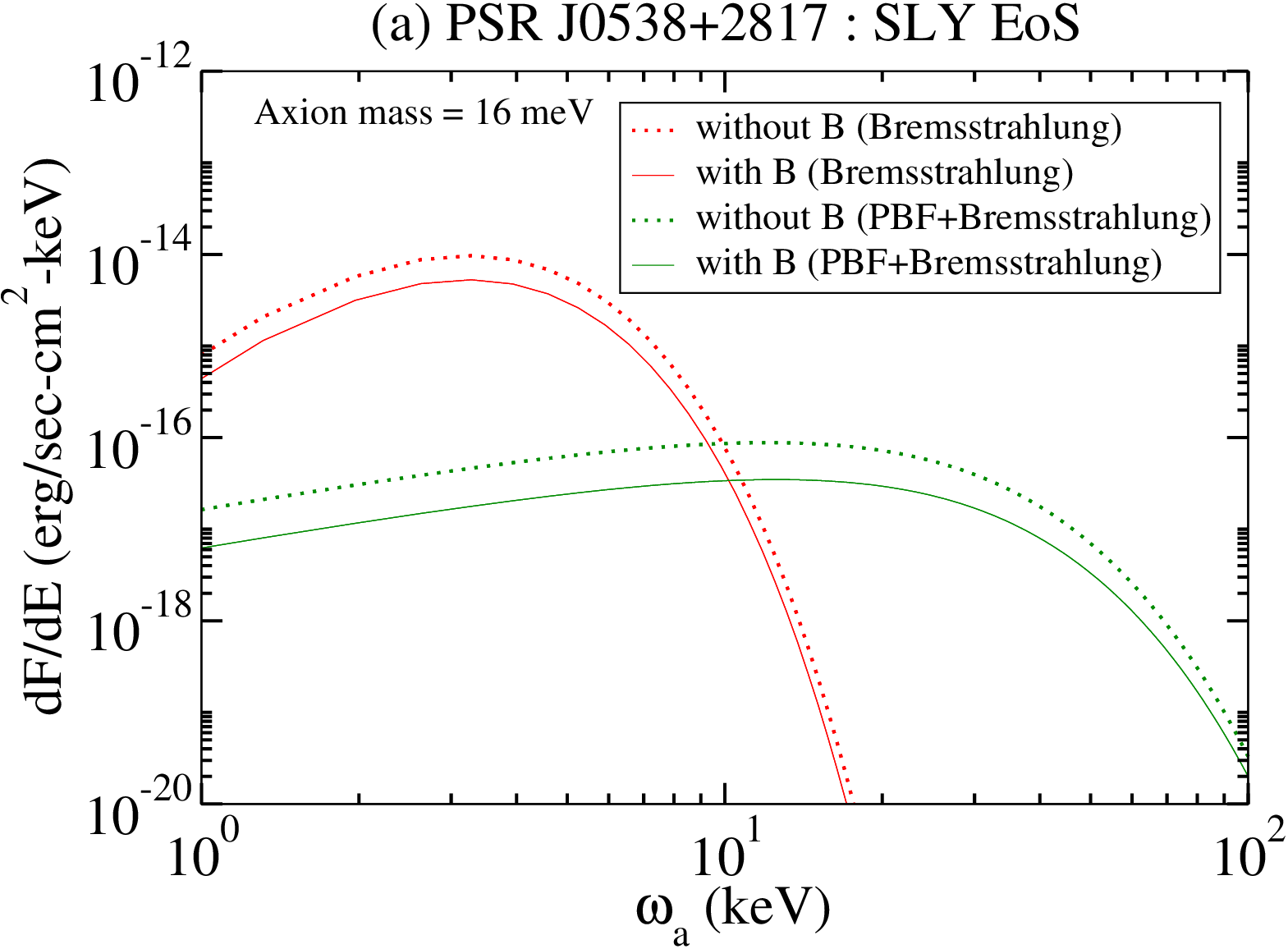} & \includegraphics[width=7.5cm]{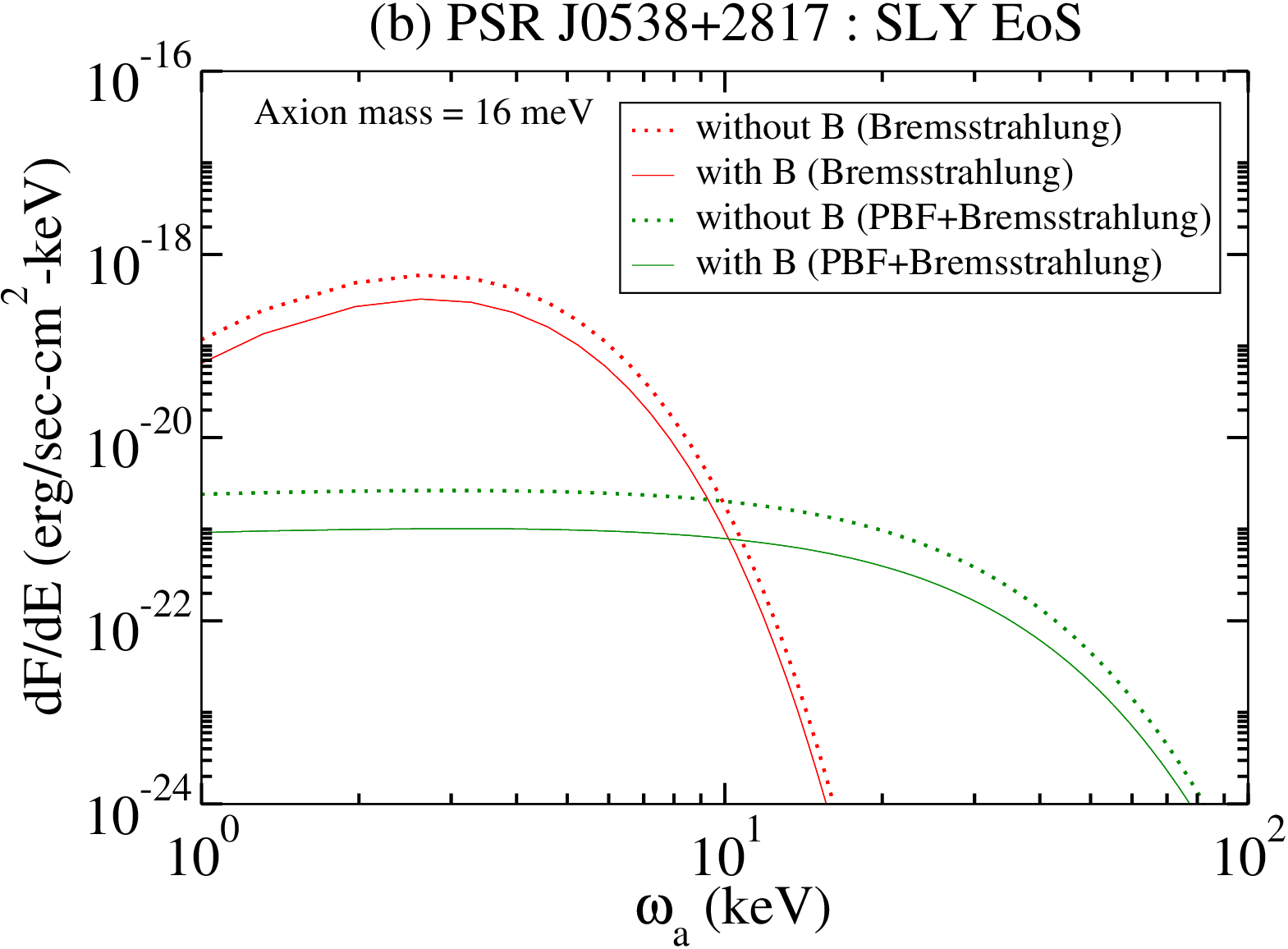}
\end{tabular}
\caption{The variation of the energy spectrum of axions (left) and the axion-converted-photon flux (right) with the axion energies by Bremsstrahlung plus PBF and only Bremsstrahlung process at an axion mass $16$ meV of star PSR J0538+2817 for SLY EoS with and without including the magnetic field.}   
\label{fig:ltimevar8}
\end{figure*}

\subsection*{Variation of the energy spectrum of axions and axion-converted-photon flux (for stars: PSR B0531+21, PSR J0538+2817) at an axion mass $16$ meV for APR \& SLY EoS}       
In this section, we have additionally shown figures for PSR B0531+21~\cite{O'Dea_2014, refId0} and PSR J0538+2817~\cite{Xu_2021,Ng_2007} at an axion mass $16$ meV for two EoSs namely APR and SLY EoSs.
The characteristic age are $9.7\times 10^{2}$ yrs~\cite{O'Dea_2014, refId0} and $3.3\times 10^{4}$ yrs~\cite{Xu_2021,Ng_2007} for PSR B0531+21 and PSR J0538+2817 stars, respectively.

Figure (\ref{fig:ltimevar5}a) and Figure (\ref{fig:ltimevar5}b) show the variation of the energy spectrum of axions and axion-converted-photon flux as a function of axion energies from Bremsstrahlung plus PBF and only Bremsstrahlung process for star PSR B0531+21 at an axion mass $16$ meV for APR EoS, respectively.

In Figure (\ref{fig:ltimevar6}a) and Figure (\ref{fig:ltimevar6}b), we have presented the variation of the energy spectrum of axions and axion-converted-photon flux as a function of axion energies from Bremsstrahlung plus PBF and only Bremsstrahlung process for star PSR J0538+2817 at an axion mass $16$ meV for APR EoS, respectively.

Figure (\ref{fig:ltimevar7}a) and Figure (\ref{fig:ltimevar7}b) depict the variation of the energy spectrum of axions and axion-converted-photon flux as a function of axion energies from Bremsstrahlung plus PBF and only Bremsstrahlung process for star PSR B0531+21 at an axion mass $16$ meV for SLY EoS, respectively.

Figure (\ref{fig:ltimevar8}a) and Figure (\ref{fig:ltimevar8}b) show the variation of the energy spectrum of axions and axion-converted-photon flux as a function of axion energies from Bremsstrahlung plus PBF and only Bremsstrahlung process for star PSR J0538+2817 at an axion mass $16$ meV for SLY EoS, respectively.
\end{document}